\documentclass[a4paper,11pt]{article}
\usepackage[utf8]{inputenc}
\pdfoutput=1 

\usepackage[makeroom]{cancel}
\usepackage{cite}
\usepackage{hyperref}
\usepackage{graphicx}
\usepackage{multirow}
\usepackage[utf8]{inputenc}
\usepackage{graphicx}
\usepackage{tikz}
\usepackage{amsmath}
\usepackage{amssymb}
\usepackage{makecell}
\usepackage{multicol,lipsum,graphicx,float}
\usepackage{color}
\usepackage{caption}
\usetikzlibrary{snakes}
\usepackage{cite}
\usepackage{blindtext}
\usepackage{hyperref}
\usepackage[bottom]{footmisc}
\usepackage[normalem]{ulem}
\usepackage{slashed}
\usepackage{epstopdf}

\newcommand{\be}{\begin{equation}}
\newcommand{\ee}{\end{equation}}
\newcommand{\bea}{\begin{eqnarray}}
\newcommand{\eea}{\end{eqnarray}}
\newcommand{\bav}{\begin{array}{cccc}}

\makeatletter

\makeatletter
\newcommand*{\rom}[1]{\expandafter\@slowromancap\romannumeral #1@}
\makeatother

\begin{document}
\begin{flushright}
HRI-RECAPP-2022-003  
\end{flushright}
\medskip
 \begin{center}
  {\Large\bf Confronting dark fermion with a doubly charged Higgs in the left-right symmetric model}
 \\
 \vskip .5cm
  {
  Shyamashish Dey$^{a,}$\footnote{shyamashishdey@hri.res.in},
  Purusottam Ghosh$^{b,a,}$\footnote{pghoshiitg@gmail.com},
  Santosh Kumar Rai$^{a,}$\footnote{skrai@hri.res.in}\\[3mm]
  {\it{
  $^a$ Regional Centre for Accelerator-based Particle Physics, Harish-Chandra Research Institute,\\
A CI of Homi Bhabha National Institute, 
Chhatnag Road, Jhunsi, Prayagraj 211019, India}\\
{\it$^b$ School Of Physical Sciences, Indian Association for the
                  Cultivation of Science,\\ 2A $\&$ 2B, Raja
                  S.C. Mullick Road, Kolkata 700032, India}
  }}
 \end{center}
 \vskip .5cm
 \begin{abstract}
 We consider a fermionic dark matter (DM) in the left-right symmetric framework by introducing a pair of vector-like (VL) 
 doublets in the particle spectrum. The stability of the DM is ensured through an unbroken $\mathcal{Z}_2$ symmetry. 
 We explore the parameter space of the model compatible with the observed relic density and direct and indirect 
 detection cross sections. The presence of charged dark fermions opens up an interesting possibility for the doubly charged 
 Higgs signal at LHC and ILC. The signal for the doubly charged scalar decaying into the dark sector is analyzed in 
 multilepton final states for a few representative parameter choices consistent with DM observations.   
 

 \end{abstract}
\section{Introduction}
\label{sec:intro}
The new era of search for particles or hints of new physics has been facing its challenges since the discovery of the 
Higgs boson\cite{Chatrchyan:2012xdj,Aad:2012tfa}. While direct searches at experiments at the Large hadron Collider (LHC) have not revealed anything 
new yet, the irrevocable hints for the existence of tiny neutrino mass and dark matter (DM) in the Universe has 
led to new efforts in non-collider experiments to establish signals for new phenomenon which may point to an extension of the
Standard Model (SM) and more crucially provide information about DM. 
Left-right symmetric models (LRSM) \cite{PhysRevD.11.2558,PhysRevD.12.1502,PhysRevD.23.165,PhysRevD.44.837,1998,Zhang:2007da} are one of the most well motivated and widely 
studied extensions of the SM as it is able to address several phenomena which are not very well understood in the framework of SM, be it 
Parity or the tiny neutrino masses which have their origin very naturally in the model. 
Parity symmetry (P) prevents one from writing P and Charge-Parity (CP) violating terms in the Quantum Chromodynamics (QCD) Lagrangian thus resolving the strong CP problem naturally without the need to introduce a global Peccei-Quinn symmetry. The gauge structure of these models force us to have a right-handed neutrino in the lepton multiplet. This right-handed neutrino can generate a light neutrino mass through the seesaw mechanism. Moreover left-right symmetry is also favored by many  scenarios of 
gauge-unification.

We have also established that a significant part of the Universe ( $\sim ~ 26\%$ of total 
energy budget) which is made of non-luminous, non-baryonic matter and interacts via gravity, popularly known as dark matter (DM) is one of the most fundamental concerns in current 
days particle physics and cosmology\cite{Spergel:2006hy,Roszkowski:2017nbc}. There are several astrophysical evidences like the rotational curve of the galaxy, bullet 
cluster, gravitational lensing, anisotropy in cosmic microwave background (CMB), etc.\cite{Rubin:1970zza,Hu:2001bc,Bertone:2004pz,Hinshaw:2012aka}, which indicate the existence of  stable DM in the present Universe.  
However we do not know much about it, the only information so far we know about DM is its relic abundance measured by 
WMAP and PLANK to be $\Omega_{\rm DM} h^2=0.120\pm 0.001$ \cite{Planck:2018vyg}. Apart from this the nature of the DM e.g. its spin, 
non-gravitational interaction and mass still remain an open question. Depending on the production mechanism in the early 
Universe, DM particles are broadly classified based on its interaction strengths into WIMP \cite{Kolb:1990vq,1996,2005}, SIMP\cite{PhysRevLett.113.171301}, FIMP\cite{2010FI}, etc. Among these 
the weakly interacting massive particle (WIMP) is one of the popular DM candidates due to its detection possibility in 
direct (XENON1T\cite{XENON:2018voc}, PANDAX 4T \cite{2021} etc.), indirect (FERMI LAT and MAGIC \cite{Ackermann_2015,2016}) and collider (LHC\cite{Abercrombie:2015wmb} and ILC) search experiments. 
The weak interaction between WIMP like DM and the visible sector can lead to thermal equilibrium in the early universe at  temperatures above the mass scale and freezes out from the thermal bath when the temperature falls below its mass \cite{Kolb:1990vq}. 
Many theoretical extension of SM have been formulated to accommodate the particle nature of DM over the past several decades.  A novel 
possibility would be its existence in the framework of left-right symmetric theories. While 
the supersymmetric versions of the LRSM \cite{2017,2019} naturally incorporate a WIMP (through conserved R-parity) in the 
form of the lightest supersymmetric particle, it is rather challenging to invoke a DM candidate in the minimal setup 
of LRSM. Therefore a natural extension could be to include new particles in the set-up in the non-supersymmetric model\cite{2008,Guo_2010,2015,2010,2012,2012WD,Bahrami_2017}.   
To accommodate DM in LRSM, we consider two vector-like fermion doublets, 
$\psi_1$ and $\psi_2$ which belong to the $SU(2)_L$ and $SU(2)_R$ respectively 
and both carry a discrete $\mathcal{Z}_2$ charge of $-1$ to ensure the stability of the lightest state. This induced dark sector in the LRSM would have enhanced interactions including its participation in both left-handed and right-handed charged currents as well as neutral currents mediated by the electroweak (EW) and new heavy gauge  bosons and the numerous scalars present in the model. Thus one expects the DM phenomenology to be very interesting and illuminating in several regions of the model parameter space. A crucial hint of LRSM is the presence of two doubly charged scalars which independently couple to the gauge bosons of $SU(2)_L$ and $SU(2)_R$ and therefore have different production strengths. With the inclusion of the dark sector which contains new fermions, the search for these exotics can be a lot different at collider experiments. We shall consider this interesting possibility in our work by studying the 
signal for the doubly charged scalars at LHC as well as the proposed International
Linear Collider (ILC) \cite{Ghosh:2017pxl,Ashanujjaman:2021txz,Agrawal:2018pci,Barman:2019tuo} while highlighting the DM phenomenology of the model in 
more detail.   

The rest of our work is organized as follows. We first briefly describe the proposed model in Section~\ref{sec:model}. In 
Section~\ref{sec:const}, we discuss possible theoretical and experimental constraints on the model parameters which 
would be applicable for our analysis. In Section~\ref{sec:dmpheno} we discuss the DM phenomenology where we demonstrate 
the allowed parameter space compatible with current relic density, direct and indirect search constraints. The collider signature of the doubly charged Higgs in the presence of DM  at LHC and ILC  in this setup are discussed in Section~\ref{sec:coll} and ~\ref{sec:coll1} respectively. Finally, we summarize our findings in Section~\ref{sec:concl}.

\section{Left-Right Model with Dark Doublets}
\label{sec:model}
The model is an extension of the popular left-right symmetry model~(LRSM) \cite{PhysRevD.11.2558,PhysRevD.12.1502} 
where the only addition is the introduction of two vector-like~(VL) fermion doublets~\cite{Bahrami_2017} in the particle spectrum, {\it viz.} 
$\psi_1^T= \left(\begin{matrix} \psi_1^0 & \psi_1^-  \end{matrix}\right)_{L,R}$ and  
$\psi_2^T= \left(\begin{matrix} \psi_2^0 & \psi_2^-  \end{matrix}\right)_{L,R}.$ 
The new fermions are a replication of the lepton doublets with the only difference being the VL character such that 
$\psi_1$ is a VL doublet under $SU(2)_L \otimes U(1)_{B-L}$ while $\psi_2$ is a VL doublet under $SU(2)_R \otimes U(1)_{B-L}$.  
A discrete $\mathcal{Z}_2$ symmetry is included in the model under which the new VL doublets are odd, 
that makes the lightest neutral component of each VL doublet stable. However, after gauge symmetry is 
spontaneously broken, the neutral components of $\psi_1$ and $\psi_2$ will mix leading to only one neutral component 
to remain stable and act as the DM candidate in the model. Note that all the remaining fields in the model are even under the
$\mathcal{Z}_2$ symmetry.
The charge assignment of all fermions including dark fields and scalar fields in the extended LRSM are tabulated in 
Table\ref{tab:tab1} and Table\ref{tab:tab2}, respectively. It is important to note here that the VL nature of the new fields 
do not lead to any chiral anomaly, keeping the model anomaly free. 

\begin{table}[h]
\centering
\resizebox{14cm}{!}{
 \begin{tabular}{|c|c|}
\hline 
{Fermion Fields}&  { $\underbrace{ SU(3)_C \otimes SU(2)_R \otimes SU(2)_L \otimes U(1)_{B-L}}$ $\otimes  \mathcal{Z}_2  $} \\ \hline
   $Q_L=\left(\begin{matrix} u \\ d \end{matrix}\right)_L$ & ~~~~~3 ~~~~~~~~~~~1~~~~~~~~~~~~2~~~~~~~~~~$\frac{1}{3}$~~~~~~~~~~~+\\
 \hline
$Q_R=\left(\begin{matrix} u \\ d \end{matrix}\right)_R$ & ~~~~~3 ~~~~~~~~~~~2~~~~~~~~~~~~1~~~~~~~~~~$\frac{1}{3}$~~~~~~~~~~~+\\
\hline
$L_L=\left(\begin{matrix} \nu_\ell \\ \ell  \end{matrix}\right)_L$ & ~~~~~1 ~~~~~~~~~~~1~~~~~~~~~~~~2~~~~~~~~~~-1~~~~~~~~~~~+\\
\hline
$L_R=\left(\begin{matrix} \nu_\ell \\ \ell  \end{matrix}\right)_R$ & ~~~~~1 ~~~~~~~~~~~2~~~~~~~~~~~~1~~~~~~~~~~-1~~~~~~~~~~~+\\
\hline
\hline
$\psi_1: \left(\begin{matrix} \psi_1^0 \\ \psi_1^-  \end{matrix}\right)_L,~\left(\begin{matrix} \psi_1^0 \\ \psi_1^-  \end{matrix}\right)_R$ & ~~~~~1 ~~~~~~~~~~~1~~~~~~~~~~~~2~~~~~~~~~~-1~~~~~~~~~~~-\\
\hline
$\psi_2: \left(\begin{matrix} \psi_2^0 \\ \psi_2^-  \end{matrix}\right)_L,\left(\begin{matrix} \psi_2^0 \\ \psi_2^-  \end{matrix}\right)_R$ & ~~~~~1 ~~~~~~~~~~~2~~~~~~~~~~~~1~~~~~~~~~~-1~~~~~~~~~~~-\\
\hline
\end{tabular}
}
\caption{\it Charge assignment of fermion fields under the left-right (LR) gauge symmetry 
$\mathcal{G}~\equiv ~SU(3)_C \otimes SU(2)_R \otimes SU(2)_L \otimes U(1)_{B-L}$ augmented with a 
discrete $\mathcal{Z}_2$ symmetry.  The electric charge is defined as $Q= I_{3L} +I_{3R}  +\frac{B-L}{2}$, where  
$I_3$ is the third component of isospin and $B$ and $L$ represent the baryon and lepton numbers respectively. }
    \label{tab:tab1}
\end{table}

\begin{table}[h]
\centering
\resizebox{14cm}{!}{
 \begin{tabular}{|c|c|}
\hline 
{Scalar Fields}&  { $\underbrace{ SU(3)_C \otimes SU(2)_R \otimes SU(2)_L \otimes U(1)_{B-L}}$ $\otimes  \mathcal{Z}_2  $} \\ \hline
   $\Phi=\left(\begin{matrix} \phi_1^0 & \phi_2^{+} \\ \phi_1^{-} & \phi_2^0 \end{matrix}\right)$ & ~~~~~1 ~~~~~~~~~~~2~~~~~~~~~~~~2~~~~~~~~~~0~~~~~~~~~~~+\\
   \hline
   $\Delta_L=\left(\begin{matrix} \frac{\Delta^+}{\sqrt{2}} & \Delta^{++} \\ \Delta^0 & -\frac{\Delta^+}{\sqrt{2}} \end{matrix}\right)_L$ & ~~~~~1 ~~~~~~~~~~~1~~~~~~~~~~~~3~~~~~~~~~~2~~~~~~~~~~~+\\
   \hline
   $\Delta_R=\left(\begin{matrix} \frac{\Delta^+}{\sqrt{2}} & \Delta^{++} \\ \Delta^0 & -\frac{\Delta^+}{\sqrt{2}} \end{matrix}\right)_R$ & ~~~~~1 ~~~~~~~~~~~3~~~~~~~~~~~~1~~~~~~~~~~2~~~~~~~~~~~+\\
   \hline
\end{tabular}
}
\caption{\it Charge assignment of scalar fields under the extended gauge group 
$\mathcal{G}~\equiv ~SU(3)_C~\otimes~SU(2)_R \otimes~SU(2)_L \otimes U(1)_{B-L} \otimes \mathcal{Z}_2$.}
    \label{tab:tab2}
\end{table}

The Lagrangian for this model can be written in two separate components:
\bea
\label{eq:lag}
\mathcal{L}=\mathcal{L}^{\rm LRSM} + \mathcal{L}^{\rm FDM}~~~ . 
\eea

The first part of the Lagrangian in Eqn.\ref{eq:lag} represents the $\mathcal{Z}_2$ even LRSM Lagrangian 
while the second part of the Lagrangian represents the proposed dark sector in this setup. 
Since the left-right symmetric Lagrangian, $\mathcal{L}^{\rm LRSM}$ has been comprehensively studied in the literature
we do not discuss it in great detail here and refer the readers to Refs.~\cite{PhysRevD.11.2558,PhysRevD.12.1502,PhysRevD.23.165,PhysRevD.44.837,1998,Zhang:2007da,Bonilla_2017,Chakrabortty:2016wkl,Dev:2016dja,Chauhan:2018uuy,BhupalDev:2018xya,Chauhan:2019fji}. 
Our main motivation in this work is to study the phenomenology of DM in the extended LRSM setup. We, therefore, restrict ourselves to only the relevant part of the Lagrangian 
of LRSM for our analysis.

\subsubsection*{\underline{\bf The scalar potential of LRSM} }

The most general $\mathcal{C} \, (\rm charge)-\mathcal{P} \, (\rm parity)$ invariant scalar potential in the LRSM, 
invariant under the gauge symmetry $SU(3)_C \otimes SU(2)_R \otimes SU(2)_L \otimes U(1)_{B-L}$ reads 
as \cite{PhysRevD.44.837,Bonilla_2017} 
\begin{eqnarray}
\label{LRSM}
V_{\rm LRSM} &=& - \mu_1^2 {\rm Tr} (\Phi^{\dag} \Phi) - \mu_2^2
\left[ {\rm Tr} (\tilde{\Phi} \Phi^{\dag}) + {\rm Tr} (\tilde{\Phi}^{\dag} \Phi) \right]
- \mu_3^2 \left[ {\rm Tr} (\Delta_L \Delta_L^{\dag}) + {\rm Tr} (\Delta_R
\Delta_R^{\dag}) \right] \nonumber
\\
&&+ \lambda_1 \left[ {\rm Tr} (\Phi^{\dag} \Phi) \right]^2 + \lambda_2 \left\{ \left[
{\rm Tr} (\tilde{\Phi} \Phi^{\dag}) \right]^2 + \left[ {\rm Tr}
(\tilde{\Phi}^{\dag} \Phi) \right]^2 \right\}+
\lambda_3 {\rm Tr} (\tilde{\Phi} \Phi^{\dag}) {\rm Tr} (\tilde{\Phi}^{\dag} \Phi)\nonumber \\
&& +
\lambda_4 {\rm Tr} (\Phi^{\dag} \Phi) \left[ {\rm Tr} (\tilde{\Phi} \Phi^{\dag}) + {\rm
Tr}
(\tilde{\Phi}^{\dag} \Phi) \right]\nonumber \\
&& + \rho_1 \left\{ \left[ {\rm Tr} (\Delta_L \Delta_L^{\dag}) \right]^2 + \left[ {\rm
Tr} (\Delta_R \Delta_R^{\dag}) \right]^2 \right\} \nonumber \\ 
&& + \rho_2 \left[ {\rm
Tr} (\Delta_L \Delta_L) {\rm Tr} (\Delta_L^{\dag} \Delta_L^{\dag}) + {\rm Tr} (\Delta_R
\Delta_R) {\rm Tr} (\Delta_R^{\dag} \Delta_R^{\dag}) \right] \nonumber
\\
&&+ \rho_3 {\rm Tr} (\Delta_L \Delta_L^{\dag}) {\rm Tr} (\Delta_R \Delta_R^{\dag})+
\rho_4 \left[ {\rm Tr} (\Delta_L \Delta_L) {\rm Tr} (\Delta_R^{\dag} \Delta_R^{\dag}) +
{\rm Tr} (\Delta_L^{\dag} \Delta_L^{\dag}) {\rm Tr} (\Delta_R
\Delta_R) \right]  \nonumber \\
&&+ \alpha_1 {\rm Tr} (\Phi^{\dag} \Phi) \left[ {\rm Tr} (\Delta_L \Delta_L^{\dag}) +
{\rm Tr} (\Delta_R \Delta_R^{\dag})  \right] \nonumber
\\
&&+ \left\{ \alpha_2 e^{i \delta_2} \left[ {\rm Tr} (\tilde{\Phi} \Phi^{\dag}) {\rm Tr}
(\Delta_L \Delta_L^{\dag}) + {\rm Tr} (\tilde{\Phi}^{\dag} \Phi) {\rm Tr} (\Delta_R
\Delta_R^{\dag}) \right] + {\rm h.c.}\right\} \nonumber
\\
&&+ \alpha_3 \left[ {\rm Tr}(\Phi \Phi^{\dag} \Delta_L \Delta_L^{\dag}) + {\rm
Tr}(\Phi^{\dag} \Phi \Delta_R \Delta_R^{\dag}) \right] + \beta_1 \left[ {\rm Tr}(\Phi
\Delta_R \Phi^{\dag} \Delta_L^{\dag}) +
{\rm Tr}(\Phi^{\dag} \Delta_L \Phi \Delta_R^{\dag}) \right] \nonumber \\
&&+ \beta_2 \left[ {\rm Tr}(\tilde{\Phi} \Delta_R \Phi^{\dag} \Delta_L^{\dag}) + {\rm
Tr}(\tilde{\Phi}^{\dag} \Delta_L \Phi \Delta_R^{\dag}) \right] + \beta_3 \left[ {\rm
Tr}(\Phi \Delta_R \tilde{\Phi}^{\dag} \Delta_L^{\dag}) + {\rm Tr}(\Phi^{\dag} \Delta_L
\tilde{\Phi} \Delta_R^{\dag}) \right] \,. \nonumber \\
\end{eqnarray}

The neutral scalar fields in the multiplets acquire non-zero vacuum expectation value~(vev)
leading to the symmetry breaking pattern: $SU(3)_C \otimes SU(2)_R \otimes SU(2)_L \otimes U(1)_{B-L} \rightarrow SU(3)_C \otimes SU(2)_L \otimes U(1)_Y  \rightarrow SU(3)_C \otimes U(1)_Q$ \cite{PhysRevD.44.837,Bonilla_2017} as: 
\bea
\langle \Delta_R^0 \rangle =\frac{v_R}{\sqrt2},~~\langle \Delta_L^0 \rangle =\frac{v_L}{\sqrt2}, ~~~
\langle \phi_1^0 \rangle =\frac{v_1}{\sqrt2},~~\langle \phi_2^0 \rangle =\frac{v_2}{\sqrt2}~.
\eea
The vevs of the scalar fields are parametrized as: 
\bea
v_1=v \cos\beta,~~v_2=v \sin\beta,~~~ \tan\beta=\frac{v_2}{v_1}~,
\eea
where $v$ can be identified as SM vev and is given by $v=\sqrt{v_1^2+ v_2^2}=246$ GeV, 
with $v_L \ll v~(v_2 \ll v_1) \ll v_R$. The parity symmetry implies $g_L=g_R$. Without any 
loss in generality of the BSM phenomenology, the above scalar potential can be simplified 
by considering  $\beta_i=0,~\alpha_2=0 ,~\lambda_4 =0~{\rm and} ~v_L \to 0$. Under these 
assumptions, the masses and corresponding eigenstates of the scalar and the gauge 
bosons are tabulated in Table\ref{tab:tabES}. Here we have categorized the different scalar 
types according to their $\mathcal{CP}$ properties and electric charge without going into 
the details, as these have been discussed in the literature \cite{Zhang:2007da}. 

We shall work in the limit of parameter choices in the scalar sector which helps us with a
favorable DM phenomenology. This will be obvious when we calculate the observables relevant for DM abundance and its correlation with the scalar spectrum.
\begin{table}[H]
\resizebox{\linewidth}{!}{
 \begin{tabular}{|c|c|}
\hline
 {\it Physical State}  &     {\it Mass}  \\   \hline \hline
  $h\simeq\sqrt{2}~ {\rm Re}[{\phi_1^0}^* + \frac{v_2}{v_1} e^{-i\alpha}\phi_2^0]$ & $M_h \simeq \sqrt{2 \lambda_1 v_1^2}~(\equiv 125 {\rm~GeV,~ SM~ like~ Higgs})$  \\ \hline
    $H\simeq \sqrt{2}~ {\rm Re}[-\frac{v_2}{v_1} e^{i\alpha} {\phi_1^0}^* +  \phi_2^0]$ & $M_{H} \simeq \sqrt{2 v^2 (2 \lambda_2 +\lambda_3)+\frac{1}{2} \alpha_3 v_R^2}$  \\ \hline
    $H_R \simeq \sqrt{2}~ {\rm Re}[\Delta_R^0]$ & $M_{H_R} \simeq \sqrt{2 \rho_1  v_R^2}$  \\ \hline
    $H_L \simeq \sqrt{2}~ {\rm Re}[\Delta_L^0]$ & $M_{H_L} \simeq \sqrt{\frac{1}{2}(\rho_3 -2 \rho_1) v_R^2}$  \\ \hline \hline   
$A \simeq\sqrt{2}~ {\rm Im}[-\frac{v_2}{v_1} e^{i\alpha} {\phi_1^0}^* +  \phi_2^0]$ & $M_A \simeq \sqrt{2 v^2 (2 \lambda_2 -\lambda_3)+ \frac{1}{2}\alpha_3 v_R^2}$  \\ \hline
    $A_L \simeq \sqrt{2}~ {\rm Im}[\Delta_L^0]$ & $M_{A_L} \simeq \sqrt{\frac{1}{2}(\rho_3 -2 \rho_1) v_R^2}$  \\ \hline  \hline 
    $H^+ \simeq \phi_2^+ + \frac{v_2}{v_1} e^{i\alpha} \phi_1^+ + \frac{1}{\sqrt{2}} \frac{v_1}{v_R} \Delta_R^+ $ & $M_{H^+} \simeq \sqrt{\frac{1}{4} \alpha_3 (v^2 + 2 v_R^2)}$  \\ \hline      
    $H_L^+ \simeq \Delta_L^+ $ & $M_{H_L^+} \simeq \sqrt{\frac{1}{2}(\rho_3 -2 \rho_1) v_R^2}$  \\ \hline  \hline     
 $H_L^{++} \simeq \Delta_L^{++} $ & $M_{H_L^{++}} \simeq \sqrt{\frac{1}{2}(\rho_3 -2 \rho_1) v_R^2+\frac{1}{2} \alpha_3 v^2 }$  \\ \hline  
 $H_R^{++} \simeq \Delta_R^{++} $ & $M_{H_R^{++}} \simeq \sqrt{2 \rho_2 v_R^2 +\frac{1}{2} \alpha_3 v^2}$  \\ \hline \hline       
$W^\pm = \frac{1}{\sqrt{2}}\big(W_L^1 \mp i W_L^2 \big) $ & $M_{W} \simeq \sqrt{\frac{g_L^2}{2} v^2} ~~(\simeq 80 {\rm GeV})$ (SM like) \\ \hline  
$W_R^\pm = \frac{1}{\sqrt{2}}\big(W_R^1 \mp i W_R^2 \big) $ & $M_{W_R} \simeq \sqrt{\frac{g_R^2}{2} v_R^2} $  \\ \hline  
$Z = -c_W W_L^3+ s_W s_Y W_R^3 + s_W c_Y B $ & $M_{Z}  = \frac{M_W}{c_W} ~~(\simeq 91 {\rm GeV}) $ (SM like ) \\ \hline  
$Z_R =  -c_Y W_R^3 + s_Y B $ & $M_{Z_R} \simeq \sqrt{\big(g_R^2 + g_{BL}^2\big)v_R^2} $  \\ \hline
$\gamma = s_W W_L^3+ c_W s_Y W_R^3 + c_W c_Y B $ & $M_{\gamma}  = 0  $ (SM like) \\ \hline  
\hline
\end{tabular}
}
\caption{\it Physical states and their masses are tabulated under the assumption 
$v_L \ll v~(v_2 \ll v_1) \ll v_R$ where $v=\sqrt{v_1^2+v_2^2}$. The mixing angles are identified as: $s_W=\sin\theta_W=\frac{g_{\rm BL}}{\sqrt{g^2+2 g_{\rm BL}^2}},~c_W=\cos\theta_W=\frac{\sqrt{g^2+ g_{\rm BL}^2}}{\sqrt{g^2+2 g_{\rm BL}^2}},~s_Y=\sin\theta_Y=\frac{g_{\rm BL}}{\sqrt{g^2+ g_{\rm BL}^2}}$ and $c_Y=\cos\theta_Y=\frac{g}{\sqrt{g^2+ g_{\rm BL}^2}}$ where $g=g_L=g_R$.}
    \label{tab:tabES}
\end{table}

\subsubsection*{\underline{\bf Dark Sector Lagrangian:}}

In this set-up the lightest neutral state which is an admixture of the neutral component of  
$SU(2)_L$ fermion doublet ($\psi_1$) and $SU(2)_R$ fermion doublet 
($\psi_2$) after symmetry breaking, gives rise to a viable candidate of DM due to 
the unbroken discrete symmetry $\mathcal{Z}_2$. The  Lagrangian of the dark sector in the extended LRSM can be written as,
\bea
\mathcal{L}^{\rm FDM} &=& \overline{\psi_1} \Big[~i \gamma^\mu\Big(\partial_\mu -i g_L \frac{\sigma^{a}}{2} W_{L\mu}^a -i g_{\rm BL} \frac{Y_{\rm BL}}{2} B_\mu\Big) - M_{L}\Big]\psi_1  \nonumber \\
&& + \overline{\psi_2}\Big[ ~i \gamma^\mu\Big(\partial_\mu -i g_R \frac{\sigma^{a}}{2} W_{R\mu}^a -i g_{\rm BL} \frac{Y_{\rm BL}}{2} B_\mu\Big) -M_{R} \Big]\psi_2  \nonumber \\ 
&&-\Big\{ \Big(Y_1 \overline{\psi_1} \Phi \psi_2 + Y_2 \overline{\psi_1} \tilde{\Phi} \psi_2 \Big) + h.c \Big\} \nonumber \\
&& - y_L \Big( \overline{\psi_1} \Delta_L^\dagger i\sigma_2 \psi_1^c + h.c \Big)- y_R \Big( \overline{\psi_2} \Delta_R^\dagger i\sigma_2 \psi_2^c + h.c \Big).
\label{eq:lag1}
\eea

A very similar extension of the LRSM was studied in Ref.\cite{Bahrami_2017}. Although our model 
is the same with similar particle content, our study differs in how the 
Yukawa structure of the model has been assigned. We chose a uniform Yukawa 
coupling strength between the VL fermion doublets with the scalar sector. It is worth noting 
that the earlier work treats the individual left-handed and right-handed projections of the 
$SU(2)_L$ VL doublet differently and therefore invokes more Yukawa couplings 
in the model. A similar structure is assumed for the $SU(2)_R$ VL doublet. While the
choice is a viable phenomenological option as it does not alter the VL doublet's 
gauge interactions, it does give additional freedom to treat the dark sector 
fermions independently. We consider 
a more natural and aesthetic approach by choosing the Yukawa couplings to be 
identical and therefore allows us to correlate the nature of the DM based on its composition. 
We note that this also allows a more interesting signature for the doubly charged scalar
in the model. Our choice also leads to a very different DM analysis and allowed parameter
space for the model.

In the above Lagrangian (in Eqn. \ref{eq:lag1}) $M_L$ and $M_R$ are the bare masses of $\psi_1$ and $\psi_2$ respectively and $Y_1$ and $Y_2$ are the Dirac Yukawa couplings. The other two Majorana type Yukawa couplings, $y_L$ and $y_R$ are responsible for generating the mass splitting between the physical eigenstates after mixing. In addition the non-zero values of both $y_L$ and $y_R$ lead to interesting collider signatures of the doubly charged scalars ($H_{L,R}^{\pm\pm}$) in the model that 
can affect the search strategies at LHC and ILC which forms a major motivation to study this model.    

After symmetry breaking, the dark sector Lagrangian in Eqn.\ref{eq:lag1} leads to mixing between the neutral components 
$\psi_1^0$ and  $\psi_2^0$ and also leads to the mixing among the charged components $\psi_1^\pm$ and  $\psi_2^\pm$  
thanks to the Yukawa interactions: $ Y_1 \overline{\psi_1} \Phi \psi_2 + Y_2 \overline{\psi_1} \tilde{\Phi} \psi_2 $. The mass matrices of the  neutral  and charged fermion states can be expressed in the interactions basis of $X_N= \left(\begin{matrix} (\psi_{1L}^0)^c & \psi_{1R}^0 & (\psi_{2L}^0)^c & \psi_{2R}^0 \end{matrix} \right)^T$ and $X_C =\left(\begin{matrix} \psi_1^- & \psi_2^-   \end{matrix} \right)^T$ respectively as,

\bea
\label{eq:massmat}
\mathcal{M}^N = 
       \left( \begin{matrix} \sqrt2 y_L v_L && M_{L}  && 0 && \alpha \\
        M_{L} && \sqrt2 y_L v_L && \alpha && 0 \\
        0 && \alpha && \sqrt2 y_R v_R &&  M_{R} \\
        \alpha && 0 && M_{R} && \sqrt2 y_R v_R 
                        \end{matrix} \right)~~{\rm and}
~~~
        \mathcal{M}^C   
        = 
        \left( \begin{matrix} M_L && \alpha^\prime  \\
         \alpha^\prime && M_R 
                        \end{matrix} \right)   ;               
\eea
where $\alpha=\frac{Y_1 v_1+ Y_2 v_2}{\sqrt2} \simeq \frac{Y_1 v_1}{\sqrt2} $ and $\alpha^\prime=\frac{Y_1 v_2+ Y_2 v_1}{\sqrt2} \simeq \frac{Y_2 v_1}{\sqrt2}$ in the limit of $\tan\beta \to 0$. The phenomenology of DM depends on the following parameters in dark sector: 
\bea
\{M_L,~M_R,~y_L,~y_R,~Y_1,~Y_2\} .
\eea
along with the free parameters of LRSM. The nature of DM i.e. whether the DM is $SU(2)_L$ type or $SU(2)_R$ type or an admixture of them is mainly decided by the choice of the above parameters. Depending on these parameter choices, the model offers three different type of DM scenarios which we shall discuss now. 

\subsubsection*{\underline{\bf $SU(2)_L$ like DM} ($M_L \ll  M_R$)}

For the given mass hierarchy, $M_L \ll  M_R$, the light neutral states, $\chi_{_{1,2}}$ and the light 
charged state, $\chi_{_1}^\pm$ dominantly behave like the fermion doublet, $\psi_1$.  The presence of 
Majorana type  Yukawa interaction with $\psi_1$ $(y_L \overline{\psi_1} \Delta_L^\dagger i\sigma_2 \psi_1^c ~)$, leads to the mass splittings (generated after symmetry breaking) 
between the light neutral and charged fermion states, $\chi_{_{1,2}}$ and  $\chi_{_1}^\pm$ as:
\bea
\chi_{_1}&\simeq& -\frac{i}{\sqrt{2}} \big( \psi_1 -\psi_1^c\big)  ~~~~    {\rm of ~mass}~~~ M_1= M_L -\sqrt{2}~ v_L~ y_L   ~(\equiv m_{\rm DM})\nonumber \\
\chi_{_2}&\simeq& \frac{1}{\sqrt{2}} \big( \psi_1 + \psi_1^c\big)  ~~~~~~    {\rm of ~mass}~~~ M_2= M_L +\sqrt{2} ~v_L ~y_L \nonumber\\
\chi_{_1}^\pm &\simeq& \psi_1^\pm ~~~~~~~~~~~~~~~~~~  ~~  {\rm of ~mass}~~~ M_1^\pm= M_L ,
\eea
whereas the remaining physical states $\chi_{_{3,4}}$ and $\chi_{_2}^\pm$ are very heavy $\left(\mathcal{O}(M_R)\right)$ 
and do not play any role in contributing to number density of DM. Note here that the light neutral and charged states are 
nearly degenerate $\left(\mathcal{O}(M_L)\right)$ as  $v_L$ is small ($ \lesssim 8$ GeV \cite{ParticleDataGroup:2020ssz}, constrained from 
$\rho$ parameter). In this setup the light dark states interact mostly with the $SU(2)_L$ fields of LRSM. 

\subsubsection*{\underline{\bf $SU(2)_R$ like DM} ($M_L \gg  M_R$)}

Unlike the previous case, here the right triplet ($\Delta_R$) vev, $v_R$ plays a significant role. In addition, for the reversed case where $M_L \gg  M_R$, the light neutral and charged physical states, $\chi_{_1,_2}$ and  $\chi_{_1}^\pm$ mostly behave like the second fermion doublet, $\psi_2$ belonging to $SU(2)_R$. Similar to the 
previous scenario, $\psi_2$ interacts with the $SU(2)_R$ triplet ($\Delta_R$) with the Majorana type Yukawa interactions: $y_R \overline{\psi_2} \Delta_R^\dagger i\sigma_2 \psi_2^c ~$ which leads to the mass splitting between light neutral and charged physical states, $\chi_{_1},~\chi_{_2}$ and $\chi_{_1}^\pm$ as:
\bea
\label{rldm}
\chi_{_1}&\simeq& -\frac{i}{\sqrt{2}} \big( \psi_2 -\psi_2^c\big)  ~~~~    {\rm of ~mass}~~~ M_1= M_R -\sqrt{2} ~v_R~ y_R  ~(\equiv m_{\rm DM}) \nonumber \\
\chi_{_2}&\simeq& \frac{1}{\sqrt{2}} \big( \psi_2 + \psi_2^c\big)  ~~~~~~    {\rm of ~mass}~~~ M_2= M_R +\sqrt{2} ~v_R ~y_R \nonumber\\
\chi_{_1}^\pm &\simeq& \psi_2^\pm ~~~~~~~~~~~~~~~~~~  ~~  {\rm of ~mass}~~M_1^\pm= M_R~,
\eea
whereas the remaining physical states $\chi_{_{3,4}}$ and $\chi_{_2}^\pm$ are much heavier ($\mathcal{O}(M_L)$), which mostly decay into light states before the time of DM freeze-out from thermal bath. Hence they do not affect todays DM density. It is important to mention here that the right triplet vev, $v_R$, can easily generate a large mass splitting between the light physical states, obvious from the mass expressions in Eqn.\ref{rldm}. The light physical states in this case are $SU(2)_R$ like in nature and dominantly interact with $SU(2)_R$ fields.

\subsubsection*{\underline{\bf Mixed DM} ($M_L \sim M_R $ )}

When both $M_L$ and $M_R$  are of similar order of magnitude and the Yukawa couplings, $y_R,Y_1$ and $Y_2$ have 
non-zero values, the physical dark states are admixture of both $\psi_1$ and $\psi_2$. We henceforth refer this as {\it mixed~DM} scenario. In order to obtain the physical neutral states, $\chi_{_i} (i=1,2,3,4)$ one needs to diagonalize the mass matrix 
$\mathcal{M}^N$ in Eqn.\ref{eq:massmat}  by a unitarity matrix, ${\mathcal{U}_N}_{4\times4}$. The mass diagonalization 
leads to a relation between the physical and interactions states given by
 \bea
 {\mathcal{U}_N}^\dagger \mathcal{M}^N \mathcal{U}_N &= & {\rm diag}( M_1,~M_2,~M_3,~M_4), \nonumber \\
 \Big(\chi_{1},~\chi_{2},~\chi_{3},~\chi_{4} \Big) &=&{\mathcal{U}_N}_{4\times 4} ~\Big( (\psi_{1}^0)^c,~ \psi_{1}^0,~ (\psi_{2}^0)^c ,~ \psi_{2}^0  \Big) ,
 \eea
 where $\chi_{_1},~\chi_{_2},~\chi_{_3}$ and $\chi_{_4}$ are the physical eigenstates with mass $M_1,~M_2,~M_3$ and $M_4$ respectively following the mass hierarchy $|M_1| < |M_2| < |M_3| < |M_4|$. Thus the lightest neutral state, $\chi_1$ of mass $M_1$ $(\equiv m_{\rm DM})$ is the stable DM candidate in this setup. The mass eigenvalues of the mass matrix, $\mathcal{M}^N$ are given in the limit of $v_L \to 0,~\tan\beta \to 0$ as 
 \bea
 \label{masseigen1}
   \lambda_{1,2}& =&  \frac{\sqrt{2} y_R v_R -(M_L + M_R)}{2} \pm \frac{\left( \left(  \sqrt{2} y_R v_R +(M_L - M_R)\right)^2 + 2 (Y_1 v_1)^2 \right)^{1/2}}{2}  \nonumber \\
 \lambda_{3,4}  & =& \frac{ \sqrt{2} y_R v_R +(M_L + M_R)}{2} \mp  \frac{\left( \left(  \sqrt{2} y_R v_R - (M_L - M_R)\right)^2 + 2 (Y_1 v_1)^2 \right)^{1/2} }{2}  .
\eea
The order of magnitude of the above eigenvalues depend on the model parameters. 
Here all the neutral physical states, $\chi_{_i}~(i=1,2,3,4)$  behave like Majorana states and are defined as $\chi_{_i}=\frac{1}{2} \left(\chi_{_i} + \chi_{_i}^c \right)$. Similarly the mass matrix, $\mathcal{M}^C$ for the charged fermion states mentioned in 
Eqn.\ref{eq:massmat} can be diagonalized by a unitarity matrix, ${\mathcal{U}_C}_{2\times2}$. The 
corresponding mass diagonalization relation and the relation between physical and interaction states are expressed as  
 \bea
 {\mathcal{U}_C}^\dagger \mathcal{M}^C \mathcal{U}_C &=&  {\rm diag}(M_1^\pm,~M_2^\pm ) \nonumber \\
 \Big(\chi_{1}^\pm,~\chi_{2}^\pm \Big) &=& {\mathcal{U}_C}_{2\times 2} ~\Big(\psi_{1}^\pm,~\psi_{2}^\pm \Big) .
 \eea
Here $\chi_{_1}^\pm$ and $\chi_{_2}^\pm$ are the physical charged eigenstates with mass $M_1^\pm$ and $M_2^\pm$ respectively with mass hierarchy $|M_1^\pm | < |M_2^\pm|$. The eigenvalues of the mass matrix, $\mathcal{M}^C$ in the limit of $v_L \to 0,~\tan\beta\to 0$ are given as  
\bea
 \label{masseigen2}
 M_{1,2}^{\pm}&=&\frac{1}{2} \left( \big(M_L+M_R\big) \mp \sqrt{(M_L - M_R)^2+2 (v_1 Y_2)^2}\right) .
\eea
In our analysis we have used the numerical tool {\tt SPheno}\cite{Porod_2012} to diagonalize the mass matrices and generate the mass spectrum. 
\section{Constraints}
\label{sec:const}
In this section we briefly address the existing theoretical and experimental constraints on the model parameters which become relevant for 
our analysis. 

{\bf Perturbativity:}  
The quartic couplings in the scalar potential, gauge couplings and the Yukawa couplings are bounded from above as: 
\bea
|\lambda_{1,2,3,4}| \lesssim 4\pi,  ~~~~|\rho_{_{1,2,3,4}}| \lesssim 4\pi, ~~~~|\alpha_{_{1,2,3}}| \lesssim 4\pi, ~~~|\beta_{_{1,2,3}}| \lesssim 4\pi ~~;  \nonumber \\ 
~|g_{_{1,L,R}}| \lesssim \sqrt{4\pi} ~~; \nonumber \\
~~~|y_{_{1,2,L,R}}| \lesssim \sqrt{4\pi}~~. 
\eea

{\bf Relic and Direct Search:} The current observation from PLANCK \cite{Planck:2018vyg} puts a stringent bound on DM number density in the Universe : 
\bea
\Omega_{\rm DM} h^2 = 0.120 \pm 0.001~~~{\rm ~at ~90\% ~CL} .
\eea
We will implement this constraint on our model parameter space. Along with this astrophysical observation, the DM-nucleon scattering cross-section also faces severe constraints from non-observation of DM at direct search experiments like XENON 1T \cite{XENON:2018voc} and PANDAX 4T \cite{2021}. We also include this limit on the model parameter space when we scan over the free parameters.  

{\bf Higgs invisible decay:} When the DM mass is below $M_h/2$, the SM Higgs can decay to DM. The invisible decay width of the SM Higgs is measured at LHC \cite{Aaboud_2019} which gives a constraint on the parameter space that leads to $m_{\rm DM} < \frac{M_h}{2}$ as well as 
the coupling strength of DM with the SM like Higgs boson through an upper bound:
\bea
{\rm BR}(h \to {\rm DM~DM}) \lesssim 13 \% .
\eea

{\bf LEP constraint:}
LEP has excluded exotic charged fermion masses below $\sim 102.7$ GeV \cite{2003}. We implement this constraint on the dark charged 
fermions ($\chi_{1,2}^\pm$) mass $M_{1,2}^{\pm} > 102.7$ GeV. 

{\bf FCNC constraint:}
The bi-doublet structure of the scalar gives rise to tree-level flavor changing neutral current (FCNC) interactions with SM quark in LRSM which 
is mediated by the heavy neutral scalars $H,~A$ and can contribute to flavor observables such as 
$K_0-\overline{K_0}$, $B_d-\overline{B_d}$ and $B_s-\overline{B_d}$ mixings \cite{Zhang:2007da,Maiezza:2010ic}. The flavor observable data puts very stringent lower bound on heavy 
neutral scalars in the model given by $M_{H,A} \gtrsim 15$ TeV. This upper bound further translates on to the 
quartic coupling $\alpha_3$ in the scalar potential and on the triplet scalar vev  $\langle \Delta_R^0 \rangle = v_R$. Using the approximate form of the heavy neutral scalar mass
$M_{H}$,  
the FCNC constraint can be expressed as: 
\bea
\frac{1}{2}\alpha_3 \Big(\frac{v_R}{{\rm TeV}}\Big)^2 \gtrsim \Big(\frac{15}{{\rm TeV}}\Big)^2~.
\eea
{\bf Collider constraint on $M_{W_R}$:} The dominant bound on heavy charged gauge boson $W^\prime$ come 
from its decay to dijet. However, in the LRSM model the presence of a heavy right-handed neutrino leads to the 
possibility of $W_R$ decay to a charged lepton and a heavy $N_R$ when kinematically allowed. This leads to a final 
state with same-sign leptons and jets, which has suppressed SM background and leads to stronger bounds.
The decay of $W_R$ into a boosted right handed neutrino $N_R$ yields same-sign lepton pair plus jets final states 
at collider as $W_R \to N_R~ \ell~ \to ~{\rm jets} +2\ell $. The current search by ATLAS \cite{ATLAS:2019isd} with 
integrated luminosity of $80 ~{\rm fb}^{-1}$ for $\sqrt{s}=13$ TeV, excludes $M_{W_R}$ smaller than $5$ TeV. 
This lower bound can be also expressed in terms of $v_R$ as:
\bea
v_R >  \sqrt2 \, \Big( \frac{M_{W_R}}{5~{\rm TeV}}\Big) \Big(\frac{0.625}{g_R}\Big) \simeq  11.31~{\rm TeV} .
\eea

In addition the doubly charged scalars, $H_{L,R}^{\pm\pm}$ are also constrained from existing searches at LHC which 
we discuss in detail in the collider analysis section. For our analysis we pick up a set of benchmark points (BPs) which are consistent with the above mentioned constraints. The BPs are tabulated in Table-\ref{tab:tab4} and \ref{tab:tabbp} 
\begin{table}[H]
\resizebox{\linewidth}{!}{
 \begin{tabular}{|c|c|c|}
\hline
BPs & Input parameters  &     Mass spectrum generated from SPheno (in GeV) \\   \hline \hline
 BP1&  \makecell{$v_R=12~{\rm TeV}$ \\ \\   $\lambda_1=0.129, ~\lambda_2=0.0,~ \lambda_3=0.1,~~ \lambda_4=0.0$~, \\ $\beta_1=\beta_3=0,$ \\$ ~\alpha_{1,2}=0,~\alpha_3= 3.4 $ \\ $\rho_1=0.04,~\rho_2=0.00020,~\rho_3=0.08357,~\rho_4=0.0$  \\ \\ ${M_{D}}_{ii}=50$ keV ${M_{D}}_{i\neq j}=0$ \\ ${Y_{\Delta R}}_{ii}=0.00147314$ ${Y_{\Delta R}}_{i\neq j}=0$ }&   \makecell{$ M_{H^{\pm\pm}_R}=400 ~{\rm GeV},~M_{H^{\pm\pm}_L}=600~ {\rm GeV}$ \\~\\ $M_h =125.06,~M_{H_L}=507.00,~M_{H_R}=3394.11,~M_{H}=15646.47$ \\ $~M_{A_L}=507.00,~M_{A}=15646.47 $ \\ $~M_{H_L^\pm}= 555.50,~M_{H^\pm}=15647.73$  \\ $M_{W}=80.35,~~M_{Z}=91.18$ \\  $M_{W_R}=5.62 \times 10^3,~~M_{Z_R}=9.43\times 10^3$ \\ \\ $m_{\nu}=0.1$ eV ~~$M_N=25$ GeV  ~~$V_{\ell N}=2.00 \times 10^{-6}$} \\ \hline
\hline
\end{tabular}
}
\caption{\it The benchmark points in the scalar sector are considered for further analysis. 
Other parameters are kept fixed as: $v_L=1$ GeV and $\tan\beta=10^{-4}$. Here $V_{\ell N}$ represents the 
mixing term in the neutrino sector.}
    \label{tab:tab4}
\end{table}
\section{DM Phenomenology}
\label{sec:dmpheno}
We now discuss the phenomenology of our proposed fermionic dark matter in the extended LRSM. The lightest neutral Majorana state $\chi_{_1}$ which can be an admixture of $SU(2)_L$ like fermion ($\psi_1$) and $SU(2)_R$ like fermion ($\psi_2$) or purely $SU(2)_L$ like fermion $\psi_1$ or purely $SU(2)_R$ like fermion $\psi_2$, is the stable DM candidate under the extended symmetry group $\mathcal{G}$. In this section we review the region of parameter space which is allowed by observed DM density from WMAP-PLANCK data\cite{Hinshaw:2012aka,Planck:2018vyg}, latest upper bound on DM-nucleon scattering cross-section from direct search experiments \cite{XENON:2018voc,2021} and also from indirect search constraints\cite{Ackermann_2015,2016}.  

The lightest neutral Majorana state, $\chi_{_1}$ assumed to be the viable candidate of DM can be produced at early time of the Universe via thermal freeze-out mechanism \cite{Kolb:1990vq}. The dark sector particles were connected with visible sector via gauge and scalar mediated interactions of the LRSM and freezes out when the interaction rate ($\Gamma=\langle \sigma v \rangle n_{\rm DM}$) falls below the expansion rate of the universe ($\mathcal{H}$). Apart from $\chi_{_1}$ the dark sector also has heavy neutral and charged fermion states, $\chi_{_i} (i=2,3,4)$  and  $~\chi_{_j}^\pm (j=1,2)$ respectively. When the mass of these heavy states lie close to the DM mass, the number density of DM also gets affected by the number changing processes due to these heavy states. So the relic density of DM, $\chi_{_1}$ is guided by the different type of gauge and Higgs mediated number changing processes as: 
\begin{align*}
{\rm DM ~annihilations}: \hspace{0.3cm}    & \chi_{_1} ~\chi_{_1} \to {\rm ~X~Y}&   \\ 
{\rm DM ~co-annihilations}: \hspace{0.3cm} & \chi_{_1} ~\chi_{_j} \to {\rm ~X~Y}  \,\,  (j=2,3,4) &\\ 
                                            & \chi_{_j} ~\chi_{_k} \to {\rm ~X~Y}  \,\,  (j,k=2,3,4) & \\ 
                                            & \chi_{_i} ~\chi_{_j}^\pm \to {\rm ~X~Y} \,\,  (i=1,2,3,4; \, j=1,2) & \\ 
                                            & \chi_{_i}^\pm~ \chi_{_j}^\mp \to {\rm ~X~Y} \,\, (i,j=1,2) &
\end{align*}                                            
where X and Y are the light states in LRSM. The evolution of DM number density with time can be described by solving the Boltzmann equation given by\cite{Kolb:1990vq}: 
\begin{equation}
	\frac{dn_{\rm DM}}{dt} + 3Hn_{\rm DM} = -\langle\sigma v\rangle_{\rm eff}\Big(n_{\rm DM}^2 - n^2_{\rm eq}\Big),
\end{equation}
where $n_{_{\rm DM}} \simeq n_{\chi_{_1}}$ denotes the number density of DM and $n_{\rm eq}= g_{_{\rm DM}}(\frac{m_{_{\rm DM }} T}{2\pi})^{3/2}\exp(-m_{_{\rm DM}}/T)$ is the equilibrium density. The mass of DM is defined as $m_{\rm DM}$  (i.e $m_{\rm DM}= M_1$). $\langle\sigma v\rangle_{\rm eff}$ denotes the effective thermal averaged cross-section where all annihilation and co-annihilation type number changing processes are taken into account \cite{PhysRevD.43.3191,Edsj1997} and which can be expressed as follows:
\bea
\label{sigveff}
		\langle\sigma v\rangle_{eff} &=& \frac{g^2_1}{g^2_{\rm eff}}\langle\sigma v\rangle_{\chi_{_1}\chi_{_1}}+ \frac{2 g_1 g_i}{g^2_{\rm eff}}\langle\sigma v\rangle_{\chi_{_1}\chi_{_i}} \Big(1+\delta^0_i\Big)^{\frac{3}{2}} \, e^{-x \delta^0_i} 
		+ \frac{2 g_1 g_k^\pm}{g^2_{\rm eff}}\langle\sigma v\rangle_{\chi_{_1}\chi_{_k}^\pm}  
		         \Big(1+\delta^{\pm}_k\Big)^{\frac{3}{2}} \, e^{-x \delta^{\pm}_k}\nonumber \\
		&& + \frac{2 g_i g_k^\pm}{g^2_{\rm eff}}\langle\sigma v\rangle_{\chi_{_i}\chi_{_k}^\pm} \Big(1+\delta^{\pm}_k\Big)^3 \, e^{-2x \delta^{\pm}_k}
		+ \frac{2 g_i g_j}{g^2_{\rm eff}}\langle\sigma v\rangle_{\chi_{_i}\chi_{_j}}\Big(1+\delta^{\pm}_k\Big)^3 \, e^{-2x \delta^{\pm}_k} \nonumber \\
		&& + \frac{2 g_k^\pm g_l^\pm}{g^2_{\rm eff}}\langle\sigma v\rangle_{\chi_{_k}^+ \chi_{_l}^-} \Big(1+\delta^0_k\Big)^3 e^{-2 x \delta^{\pm}_k},
\eea	
where
\bea
g_{\rm eff}= g_1 + g_i \Big(1+\delta^0_i\Big)^{\frac{3}{2}} \, e^{-x \delta^0_i}+ g_k^\pm \Big(1+\delta^{\pm}_k\Big)^{\frac{3}{2}} \, e^{-x \delta^{\pm}_k}, \nonumber
\eea
$\delta^0_k=\frac{M_k-M_1}{M_1}$, $\delta^{\pm}_k=\frac{M_k^\pm-M_1}{M_1}$ and $\delta^0_i=\frac{M_i^\pm-M_1}{M_1}$.
The $g_1$, $g_i$ and $g_k$ are the internal degrees of freedom of $\chi_{_1},~\chi_{_i}$ and $\chi_{_k}^\pm$ state respectively and $i,j=2,3,4$; $k,l=1,2$. The parameter $x$ is defined as $x=\frac{m_{\rm DM}}{T} \equiv \frac{M_1}{T}$ where T is the thermal bath temperature. Using the above expression of $\langle\sigma v\rangle_{\rm eff}$, one can express the number density of DM approximately as \cite{PhysRevD.43.3191,Edsj1997}: 
	\begin{equation}
		\Omega_{\chi_{_1}}h^2 = \frac{1.09\times 10^9 ~{\rm GeV}^{-1}}{\sqrt{g_*} M_{Pl}}\frac{1}{\int_{x_f}^{\infty} \frac{\langle\sigma v\rangle_{\rm eff}}{x^2} dx}
		\label{aprxBEQ}
	\end{equation}
where the SM degrees of freedom $g_{*}=106.7$ and $x_f= \frac{m_{_{\rm DM}}}{T_f} \simeq 20-25$\cite{Kolb:1990vq}. $T_f$ denotes 
here the freeze-out temperature of DM. The first term of Eqn.\ref{sigveff} represents the standard DM annihilation while the rest of the 
terms are part of DM co-annihilation. Note that the co-annihilation contribution reduces with large mass splitting, $\Delta M= M_i- M_1, M_i^\pm-M_1$, due to the Boltzmann suppression of ${\rm exp}(-\Delta M/ T)$. 
It is worth mentioning here the different tools/packages that have been used for our study. We first implement the model in the public 
code {\tt SARAH} \cite{Staub_2014} for generating numerical modules for {\tt SPheno} \cite{Porod_2012} and model files 
for {\tt MicrOmegas}\cite{Belanger:2018ccd}. We then use {\tt SPheno} to compute the mass spectrum and branching ratios which is passed 
on to {\tt MicrOmegas} to calculate DM relic density as well as obtain the direct and indirect search cross-section. 

We shall now illustrate the behavior of DM number density with the following independent free parameters in the dark sector which are relevant for DM phenomenology:
\bea
  \{M_L,~M_R,~Y_1,~Y_2,~y_{_L},~y_{_R}\}.
\eea
The phenomenology of DM also significantly depends on the free parameters like mass of the light states which behave as mediator between dark 
and visible sector, vevs of the scalar fields $v_R$, $v_L$ and $\tan\beta$ in LRSM. It is difficult to study the role of all the free parameters at a 
time. In addition varying all the parameters at the same time makes it difficult to determine the physics implications of any particular set. We therefore 
choose to fix the well studied LRSM model parameters which simplifies things but allows us to study the DM phenomenology based on the 
dark sector parameters. We first fix the LRSM sector as mentioned in BP1 listed in Table\ref{tab:tab4} which is consistent with existing 
theoretical and experimental constraints. The dark sector particles speak with the visible sector via light scalars ($h$, $H_L$, $A_L, H_L^\pm, H_L^{\pm\pm}, H_{R}^{\pm\pm}$ ) and the SM gauge bosons ($Z, W, \gamma$). While the model possesses many more particles in the 
spectrum, other heavy state mediated diagrams are suppressed by the mass and they give negligible contribution to DM density. 
We now consider the DM phenomenology of all three possible scenarios. 

\subsubsection*{\underline{\bf $SU(2)_L$ like DM} ($M_L \ll  M_R$)}
As stated earlier in section \ref{sec:model}, the DM ($\chi_1$) along with heavier components $\chi_2$ and $\chi_1^\pm$ behave like the $SU(2)_L$ 
doublet dark fermion ($\psi_1$) for the mass parameter $M_L \ll  M_R$. Since the physical states belong to the 
$SU(2)_L$ doublet, the relic density of DM is mainly governed by SM gauge boson mediated interactions along 
with the new scalar triplet $\Delta_L$ involved through the Yukawa 
interaction: $ y_L \overline{\psi_1} \Delta_L^\dagger i\sigma_2 \psi_1^c$ which is also responsible for generating the 
small mass splittings between $\chi_{_1}$, $\chi_{_2}$ and $\chi_{_1}^\pm$. Note here that all other states, ${\chi_{_3,_4}} , \chi_2^\pm$ 
have mass $\mathcal{O}(M_R)$ and do not play any relevant role in the DM density.  

\begin{figure}[htb!]
 $$
    \includegraphics[scale=0.27]{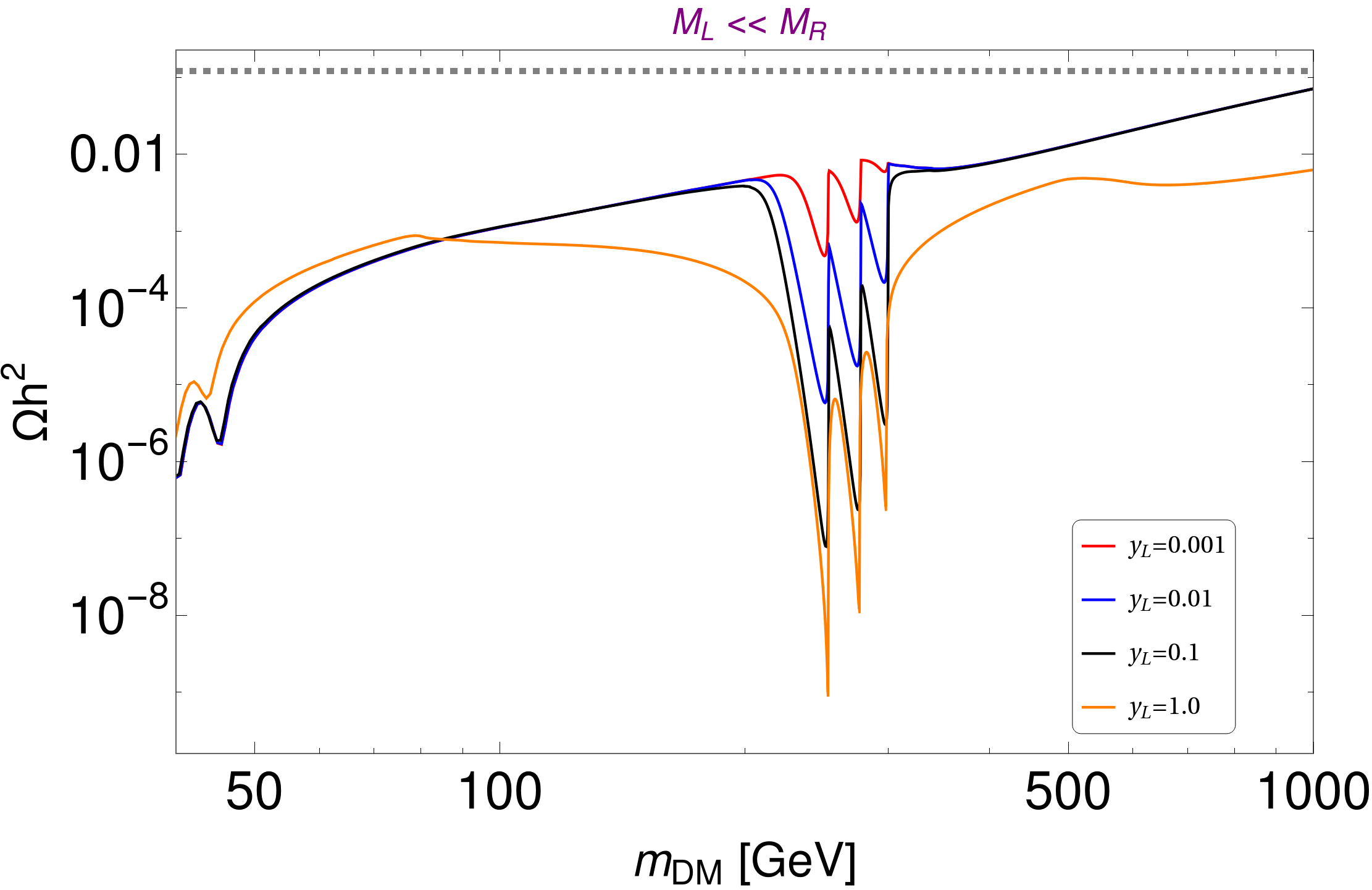}~
    \includegraphics[scale=0.28]{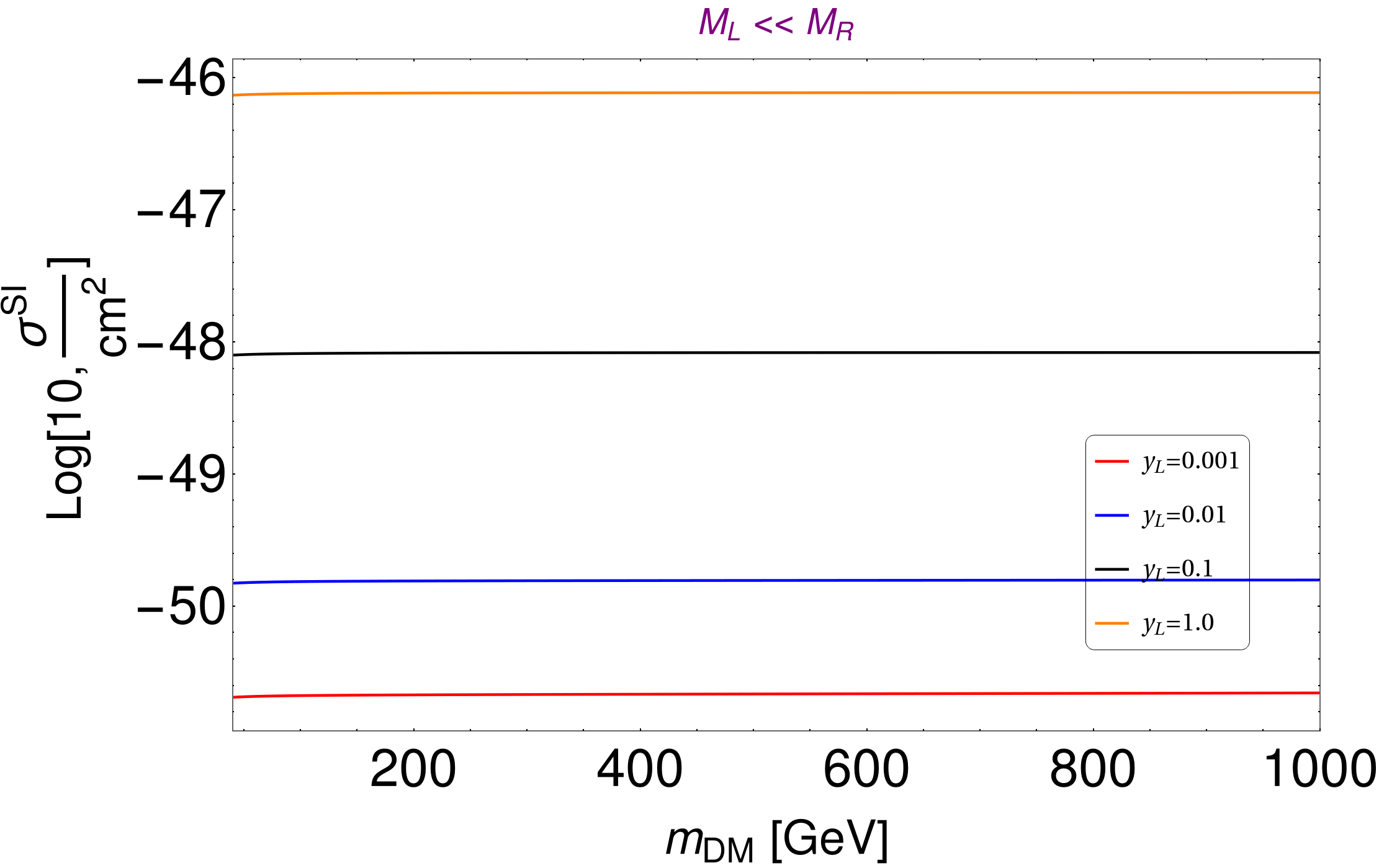}
  $$
  \caption{\it [Left] Variation of relic density as a function of DM mass for $SU(2)_L$ like DM ($M_L \ll  M_R$) with different 
  values of $y_L$. [Right] Spin independent DM-nucleon scattering cross-section as a function $m_{\rm DM}$ with different values of $y_L$.}
  \label{fig:relicdd1}
 \end{figure}
 In the left panel of Fig.\ref{fig:relicdd1} we show the variation of relic density as a function of the DM mass $m_{_{\rm DM}} (\equiv M_1$) for different 
 values of $y_L$. Correct observed relic density bound measured by WMAP-PLANCK, $(\Omega_{\rm DM}h^2=0.120 \pm 0.001)$ is shown 
 by grey dashed line in the same plane. The parameters in LRSM sector are kept fixed as listed in Table\ref{tab:tab4} for the benchmark BP1. 
 The mass splitting between the light dark states are very small leading to nearly degenerate states with splitting $\sim y_L \, v_L$. 
 As the vev of the left triplet, $v_L$  is constrained from the $\rho$ parameter as $v_L \lesssim 8$ GeV \cite{ParticleDataGroup:2020ssz}, hence the maximum splitting 
 available for this type of scenario is $\sim 8$ GeV for $y_L=1$. 
 As a result the relic becomes under abundant due to the large gauge 
 mediated co-annihilation cross-sections ($\chi_{_i} ~\chi_{_1}^\pm \to $ X Y) as depicted in the left panel of Fig.\ref{fig:relicdd1}. 
 The co-annihilation contribution will get suppressed with increase in DM mass and we find that the correct density is obtained when the DM 
 mass is around $1.2$ TeV. As we further increase $y_L$, the light scalars, 
 $H_L \, (\rm 505~GeV),~A_L \, (\rm 505~ GeV),~H_L^\pm \, (\rm 555~ GeV),~H_L^{\pm\pm} \, (\rm 600~ GeV)$ 
 mediated diagrams start contributing. 
 Therefore the relic becomes more under abundant which is shown using the orange line for $y_L=1$. Different kind of resonance dips are 
 visible near high DM mass, $m_{_{\rm DM}} \sim {M_{H_L}}/{2},~{M_{A_L}}/{2}$; $~{M_{H_L^\pm}}/{2}$ and $~{M_{H_L^{\pm\pm}}}/{2}$ 
 and these become more prominent for large values of $y_L$ due to the 
 Yukawa interaction: $ y_L \overline{\psi_1} \Delta_L^\dagger i\sigma_2 \psi_1^c$. There are two more dips observed in the low DM mass 
 region, around $m_{_{\rm DM}} \sim {M_Z}/{2}$ and  $\sim {M_h}/{2}$. 
 Note that, when the DM mass becomes larger 
 than the light triplet scalars, DM can annihilate and co-annihilate to triplet final states which further reduce the DM density. This effect is 
 observed for large value of $y_L$ as depicted in the left panel of Fig.\ref{fig:relicdd1} by the orange line for $m_{_{\rm DM}} \gtrsim 500$ GeV.   
  
The spin-independent (SI) DM-nucleon scattering cross-section with DM mass is shown in the right panel of Fig.\ref{fig:relicdd1} for 
different values of $y_L$. Due to Majorana nature of DM, the $\chi_{_1} \chi_{_1} Z$ interaction leads to a vanishing contribution to 
DM-nucleon scattering cross-section. However the presence of the Yukawa interaction 
$ y_L \overline{\psi_1} \Delta_L^\dagger i\sigma_2 \psi_1^c$ can still lead to SI DM-nucleon scattering through $\Phi-\Delta_L$ mixed t-channel 
diagrams which depend on the coupling $y_L$. The DD cross-section increases with increase of $y_L$ which can be easily seen 
in the right panel of Fig.\ref{fig:relicdd1}.   
 
\subsubsection*{\underline{\bf $SU(2)_R$ like DM} ($M_L \gg  M_R$)} 
With the mass parameter hierarchy $M_L \gg  M_R$, the DM ($\chi_{_1}$) along with $\chi_{_2}$ and $\chi_{_1}^\pm$ now behave like 
the $SU(2)_R$ dark doublet fermion, $\psi_2$. The relic density in this case is mainly governed by right handed gauged mediated interactions 
along with scalar triplet $\Delta_R$ via the Yukawa interaction: $ y_R \overline{\psi_2} \Delta_R^\dagger i\sigma_2 \psi_2^c$. The Yukawa 
interaction splits the state, $\psi_2^0$ into two physical Majorana states $\chi_{_{1,2}}$ and generates a much larger mass splitting between 
$\chi_{_1}$, $\chi_{_2}$ and $\chi_{_1}^\pm$, thanks to the large $v_R$. The other heavy states, $\chi_{_3,_4}$ and $\chi_{_2}^\pm$ which 
are $SU(2)_L$ like with mass $\mathcal{O}(M_L)$ decay into the less heavier dark states much before DM freeze-out and do not alter the observed DM density. 

\begin{figure}[htb!]
 $$
    \includegraphics[scale=0.25]{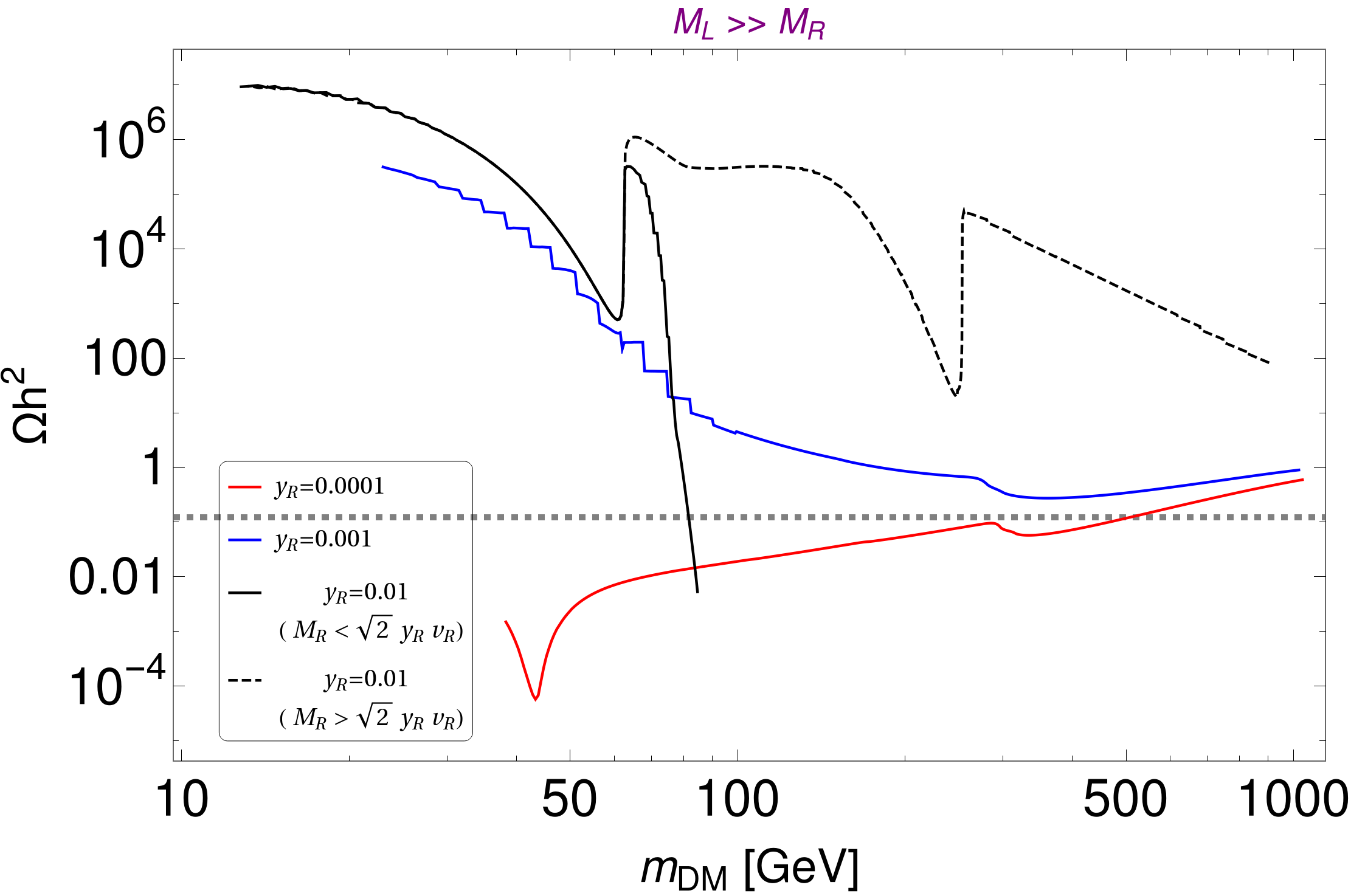}~~
    \includegraphics[scale=0.28]{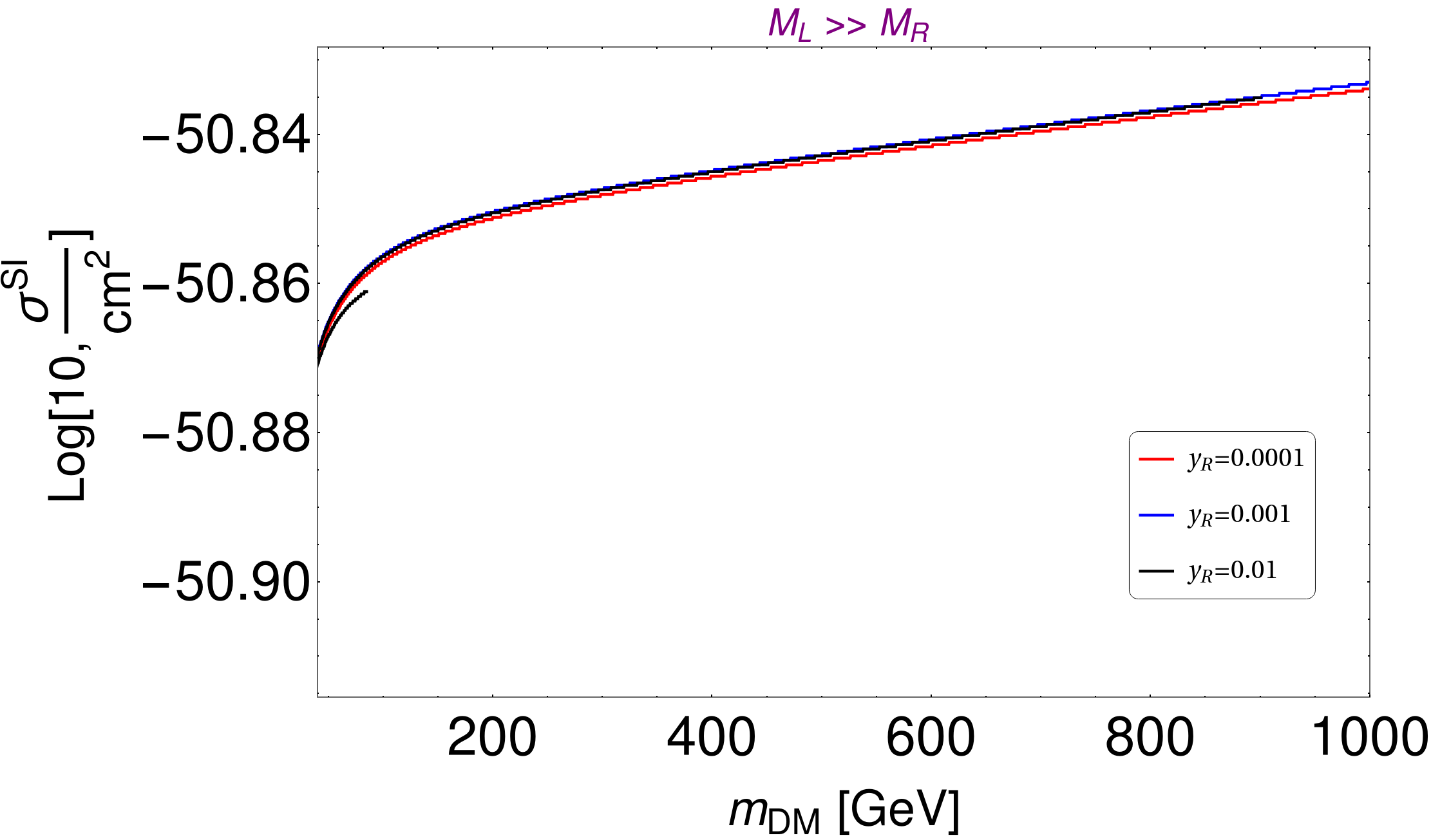}
  $$
  \caption{\it [Left] Variation of relic density as a function DM mass for $SU(2)_R$ like DM ($M_L \gg  M_R$) with different values of $y_R$. [Right] Spin-independent DM-nucleon scattering cross-section as a function of $m_{_{\rm DM}}$ for different values of $y_R$.}
  \label{fig:relicdd2}
 \end{figure}

In the left panel of Fig.\ref{fig:relicdd2} we have shown the variation of relic density as a function of DM mass, $m_{\rm DM} (\equiv M_1$) with different choices of $y_R$ shown by different colored lines. The black dashed line shows the observed relic density measured 
by WMAP-PLANCK. The vev of right triplet ($\Delta_R$) responsible for the $SU(2)_R$ breaking is taken to be quite 
large ($v_R=12\times 10^3$ GeV for BP1) which corresponds to  larger mass splitting between the dark states of around $\sim ~y_R ~v_R$. 
For a fixed value of $v_R$ the mass difference between the light dark states increases with increase of $y_R$. Therefore the 
co-annihilation contribution becomes diluted with increase of $y_R$ due to the Boltzmann suppression ($\sim$ exp$(-\Delta M/T)$). 
This results in a DM which is mediated mostly by the right-handed sector which is very heavy and therefore the relic density becomes 
over abundant as shown in the left panel of Fig.\ref{fig:relicdd2}. In addition, due to the $SU(2)_R$ like nature, the 
light dark states are 
dominantly connected with the thermal bath via the right handed heavy gauge fields ($W_R, Z_R$) and right scalar fields. 
The effective cross-section 
is suppressed due to large mediator mass  unlike the $SU(2)_L$ like DM scenario. However for $y_R=0.0001$ 
which corresponds to small mass 
splitting ($y_R v_R = 1.2$ GeV) one achieves dominant co-annihilation contributions resulting in under abundance. 
The density increases slowly 
with increase of DM mass and satisfies correct density for $m_{\rm DM} \sim 500$ GeV. For larger value of $y_R= 0.01$ 
which corresponds to 
a large mass splitting ($y_R v_R = 120$ GeV) one again gets over abundance due to suppressed co-annihilation contribution. 
Now let us try to understand the peculiar behavior of DM density observed when the DM mass is around $80$ GeV for 
$y_R= 0.01$. 
This can be understood from the DM mass formula which is defined as $M_1=M_R-\sqrt{2} y_R v_R$. 
When $M_R < \sqrt{2} y_R v_R  \, (\sim 169 ~{\rm GeV})$ the mass eigenvalue of DM decreases with increase of $M_R$ while simultaneously increasing the mass splitting between $\chi_{_1}$ and $\chi_{_1}^\pm$. An outcome of this is that the relic density 
increases because co-annihilation get suppressed. This behavior is depicted by the solid black line in left panel of 
Fig.\ref{fig:relicdd2} if one looks at lower DM mass from $m_{_{\rm DM}}\sim 80$ GeV. Again for 
$M_R = \sqrt{2} y_R v_R~(\sim 169 ~{\rm GeV})$, the DM 
mass eigenvalue becomes zero and starts increasing with increase of $M_R$ ($ > \sqrt{2} y_R v_R~\sim 169 ~{\rm GeV}$) which is shown by the dotted line starting from low DM mass ($M_1 > 0$).  

Similar to the $SU(2)_L$ like case, the $\chi_{_1} \chi_{_1} Z$ interaction again leads to vanishing DM-nucleon scattering cross-section  
due to the Majorana nature of $\chi_1$. But the presence of Yukawa interaction with 
$\Delta_R$: $ y_R \overline{\psi_2} \Delta_R^\dagger i\sigma_2 \psi_2^c$ can lead to spin-independent (SI) DM-nucleon scattering through $\Phi-\Delta_R$ mixed t-channel diagrams. 
Due to small mixing between $\Phi$ and $\Delta_R$  and large mass suppression due to heavy $H_R$ the direct search cross-section is 
much suppressed and almost independent of $y_R$ which is shown in the right panel of Fig.\ref{fig:relicdd2}, where we 
have plotted the SI DM-nucleon scattering cross-section against DM mass for different values of $y_R$. 

\subsubsection*{\underline{\bf Mixed DM} ($M_L \sim  M_R$)}
Let us now discuss the mixed scenario where the dark states, $\chi_{_{i}}$ and $\chi_{_{j}}^\pm$ are an admixture of 
both $SU(2)_L$ 
type doublet ($\psi_1$) and $SU(2)_R$ type doublet ($\psi_2$). The mixing between the neutral states, 
$\chi_{_{i}} ~(i=1,2,3,4)$ are characterized by the parameters: $M_L,~M_R,~Y_1~,y_R$ and $v_R$ whereas for 
the charged states, $\chi_{_{j}}^\pm ~(j=1,2)$, it is defined by the parameters
$M_L,~M_R$ and $Y_2$. When both the bare masses, $M_L$ and $M_R$ are of similar order in magnitude, 
the mixing between the neutral dark states 
is mainly controlled by the terms, $Y_{1} v_{_1}$ and $y_{_R} v_{_R}$. For $y_{_R} \to 0$, the mixing between the neutral component of $\psi_1^0$ and $\psi_2^0$ 
increases with increase of $Y_1$ and  it is maximal when $Y_1 \gtrsim 0.1$. Keeping $Y_{1} = 0.1$ fixed if we 
increase $y_{_R}$, the contribution 
of $\psi_2^0$ becomes dominant in DM $\chi_{_1}$ which denotes the DM becoming more $SU(2)_R$ like 
and the interaction strength 
with left like scalars and gauge fields are suppressed. The effect of $y_{_R}$ on DM becomes prominent 
when $ y_{_R} v_{_R} \gtrsim Y_1 v_{_1}$. 
On the other hand, the mixing between the charged dark states simply depend on $Y_2$ when $M_L \sim  M_R$.     

\begin{figure}[htb!]
 $$
    \includegraphics[scale=0.25]{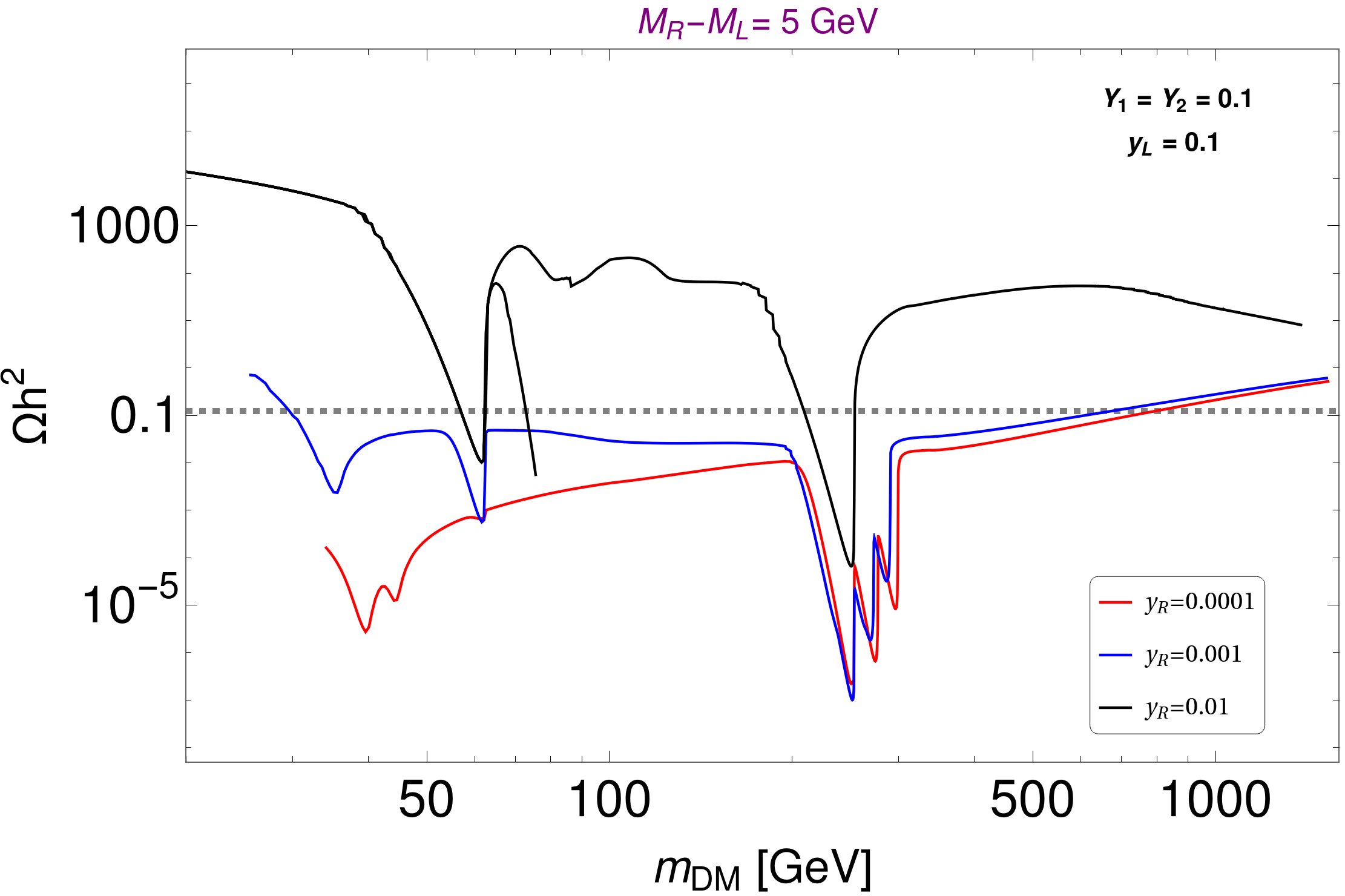}~
        \includegraphics[scale=0.25]{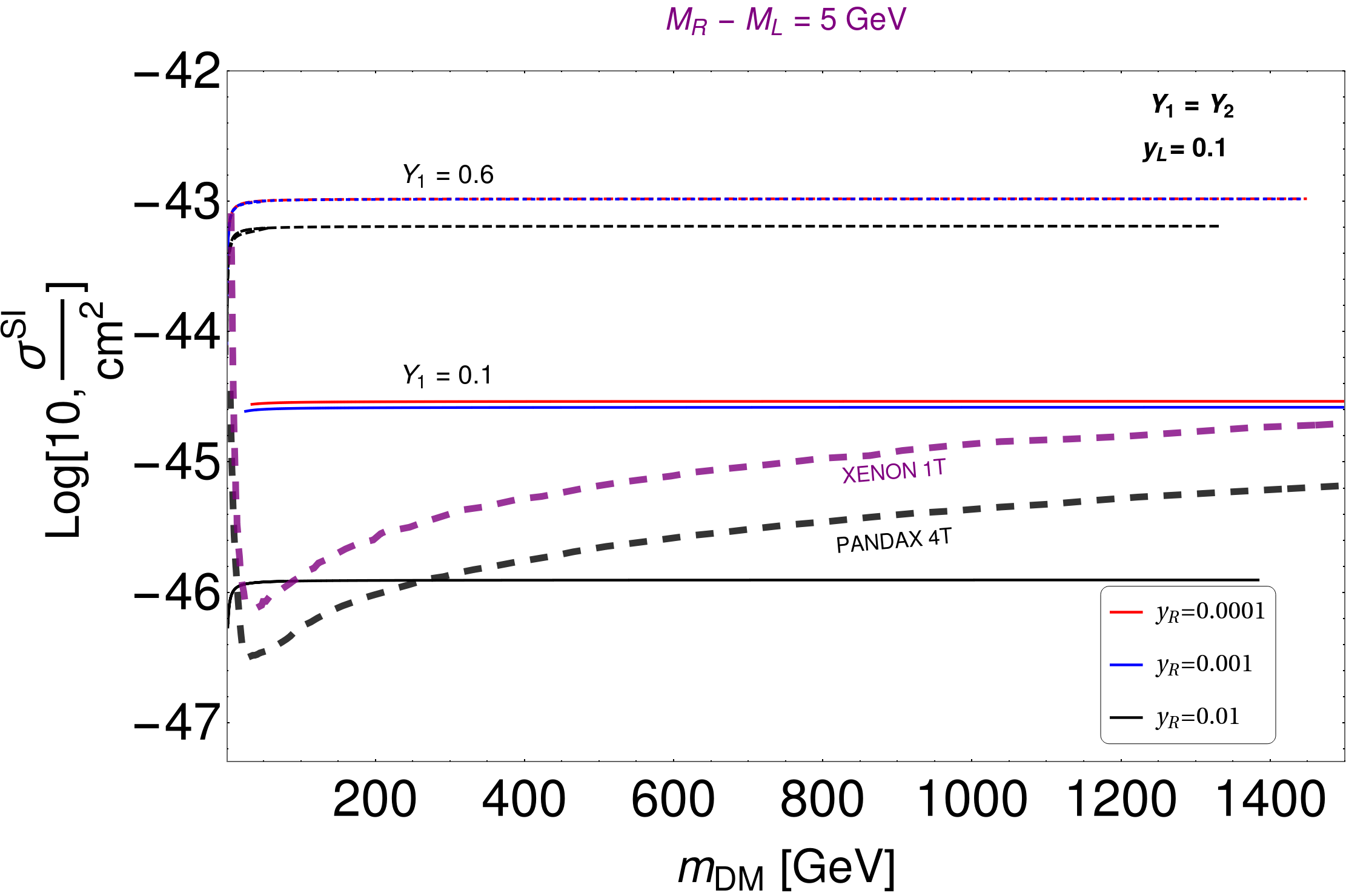}
  $$
  $$
    \includegraphics[scale=0.25]{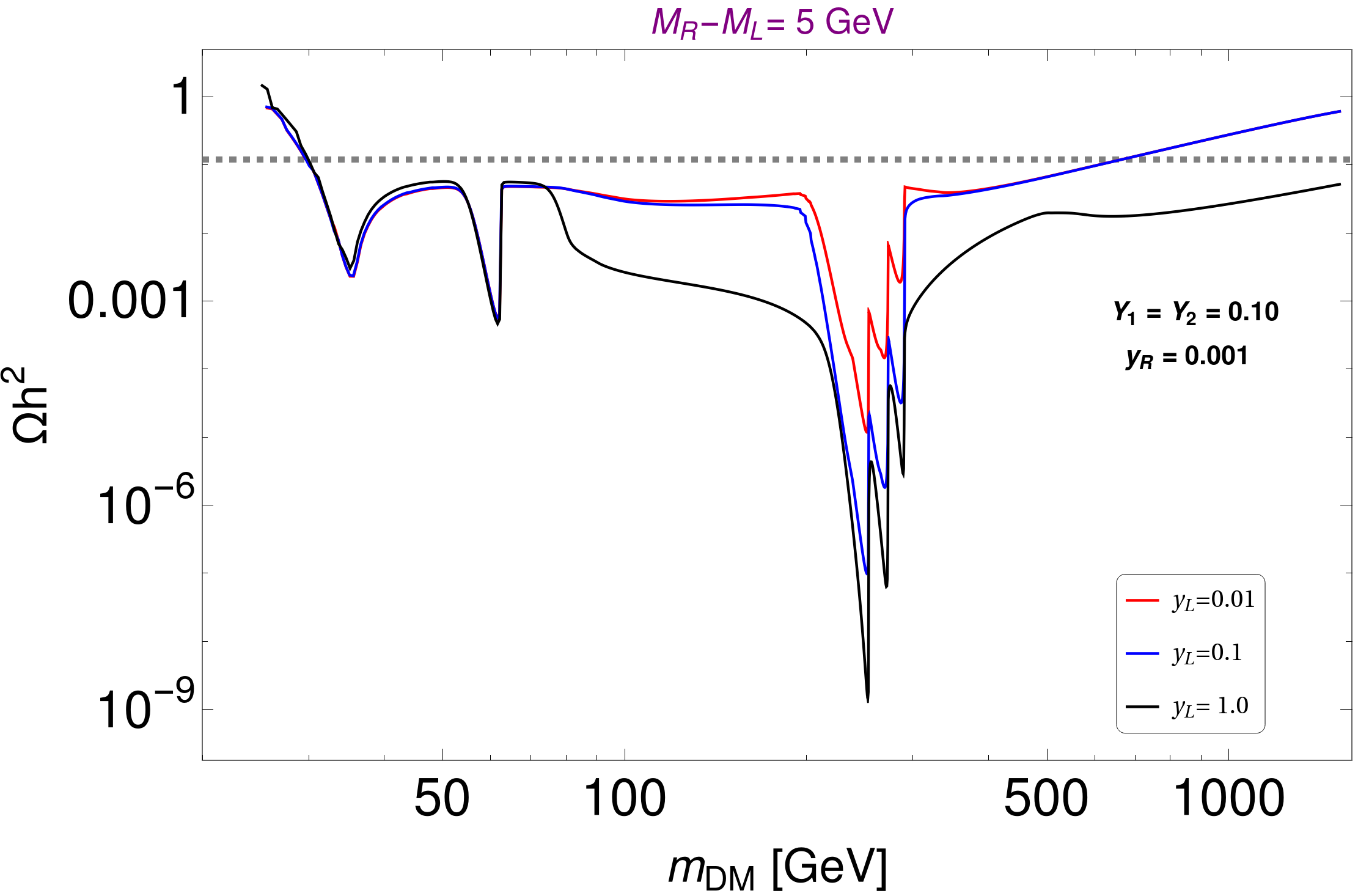}~
        \includegraphics[scale=0.25]{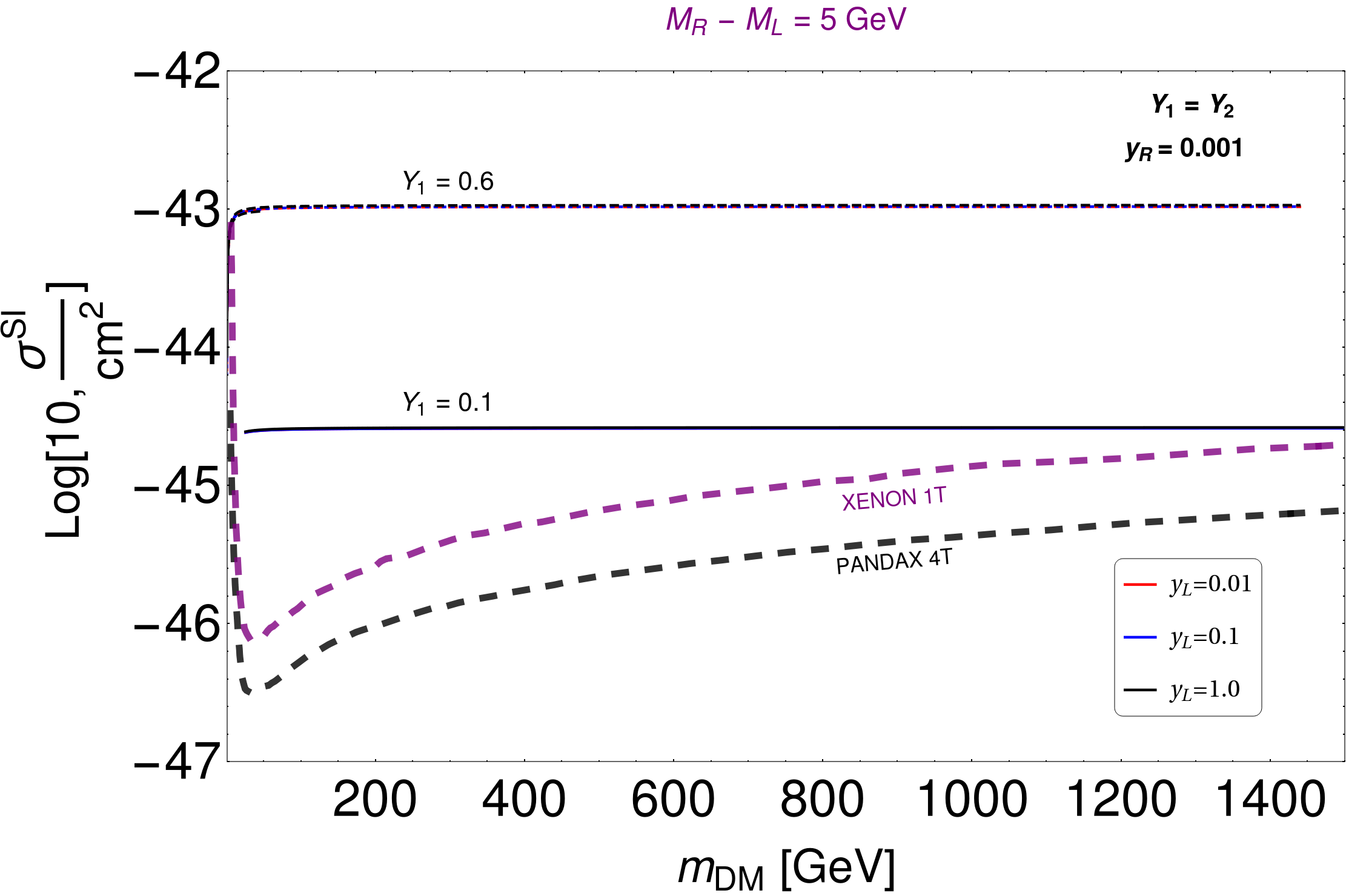}
  $$
  \caption{\it Relic density and spin-independent DM-nucleon scattering cross-section are plotted against DM mass for the mixed DM scenario ($M_L \sim  M_R$) with different values of $y_{_R}$ keeping $y_{_L}$ fixed (top panel) and with different values of $y_{_L}$ keeping $y_{_R}$ fixed (bottom panel). Other parameters are kept fixed as mentioned inset of each figure. In the right top and bottom panels, the solid lines correspond to $Y_1=0.1$ whereas the dotted lines correspond to $Y_{1,2}=0.6$. The current upper bound on the SI DD cross-section from XENON 1T \cite{XENON:2018voc} and PANDAX 4T \cite{2021} data are shown in the same plane for comparison.}
  \label{fig:relicdd3}
 \end{figure}

The mixed DM behavior is illustrated in Fig.\ref{fig:relicdd3} where we have considered $M_R-M_L=5$ GeV. In the
top-left panel of Fig.\ref{fig:relicdd3}, we have shown the variation of relic density as a function of $m_{_{\rm DM}}$ 
with different values of $y_{_R}$ mentioned in the figure inset. Other dark parameters are kept fixed as mentioned 
inset of the figure. As stated earlier, for small values of $y_{_R}$ ($\to 0$), the DM, $\chi_{_1}$ has 
maximal $\psi_1^0$ and $\psi_2^0$ mixing and the $SU(2)_R$ composition 
increases with the increase of $y_{_R}$ for a fixed value 
of $v_{_R}$. As a consequence of that, the DM becomes more $\psi_2^0$ dominant. 
Similarly with the increase of $y_R$, the splitting between DM 
and heavy state increases which corresponds to less co-annihilation contribution to DM density. Therefore the relic density spans the region between 
under abundance to over abundance with increase in $y_R$ as shown in the 
left top panel Fig.\ref{fig:relicdd3}. This is because the 
co-annihilation contribution becomes subdued and the DM interactions with the light mediators available in LRSM for BP1 are suppressed. 
This can be also understood from the resonance behavior near lower DM mass $\sim 50$ GeV. When DM has enough $SU(2)_L$ component 
$\psi_1^0$ contribution,  we observe dips near $m_{_{\rm DM}} \sim M_W/2, M_Z/2$  
which are prominent but disappear with the increase 
of $y_{_R}$. However the resonance dip due to SM Higgs, $h$ near $m_{_{\rm DM}} \sim M_h/2$ becomes more prominent with increase of 
$y_R$. Apart from the standard resonances we can also see few dips near DM mass: $m_{_{\rm DM}} \sim {M_{H_L}}/{2},~{M_{A_L}}/{2}$; $~{M_{H_L^\pm}}/{2},~{M_{H_L^{\pm\pm}}}/{2}$ and $~{M_{H_R^{\pm\pm}}}/{2}$ for the given benchmark point (BP1). 

In the bottom left panel of Fig.\ref{fig:relicdd3} we have shown the variation of relic density against DM mass with different values of $y_{_L}$ keeping $y_{_R}$ fixed ($=0.001$) as 
quoted in the figure inset. For a fixed value of $y_{_L}=0.01$ and $M_R-M_L=5$ GeV, the mixing between the 
$SU(2)_L$ component, $\psi_1^0$ and $SU(2)_R$ component, $\psi_2^0$ is almost constant in $\chi_{_1}$. With the 
increase of $y_{_L}$, the mass splitting between the light-dark states slightly increases which results in large 
DM density due to less co-annihilation contribution. The effect of DM annihilation to light triplet states start contributing 
for large values of $y_L$ ($\gtrsim 1.0$) with $m_{\rm DM} > m_{X} ~(X=H_L,~A_L,~H_L^\pm,~H_{L,R}^{\pm\pm})$.

In the right top and bottom panels of Fig.\ref{fig:relicdd3}, we have shown the SI DM-nucleon scattering 
cross-section as a function of DM mass with different choices of Yukawa couplings as mentioned inset of both figures. 
The spin-independent cross-section in the mixed scenario is strongly dependent on the Dirac like Yukawa 
couplings $Y_{1,2}$ with the bi-doublet which can be seen from the right top and bottom panels of Fig.\ref{fig:relicdd3} 
where we have considered two different values of $Y_{1,2}$ keeping $y_{_L}$ and $y_{_R}$ fixed respectively. It is also 
observed from the figures that for non-zero values of $Y_{1,2}$ the 
SI DM-nucleon scattering cross-section is almost independent of the other two Yukawa couplings $y_{_L}$ and $y_{_R}$ 
due to the small scalar mixing between $\Phi-\Delta_L$ and $\Phi-\Delta_R$ respectively.

Before going into the detail of parameter scans, we briefly point out the 
important outcomes that emerge from the three scenarios. As we have seen, due to 
the relatively larger contributions from gauge interactions and dominant co-annihilation 
due to small mass splittings in the $SU(2)_L$ like scenario when $M_L \ll M_R$, 
the relic density is always under-abundant for DM mass below a TeV. On the other hand 
for $SU(2)_R$ like DM with $M_R \ll M_L$, the observed relic density can only be 
achieved for small values of $y_R$ which leads to degenerate dark states. But the 
mixed DM scenario has rich phenomenological aspects due to its mixed nature. It can 
open up a large region of parameter space which can satisfy both relic and direct search constraints.    

 \subsection*{{\underline{Parameter space scan}}}
 Here we investigate the allowed region of DM parameter space for the mixed scenario.   
 We perform a numerical scan over the following region :
\bea
 M_L  : \{100-1000\}~{\rm GeV}~~~~~ 
 ~~M_R :  \{100-1000\}~ {\rm GeV} \nonumber  \\ 
 ~~Y \equiv Y_{1}=Y_2 : \{0.01-0.20\} ~~~~~  
 ~~y \equiv y_{_L}=y_{_R} : \{0.001,~0.01\} ;
\eea
while the remaining parameters in LRSM are kept fixed as specified in BP1. Note here 
that the choices of $Y_1=Y_2$ and $y_L=y_R$ does not affect much in DM 
phenomenology but it does matter for the collider study which we will discuss in detail 
in a later section.

\subsection*{Relic density}
 \begin{figure}[htb!]
 $$
   \includegraphics[scale=0.3]{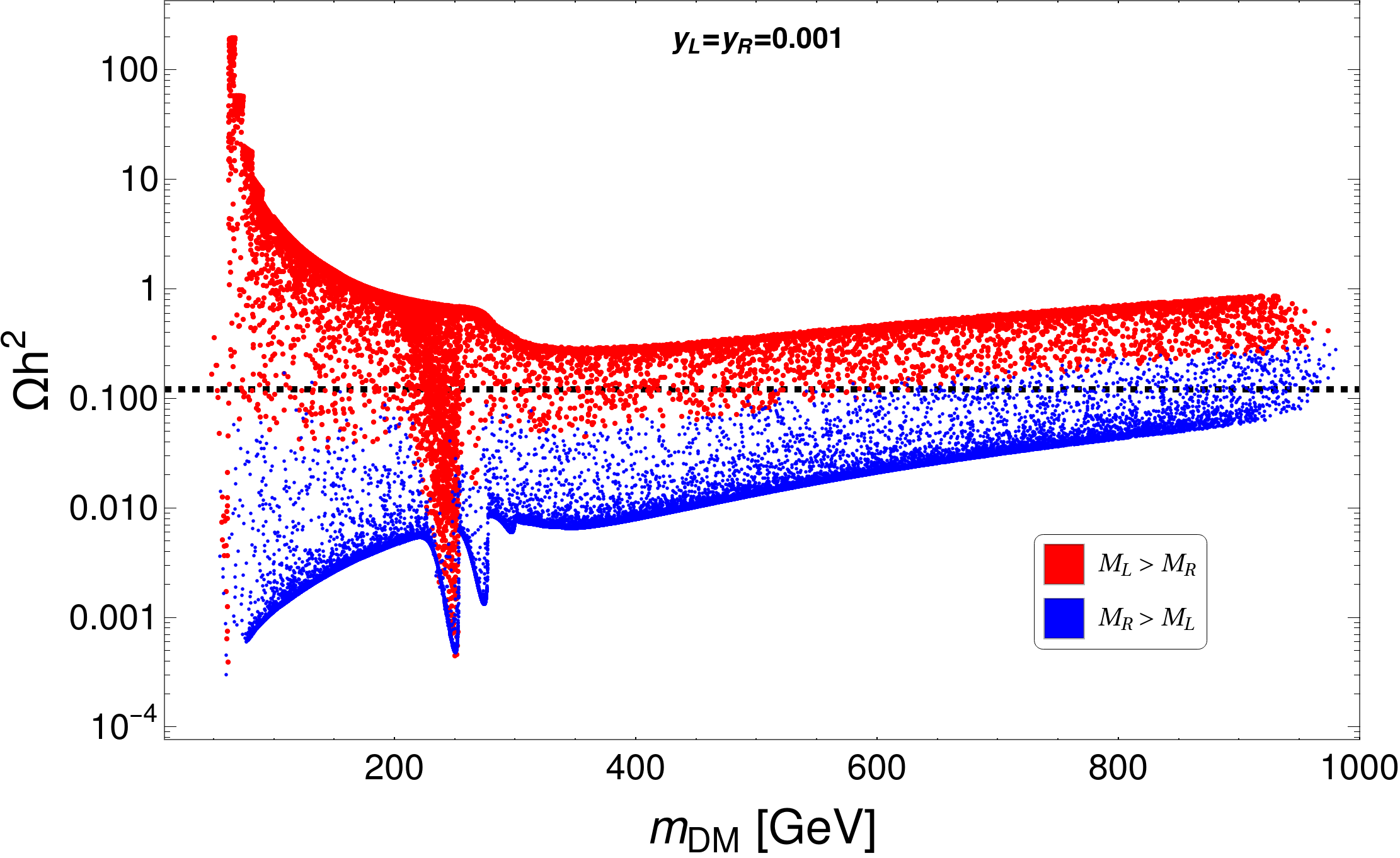}
   \includegraphics[scale=0.29]{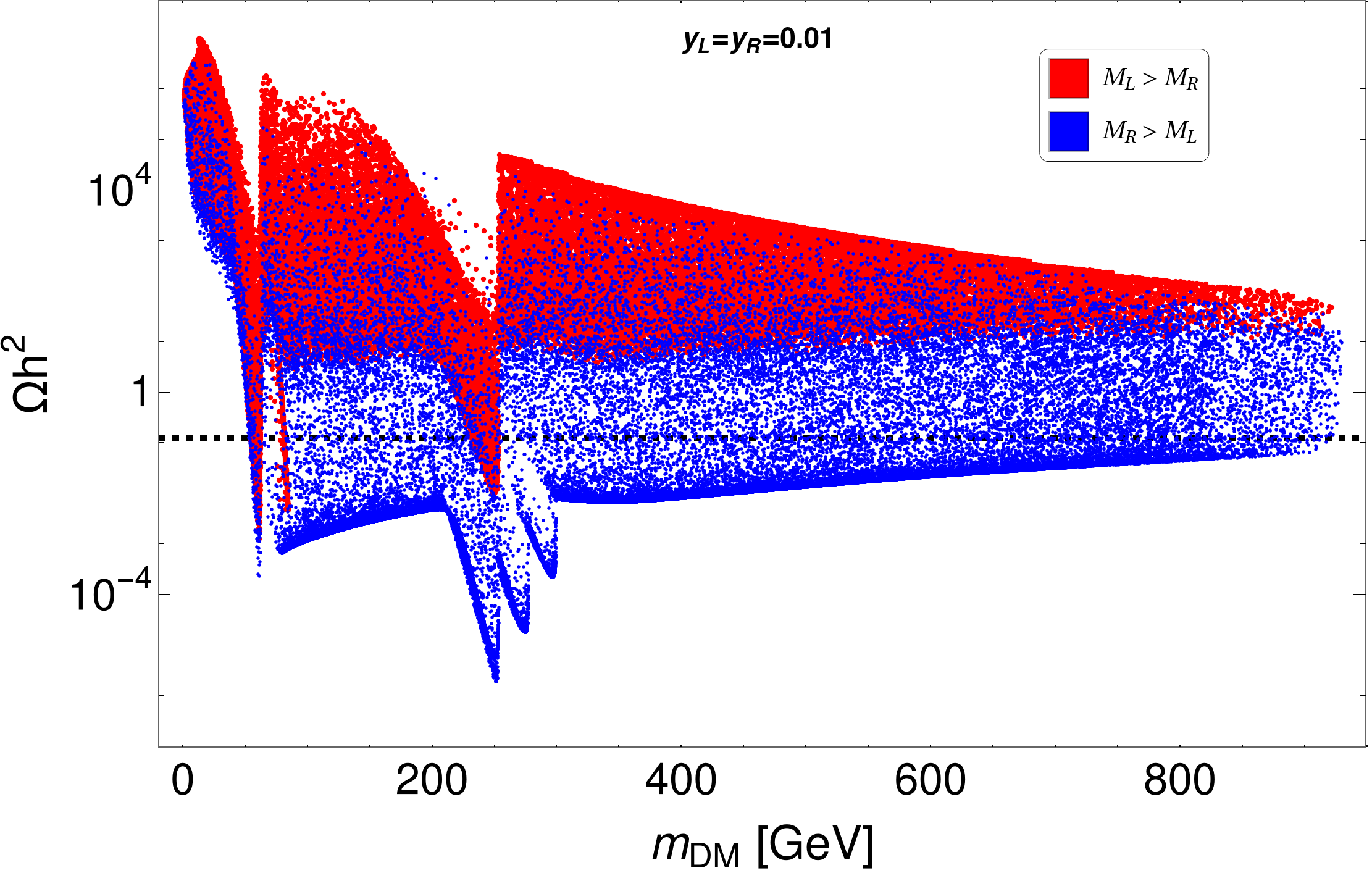}
  $$
  \caption{\it  Relic density as a function of DM mass with two different choices: $M_L > M_R$ (red points) and $M_R > M_L$ (blue 
  points). The black dotted line corresponds to the observed DM relic density from WMAP-PLANCK. We fixed $y_{_R}=0.001$ for 
  left panel figure and $y_{_R}=0.01$ for the right panel figure.}
  \label{fig:relicpar}
 \end{figure}

The variation of DM relic density as a function of DM mass is shown in  Fig.\ref{fig:relicpar} 
for two different choices of $y_{_R}=0.001$ (left) and $0.01$ (right). Two different mass 
hierarchies : $M_L > M_R$ (red points) and $M_R > M_L$ (blue points) are considered 
for each case.  The black dotted line corresponds to the central value of observed DM density \cite{Planck:2018vyg}. For a small value of $y_{_R}$ ($=0.001$), the mixing between 
the DM component of $SU(2)_L$ and $SU(2)_R$ is maximal when the bare mass parameter $M_L$ and $M_R$ are of the same order. When $M_L > M_R$ the 
$SU(2)_R$ component of the 
DM gets enhanced depending on their relative mass separation. Similarly, for the other mass hierarchy, $M_R > M_L$, the $SU(2)_L$ part of the DM gets increased. 
This phenomenon is exactly depicted in left panel of Fig.\ref{fig:relicpar} for $y_{_R}=0.001$. When DM becomes more $SU(2)_R$ like ($M_L > M_R$) relic density gets enhanced due to suppressed interaction of DM with mediator fields in LRSM. Whereas for the reverse case ($M_R > M_L$) when DM becomes more $SU(2)_L$ like, the interaction between DM and 
the mediators get enhanced resulting in less density as it is seen from the left panel 
of Fig.\ref{fig:relicpar} (blue points). If we increase the couplings $y_{_R}$ and $y_{_L}$ 
to $0.01$, the mass splitting between dark sector particles increase which leads to 
a comparably smaller co-annihilation contribution. As a result, relic density of DM 
increases which is shown in the right panel of Fig.\ref{fig:relicpar}. Due to 
co-annihilation suppression for $y_{_R}=0.01$, the correct relic density can be 
achieved throughout the DM mass region for the case $M_R > M_L$ while for the other hierarchy it only satisfies the near resonance region. There are few dips in the relic density plots which are essentially due to resonances corresponding to the gauge boson ($Z$), SM like Higgs ($h$), and the additional light scalars ($H_L, ~A_L,~H_L^\pm,~H_L^{\pm\pm}$ and $H_R^{\pm\pm}$) available in LRSM for BP1.

\subsection*{Direct search constraint}
Non-observation of DM signal at direct search experiments like XENON-1T \cite{XENON:2018voc}, PANDAX 4T\cite{2021} has set a stringent constraint on DM-nucleon scattering cross-section for WIMP like DM. Here we will apply those constraints on our model parameters space which satisfy the observed relic density constraint.

\begin{figure}[htb!]
 $$
   \includegraphics[scale=0.4]{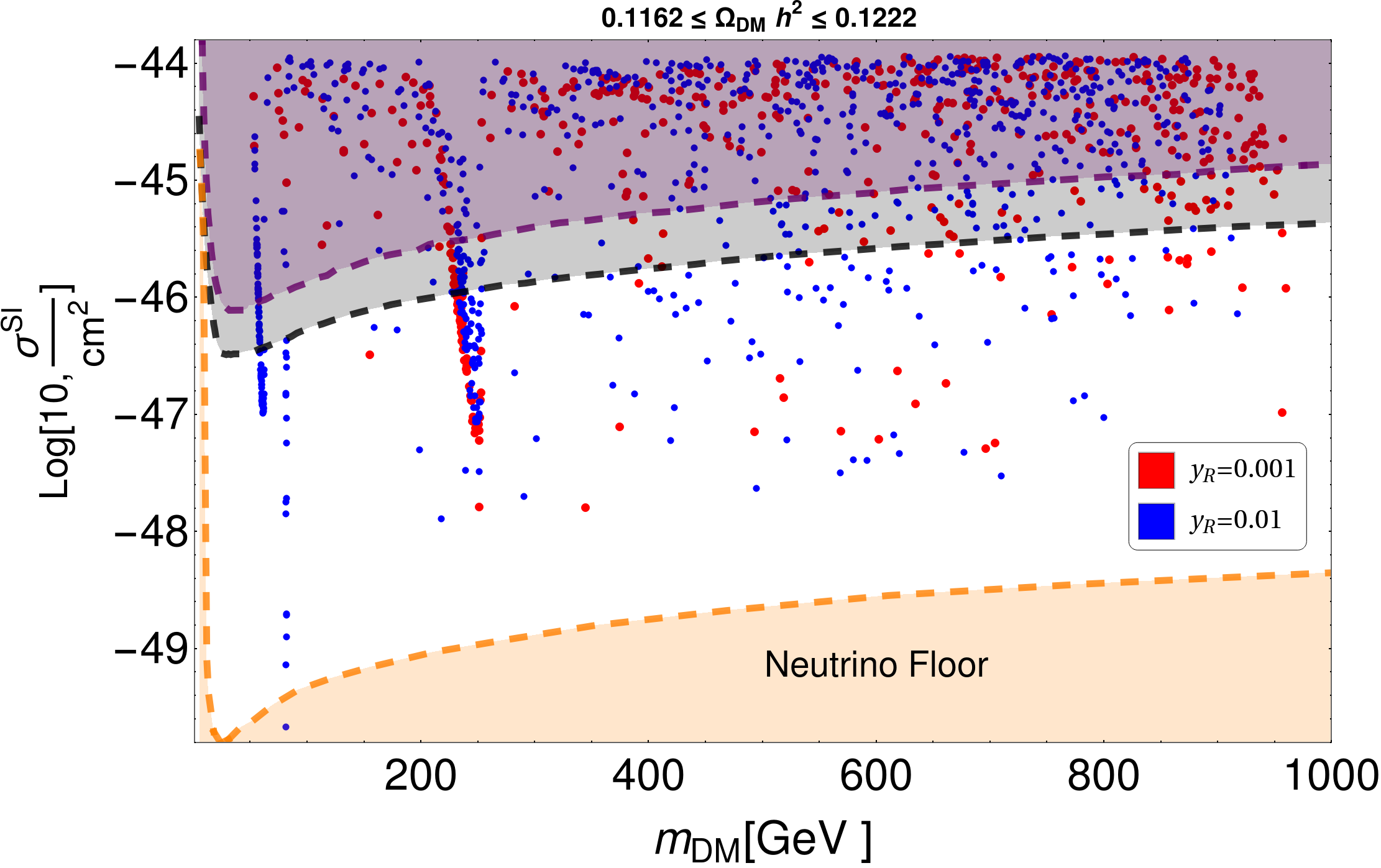}
 $$
   \caption{\it Relic density satisfied points are shown in spin-independent (SI) direct search cross-section against DM mass 
   for $y_{_R}=0.001$ (red points) and $y_{_R}=0.01$ (blue points). The direct search bounds from PANDAX-4T (black dotted line) 
   and XENON-1T (purple dotted line) are shown in the same plane for comparison purposes. The bottom shaded 
   orange region corresponds to the neutrino floor.}
  \label{fig:DD1}
 \end{figure}

 \begin{figure}[htb!]
 $$
   \includegraphics[scale=0.28]{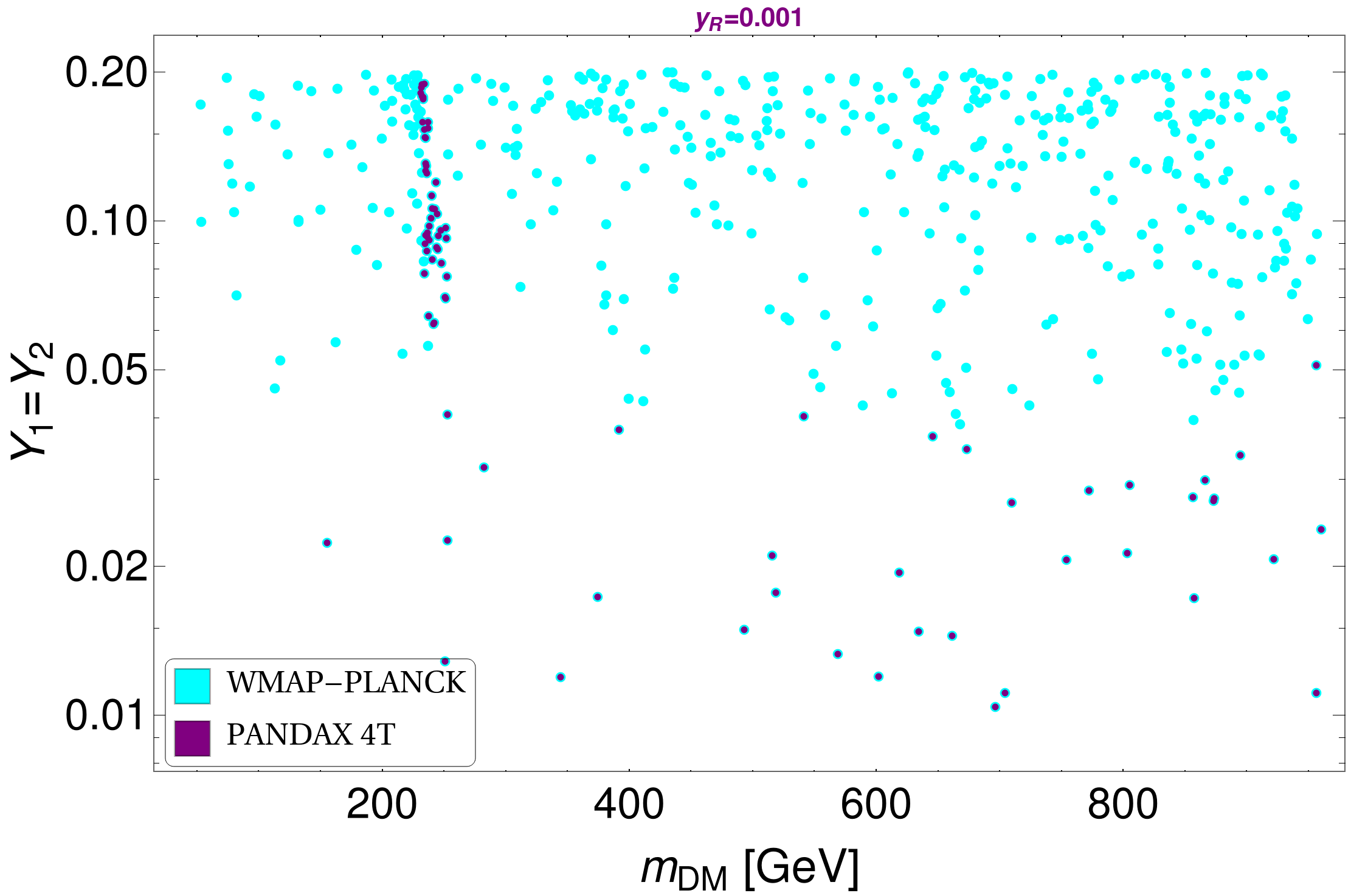}~
      \includegraphics[scale=0.28]{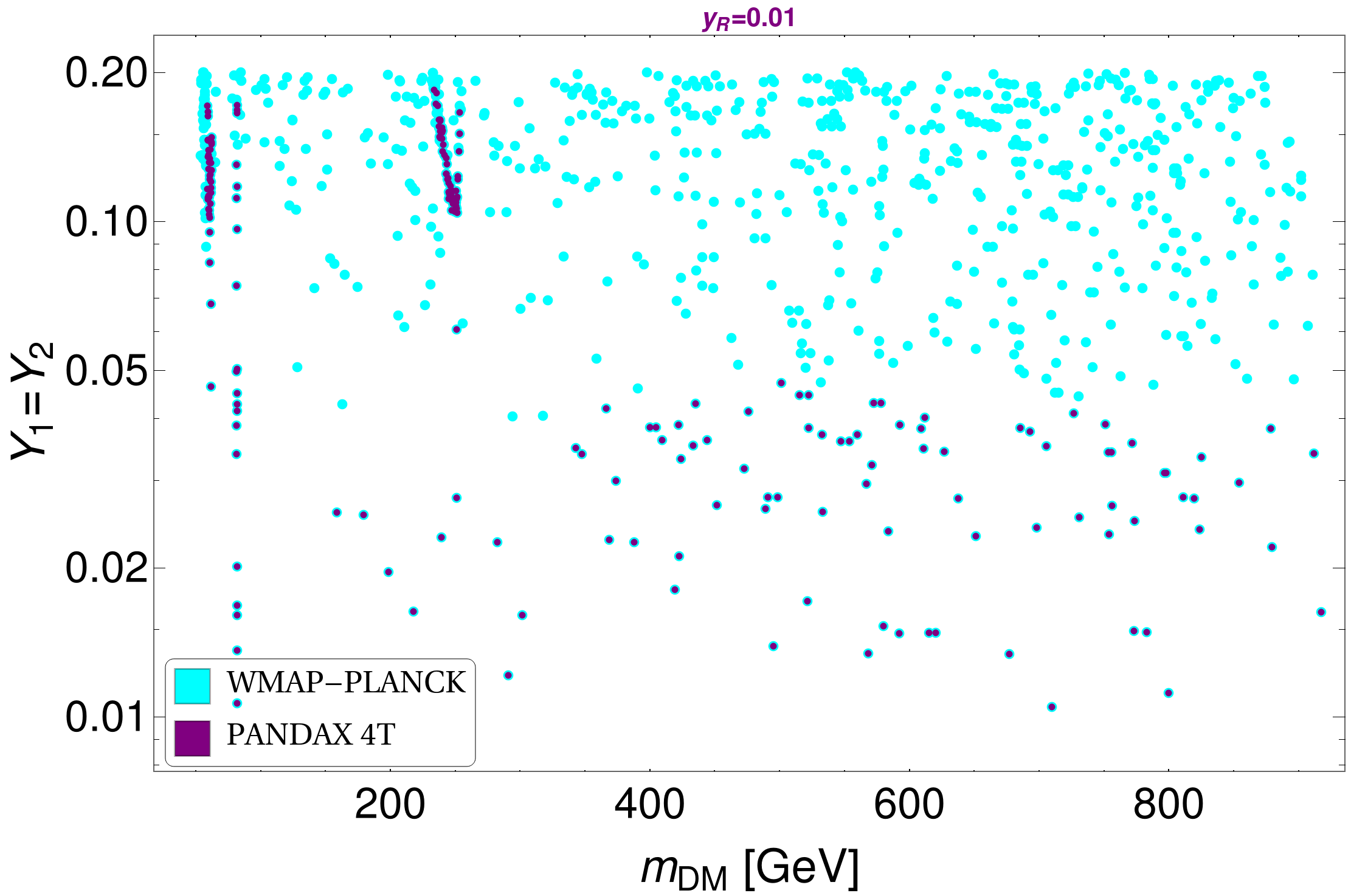}
 $$
  \caption{\it Relic density (cyan points)  and direct search (PANDAX 4T) (purple points) satisfied parameter space are shown in $m_{\rm DM} - Y$ plane for $y_{_R}=0.001$ (left) and $y_{_R}=0.01$ (right).}
  \label{fig:DD2}
 \end{figure}

 In Fig.\ref{fig:DD1} we show SI DM-nucleon cross-section for the model as a function of DM mass for the relic 
 satisfied parameters space. The red points correspond to $y_{_R}=0.001$ while the blue points are for $y_{_R}=0.01$. 
 The latest exclusion limits on DM-nucleon scattering cross-section 
 against DM mass from PANDAX-4T\cite{2021} and XENON 1T \cite{XENON:2018voc} are shown in the same plane by 
 black dashed line and purple dashed line respectively. It is worth 
 mentioning here that the parameter space below the dashed line (PANDAX -4T 
 and XENON 1T) can be allowed from the corresponding direct search experiments. We note here that more number of data points lie below the direct search exclusion limit from recent PANDAX-4T with the increase of $y_R$ from $0.001$ (red points) 
 to $0.01$ (blue points). With the increase of $y_R$, DM becoming more $SU(2)_R$ dominated resulting in a smaller DD cross-section. The orange shaded 
region in Fig.\ref{fig:DD1} corresponds to the neutrino floor due to neutrino-nucleon 
coherent elastic scattering. In the neutrino floor region, the direct search DM signal 
is not distinguishable from the neutrino background.
 
The SI DM-nucleon scattering cross-section in the mixed scenario takes place 
dominantly via t-channel scalar-mediated diagrams through the Yukawa interactions:    
$Y_1 \overline{\psi_1} \Phi \psi_2 + Y_2 \overline{\psi_1} \tilde{\Phi} \psi_2$. 
So the DM-nucleon scattering cross-section directly depends on the 
Yukawa coupling strength $Y_1$ and $Y_2$ which are considered equal here. 
Therefore the large values of $Y_1$ are strongly constrained from the recent 
PANDAX 4T data except the scalars resonance regions as shown in Fig.\ref{fig:DD2}. 
 
\subsection*{Indirect search constraint}

\begin{figure}[htb!]
 $$
   \includegraphics[scale=0.25]{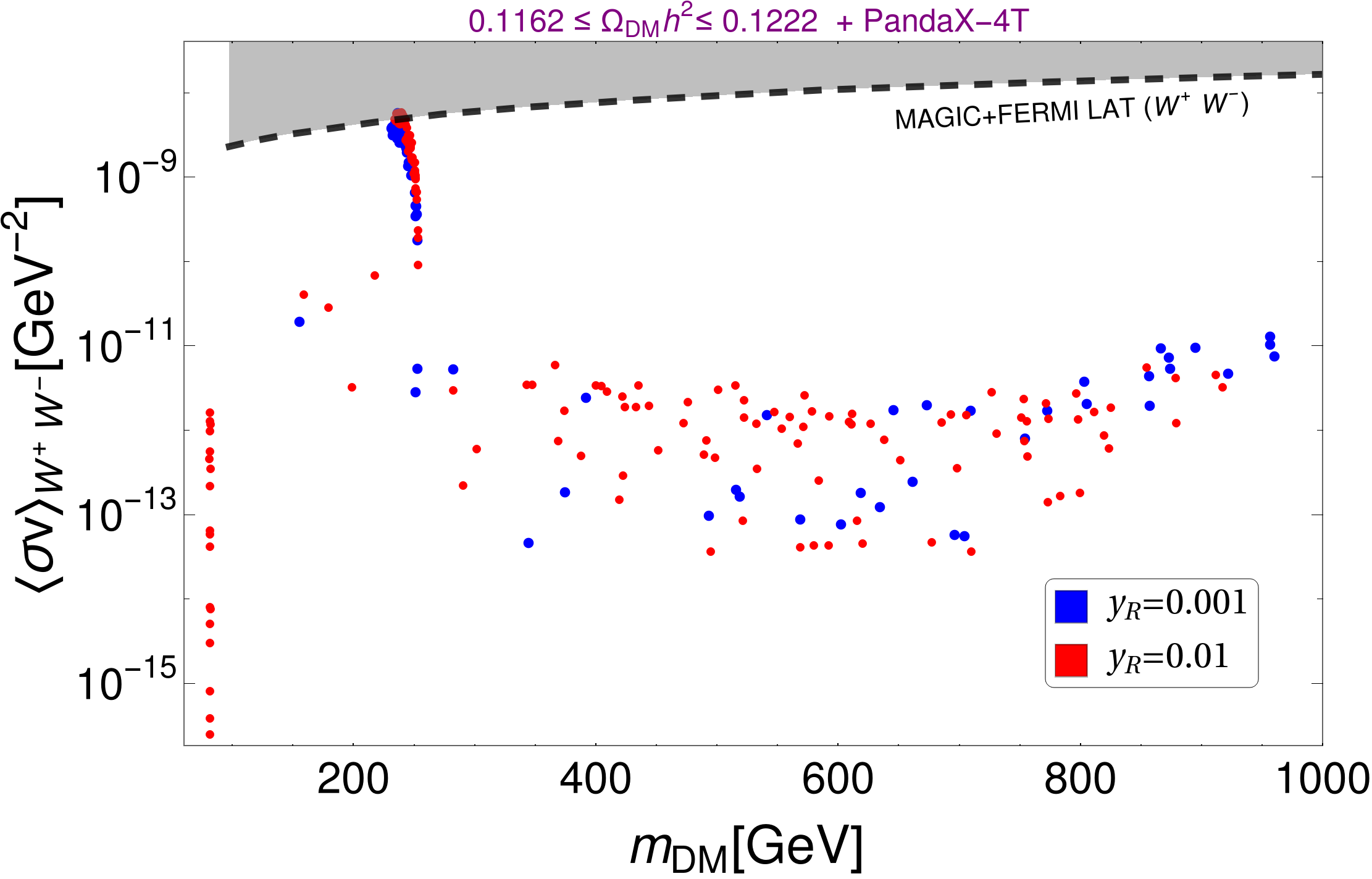}~
   \includegraphics[scale=0.25]{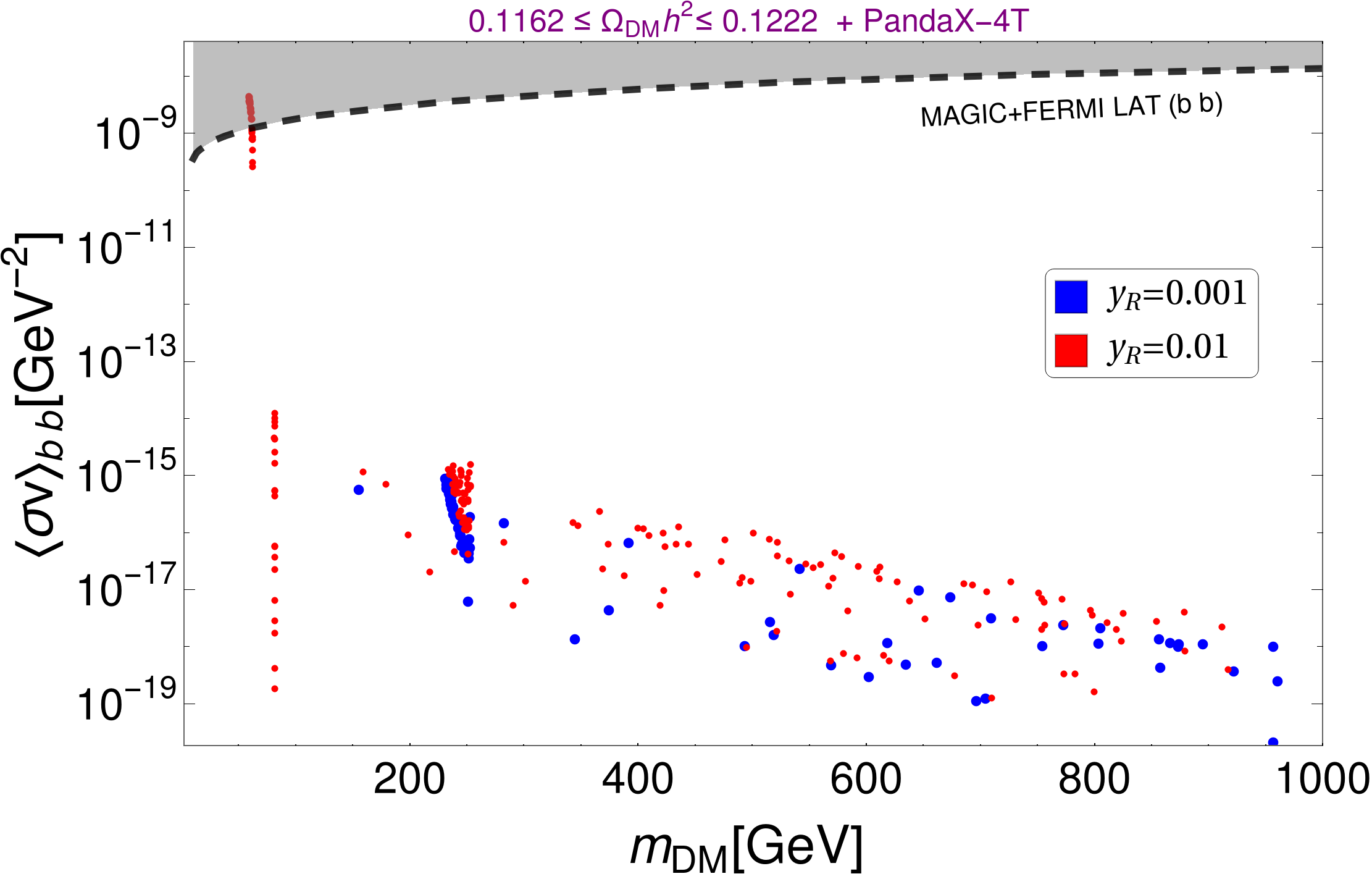}
  $$
  \caption{\it Parameter space satisfying relic density and direct search constraint is shown in the plane of indirect search cross-section as a function of $m_{\rm DM}$ for DM annihilating to $WW$ (left) and DM annihilating to $\overline{b} b$ (right) final states. The indirect search cross-section is compared with the latest exclusion bound of the corresponding channel from Fermi-LAT and MAGIC \cite{Ackermann_2015,2016} shown as shaded regions.}
  \label{fig:IDD}
 \end{figure}
 DM can also be probed at various indirect search experiments via the production of SM particles either through DM annihilation or via decay in the local Universe. Neutral stable particles like photons, neutrinos which are produced via DM annihilation or decay, can reach indirect search detectors without getting affected much by intermediate regions between the source and the detector. These photons and neutrinos are ideal messengers of DM indirect detection. Due to the WIMP nature of the 
 DM ($m_{\rm DM} \sim \mathcal{O}({\rm GeV-TeV})$), the photons emitted during DM annihilations or decay lie in the gamma-ray regime, which can be probed at 
 ground-based telescopes of 
 MAGIC (Major Atmospheric Gamma-ray Imaging Cherenkov)\cite{2016} and the space-based telescopes of Fermi-LAT (Fermi Large Area Telescope)\cite{Ackermann_2015}. It is important to note here that DM cannot interact directly with photons. But the gamma 
 rays can be produced via DM annihilation into different SM charged final states 
 like $\mu^+~\mu^-,~\tau^+~\tau^-,~b~\overline{b}$ and $W^+~W^+$ which finally decay 
 into photons. No signal for DM in indirect search experiments like Fermi-LAT and 
 MAGIC put a strong constraint on the annihilation cross-section of DM 
 into $\mu^+~\mu^-,~\tau^+~\tau^-,~b~\overline{b}$ and $W^+~W^+$. The most stringent constraint comes from the annihilation channel, DM DM $\to W^+~W^-$ for 
 $m_{\rm DM} > M_{W^\pm}$ and DM DM $\to b~ \overline{b}$ as compared to other 
 two channels, $\mu^+~\mu^-,~\tau^+~\tau^-$. In our analysis, we compare the indirect 
 search cross-section for $W^+W^-$ and $b\overline{b}$ final state with the 
 corresponding indirect search bounds arising from the Fermi-LAT and MAGIC 
 observations. In Fig.\ref{fig:IDD}, we show both relic  density and direct search 
 constraint(PANDAX 4T) satisfied points in $m_{\rm DM}-\langle \sigma v \rangle$ for annihilation of DM 
 to $WW$ (left panel) and to $b\overline{b}$ (right panel) final states. We find that most of the parameter 
 space shown in blue points for $y_{_R}=0.001$ and red points for 
 $y_{_R}=0.01$ are consistent with indirect search constraints except the light scalar resonance 
 region of DM mass around $\sim \frac{M_{h}}{2}$ and $~\sim \frac{m_{\Delta_L}}{2}$ with 
 large $Y_1$.

 \begin{figure}[htb!]
 $$
   \includegraphics[scale=0.40]{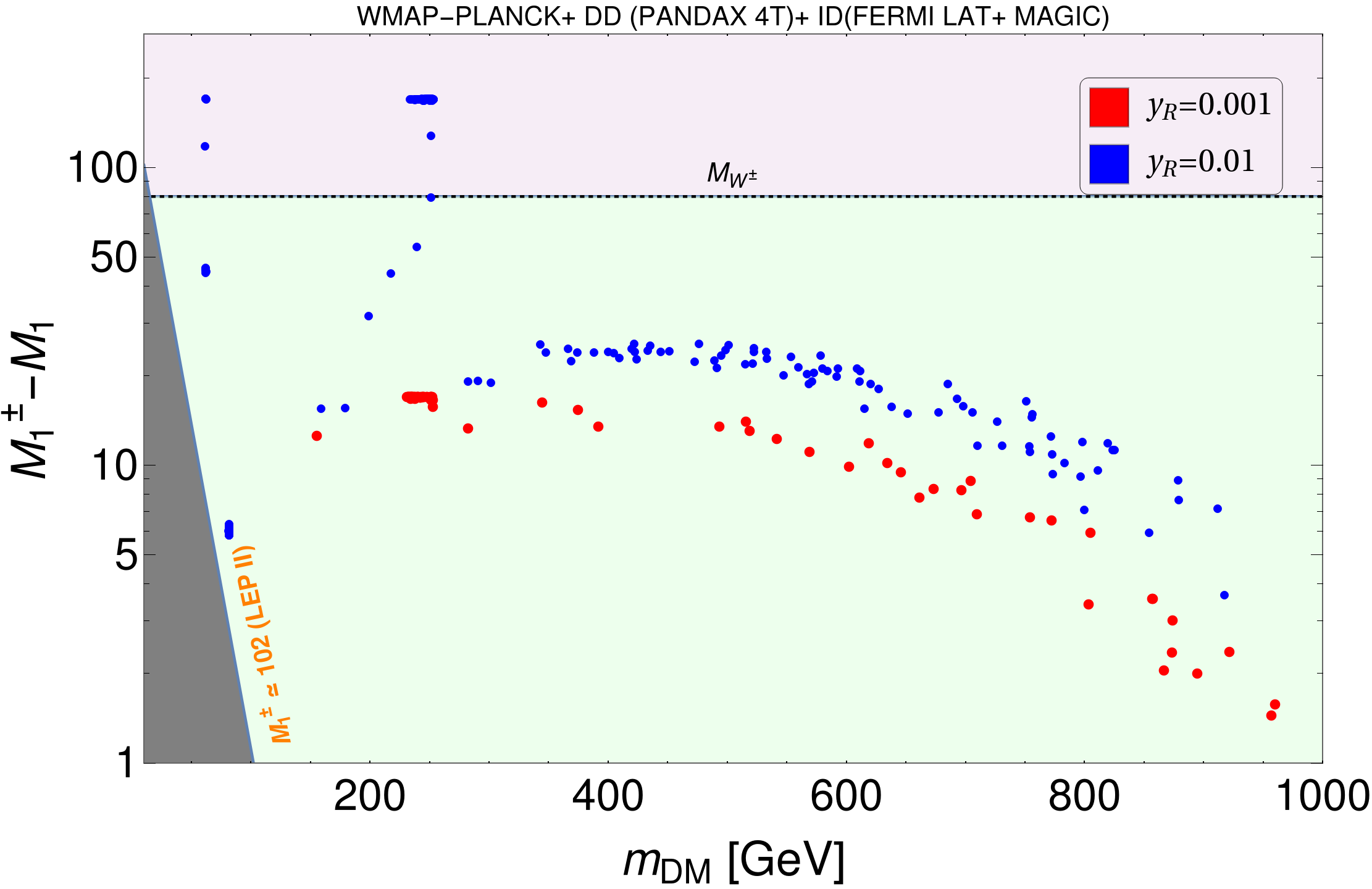}
  $$
  \caption{\it Relic density, direct and indirect search constraint allowed parameter space is summarized in 
  the $m_{DM}$ versus $\Delta M=M_1^\pm-m_{DM}$ plane for $y_R$ $=0.001$ (red points) and $=0.01$ (blue points). The grey 
  shaded region is excluded by LEP ($M_1^\pm < 102.7$). The black dotted horizontal line corresponding to $\Delta M= M_W$ which divides the parameter space into two regions: $\Delta M > M_W$ (light purple) and $\Delta M < M_W$ (light green).}
  \label{fig:DMpar}
 \end{figure}
 
 Finally, we put together all the constraints coming from relic density, direct and indirect searches and show the allowed region of parameter space in the plane 
 of $m_{\rm DM}-\Delta M (=M_1^\pm - m_{\rm DM})$ of Fig.\ref{fig:DMpar}. The red and blue points correspond to the Yukawa couplings  $y_{_R}=0.001$ and $0.01$ respectively. 
One should note here that the large $\Delta M=M_1^\pm-m_{DM}$ is only available 
near the Higgs resonance: $m_{\rm DM} \sim M_h/2$  and the the light scalar 
resonances: $\sim m_{\Delta_L}/2$. Apart from resonances the correct relic density 
only relies on the co-annihilation resulting in the small mass splitting as is shown in 
the figure. Again the large mass splitting is only available for the large Yukawa coupling $y_{_R}=0.01$ which is absent for $y_{_R}=0.001$. 
This is because the DM, $\chi_1$ becomes more $SU(2)_R$ dominated and the 
interactions between DM and left like fields in LRSM are suppressed with the increase 
of $y_{_R}$. The grey shaded region in the bottom left corner of Fig.\ref{fig:DMpar} which corresponds to $M_1^\pm > 102.7$ GeV is excluded by the LEP data\cite{2003}. We separate 
the parameter space in two regions along the $\Delta M$ direction with $\Delta M= M_W$ (black dashed line). The light purple region with $\Delta M > M_W$, the light 
charged dark fermion, $\chi_{_1}^\pm$ (of mass $M_1^\pm$) can decay to DM, $\chi_{_1}$ 
via on shell $W$. Whereas the region with $\Delta M < M_W$ is shown by light green shaded region, the light charged dark fermion, $\chi_{_1}^\pm$ can decay to 
DM ($\chi_{_1}$) via off-shell $W$. Depending on the on-shell or off-shell decay of the 
charged dark fermion to DM, the collider phenomenology will be different which we 
will discuss in the next section. So far throughout our analysis, we have only 
considered fixed particle spectrum in LRSM as mentioned in BP1. Now the obvious 
question arises, what will be the DM parameter space if one considers different particle spectrums in LRSM? 
The answer is that the DM phenomenology is almost the same except 
for the second resonance region which depends on the spectrum of the light scalars available in LRSM for a given benchmark point. That's why we do not repeat the DM analysis for 
different benchmark points in the LRSM sector. However for the collider analysis 
where we focus on the doubly charged scalar, we shall choose a different set of BP 
in the LRSM sector which we discuss in the next section.  
 \section{Collider signatures of $H^{\pm\pm}$ in presence of dark fermion doublets}
\label{sec:coll}
The LRSM gives us some unique collider signatures in the form of new gauge bosons,
a right-handed charged current interaction, heavy Majorana neutrino production, 
lepton number violations, etc. Thus each of them can be a test of the model. In addition to
the above, the presence of doubly charged scalars in the theory which when produced 
give a smoking gun signal in terms of resonances in the same sign dilepton final state. 
In fact, this signal is one of the well studied cases at LHC~\cite{ATLAS:2017xqs,CMS:2017pet,ATLAS:2021jol} which
leads to very strong bounds on the mass of the doubly charged scalar. This signal is 
however shared with other models which also predict doubly charged scalars, for example 
the Higgs Triplet model which leads to Type-II seesaw for neutrino masses. A significant 
part of the parameter space for sub-TeV doubly charged scalar is ruled out, when it decays 
dominantly in the leptonic mode. In LRSM, we have two copies of the doubly charged 
Higgs where one predominantly couples to the $Z$ while the other couples to the 
$Z_R$. This in turn affects the production rates for the two incarnates at LHC. However,
in the process, both could have a significantly off-shell $Z$ or $Z_R$ and conspire to 
give a cross section of nearly similar strengths for a given mass. The case where the Yukawa 
couplings that dictate the branching ratios in the leptonic mode being very small for the 
$H_L^{++}$ and $H_R^{++}$ makes the diboson mode ($WW$) as the 
other possibility. Since the $W_R$ is relatively heavy, the $H_R^{++}$ decays mostly via 
the Yukawa coupling to charged leptons. For the 
$H_L^{++}$ the decay is to on-shell $W$ boson. Thus, we get the possibility of 
$4W$ or $4\ell + \slashed{E_T}$ signal from the doubly charged scalar pair 
production at LHC. It is noteworthy that this decay mode also relaxes the bound on the 
doubly charged scalars significantly~\cite{ATLAS:2017xqs,ATLAS:2021jol}. In this work, we point out an additional 
channel that may open up for the $H^{\pm\pm}$ signal giving us hints of the dark sector.
As the Yukawa coupling of the Higgs triplets to the new VL doublets are constrained 
only through the allowed mass spectrum for the DM and its annihilation rates 
(as discussed in the previous sections), the doubly charged scalar can have its most 
dominant decay to the VL fermions. We focus on this signal by choosing a few 
representative points in the model parameter space which are consistent with DM 
observations and neutrino mass. Our choice of benchmark points shown in Table\ref{tab:tabbp} 
for the LHC study, are allowed from DM relic, direct search, indirect search and others 
constraints as discussed earlier.

\begin{table}[]
 \resizebox{\linewidth}{!}{
  \begin{tabular}{|c|c|c|c|}
  \hline
& {\bf BPC1}   & {\bf BPC2} & {\bf BPC3}  \\ \hline\hline  
\makecell{$\Big(M_{H_R^{\pm\pm}},~M_{H_L^{\pm\pm}},~~ v_R \Big)$ } &     \makecell{(889.2,~300.3,~~30$\times 10^3$)}   &   \makecell{(1000,~280,~~30$\times 10^3$)} & \makecell{(800,~300,~~30$\times 10^3$)}   \\ \hline  
\makecell{DM Inputs \\ Mass(GeV)}  &  \makecell{$M_L=150$ $~M_R=150$ \\ $Y_1=Y_2= 4\times 10^{-2}$ \\$y_R=2.06 \times 10^{-3}$$~~y_L=0.6$  }    &  \makecell{$M_L=142$ $~M_R=142$ \\ $Y_1=Y_2= 4\times 10^{-2}$ \\$y_R=2.08 \times 10^{-3}$$~~y_L=2.0$  } & \makecell{$M_L=120$ $~M_R=210$ \\ $Y_1=0.036$$~Y_2= 0.06$ \\$y_R=0.028$$~y_L=0.6$  }\\ \hline
\makecell{Dark Particles \\ Mass(GeV) }  & \makecell{$M_1 = 62.044, ~M_2 =149.708 ,$ \\ $~M_3= 150.291 ,~M_4= 237.955  $ \\ $M_1^\pm= 143.035 ,~M_2^\pm= 156.964 $ }      & \makecell{$M_1 = 61.118, ~M_2 =147.735 ,$ \\ $~M_3= 152.264 ,~M_4= 238.811  $ \\ $M_1^\pm= 143.035 ,~M_2^\pm= 156.964 $ } & \makecell{$M_1 = 89.8 , ~M_2 = 120.4,$ \\ $~M_3= 120.6,~M_4= 328.9  $ \\ $M_1^\pm= 118.8 ,~M_2^\pm= 211.1 $ } \\ \hline
\makecell{Relic Density, \\ Direct Detection, \\ Indirect Detection}  &  \makecell{ $\Omega_{\rm DM} h^2 =0.107$ \\ \\ $\sigma^{\rm SI}_n = 4.92 \times 10^{-48} ~{\rm cm}^2$ \\ \\ $\langle \sigma v \rangle_{\rm \mu \mu} =1.61 \times 10^{-13}~ {\rm GeV}^{-2}$\\ $\langle \sigma v \rangle_{\rm \tau\tau} = 4.54 \times 10^{-11}~ {\rm GeV}^{-2}$ \\ $\langle \sigma v \rangle_{\rm b b} = 7.50 \times 10^{-10}~ {\rm GeV}^{-2}$}    &  \makecell{ $\Omega_{\rm DM} h^2 =0.1150$ \\ \\ $\sigma^{\rm SI}_n = 1.46 \times 10^{-48} ~{\rm cm}^2$ \\ \\ $\langle \sigma v \rangle_{\rm W W} =1.19\times 10^{-12}~ {\rm GeV}^{-2}$ \\ $\langle \sigma v \rangle_{\rm \mu \mu} =4.38 \times 10^{-21}~ {\rm GeV}^{-2}$\\ $\langle \sigma v \rangle_{\rm \tau\tau} = 1.24 \times 10^{-18}~ {\rm GeV}^{-2}$ \\ $\langle \sigma v \rangle_{\rm b b} = 1.98 \times 10^{-17}~ {\rm GeV}^{-2}$} & \makecell{ $\Omega_{\rm DM} h^2 =0.0973$ \\ \\ $\sigma^{\rm SI}_n = 6.687 \times 10^{-47} ~{\rm cm}^2$ \\ \\ $\langle \sigma v \rangle_{\rm \mu \mu} =4.46\times 10^{-18}~ {\rm GeV}^{-2}$\\ $\langle \sigma v \rangle_{\rm \tau\tau} =1.26 \times 10^{-15}~ {\rm GeV}^{-2}$ \\ $\langle \sigma v \rangle_{\rm b b} = 1.07 \times 10^{-14}~ {\rm GeV}^{-2}$} \\ \hline

\makecell{Neutrino sector \\ ($m_{\nu}=0.1$ eV)}  &  \makecell{${M_{D}}_{ii}=5 \times 10^{-5}$ GeV  ${M_{D}}_{i\neq j}=0$ \\ ${Y_{\Delta R}}_{ii}=5.892\times 10^{-4}$ ${Y_{\Delta R}}_{i\neq j}=0$ \\ $M_N=25$ GeV;~$V_{\ell N} \simeq 10^{-6}$ }     &  \makecell{${M_{D}}_{ii}=5 \times 10^{-5}$ GeV  ${M_{D}}_{i\neq j}=0$ \\ ${Y_{\Delta R}}_{ii}=5.892\times 10^{-4}$ ${Y_{\Delta R}}_{i\neq j}=0$ \\ $M_N=25$ GeV;~$V_{\ell N}\simeq  10^{-6}$} & \makecell{${M_{D}}_{ii}=1 \times 10^{-4}$ GeV ${M_{D}}_{i\neq j}=0$ \\ ${Y_{\Delta R}}_{ii}=2.357\times 10^{-3}$ ${Y_{\Delta R}}_{i\neq j}=0$ \\ $M_N=100$ GeV;~$V_{\ell N}\simeq 10^{-6}$ }\\ \hline
\makecell{Doubly charged \\ Scalar} & \makecell{  $\Gamma(H_L^{\pm\pm})=1.228 \times 10^{-1}$ GeV   \\ Br($H_L^{\pm\pm} \to \chi_1^\pm~\chi_1^\pm $) $\simeq 98.67 \%$ \\ Br($H_L^{\pm\pm} \to \chi_1^\pm \chi_2^\pm $) $\simeq 0.659 \%$ \\ Br($H_L^{\pm\pm} \to W W$) $\simeq 0.665 \%$ }   &  \makecell{  $\Gamma(H_L^{\pm\pm})=7.571 \times 10^{-2}$ GeV   \\ Br($H_L^{\pm\pm} \to \chi_1^\pm~\chi_1^\pm $) $\simeq 99.08 \%$ \\ Br($H_L^{\pm\pm} \to W W $) $\simeq 0.9195 \%$ } & \makecell{   $\Gamma(H_L^{\pm\pm})= 3.835 $ GeV  \\  Br($H_L^{\pm\pm} \to \chi_1^\pm~\chi_1^\pm $)$ \simeq 99.9 \%$  } \\ \hline
\makecell{Dark charged \\ Fermion }  & \makecell{$\Gamma(\chi_1^{\pm})=4.005\times 10^{-6}$ GeV  \\ Br($\chi_1^{\pm} \to \chi_1 W^{\pm} $)$\simeq  100\%$} &   \makecell{$\Gamma(\chi_1^{\pm})=1.462\times 10^{-5}$ GeV  \\ Br($\chi_1^{\pm} \to \chi_1 W^{\pm} $)$\simeq  100\%$  } & \makecell{$\Gamma(\chi_1^{\pm})=9.39 \times 10^{-7}$ GeV  \\   
Br($\chi_1^{+} \to \chi_1 u_i \bar{d_j}^\prime $)$\simeq  66.6\%$  \\ Br($\chi_1^{+} \to \chi_1 \ell^+ \nu_\ell  (\ell= e, \mu)$)$\simeq  22.2\%$ \\ Br($\chi_1^{+} \to \chi_1 \ell^+ \nu_\ell  (\ell=\tau)$)$\simeq  11.1\%$  } \\ \hline
\makecell{Cross-section \\  
$\sqrt{s}=14$ TeV (LHC)} &  \makecell{$\sigma(p p \to H_L^{++} H_L^{--})= 13.953$ fb  }  &  \makecell{ $\sigma(p p \to H_L^{++} H_L^{--})=16.58$ fb  }& \makecell{$\sigma(p p \to H_L^{++} H_L^{--})=13.9$ fb   \\ $\sigma(e^+ e^- \to H_L^{++} H_L^{--})= 58.22$ fb \\ (ILC: $\sqrt{s}=1$ TeV)}  \\ \hline
 \end{tabular}
}
\caption{\it The above benchmark points are considered for collider analysis in our model. The mass of dark sector particles, corresponding Yukawa couplings, relic density, direct and indirect search cross-section for DM are tabulated. 
Dark fermion decay branching ratios and the decay branching ratios of doubly charged scalars are also shown for the
BP's. Other dark sector parameters are kept fixed as mentioned earlier.}
\label{tab:tabbp}
\end{table}
The production of doubly charged scalar at LHC yields the attractive collider signature 
in presence of DM in our model. The doubly charged scalar decays promptly to a 
pair of charged component of the dark doublets ($\chi_{_i}^\pm$) which then further 
decay to DM ($\chi_{_1}$) and $W^\pm$.  This yields a $W^+ W^+ W^- W^-$ plus DM 
final state from a pair produced doubly charged scalar. The very weakly interacting 
DM particles escape the detector leaving their imprint in the form of missing energy. 
We look at the following signal subprocesses:       
\bea
{\rm ~Signal ~:}
~~ p~p \rightarrow H_L^{++},H_L^{--}, ~(H_{L}^{\pm\pm} \rightarrow \chi_1^\pm~\chi_1^\pm ),~~(\chi_1^\pm \to \chi_1 ~ W^\pm).
\eea

\begin{figure}[htb!]
 $$
 \includegraphics[scale=0.45]{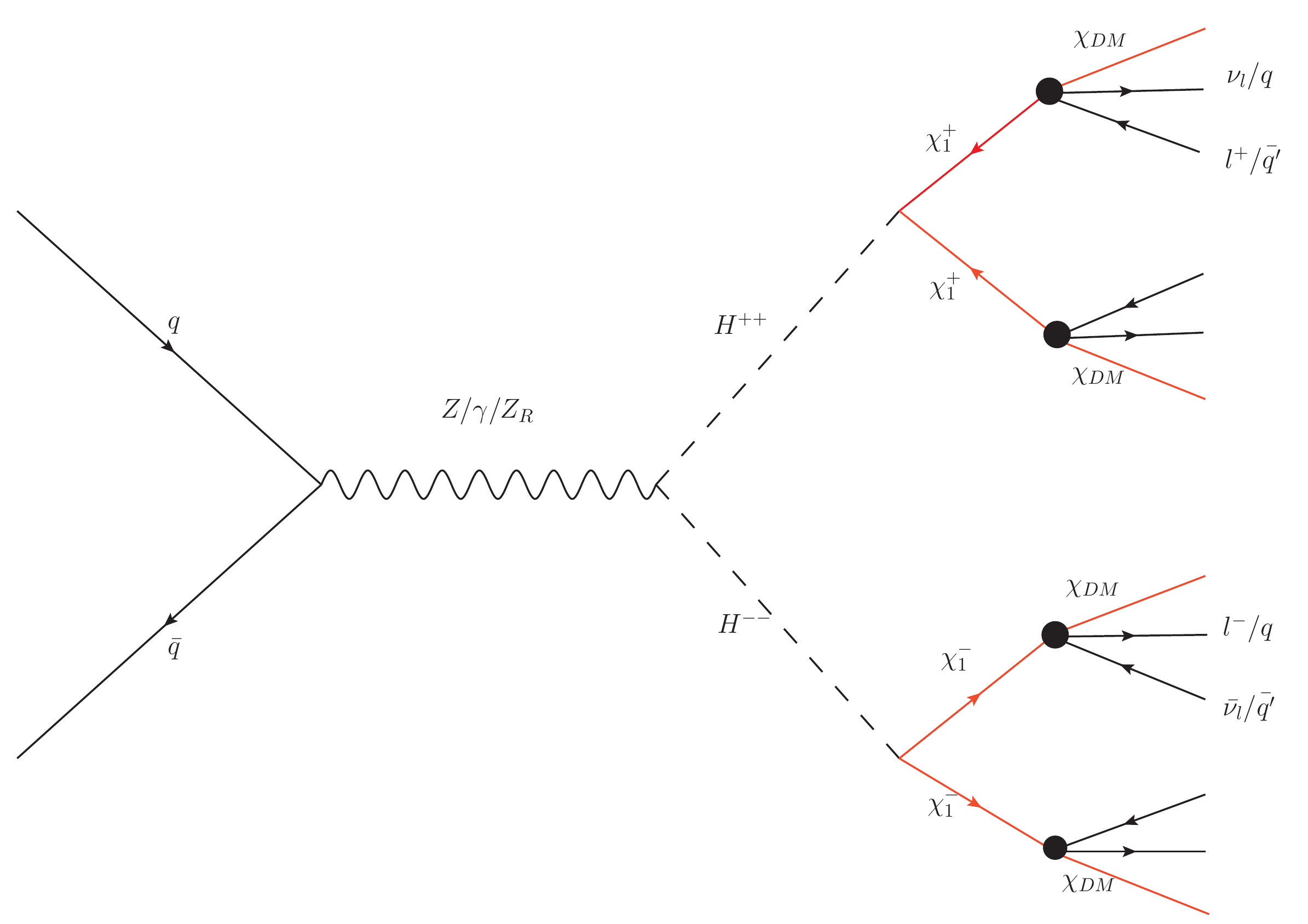}
 $$
  \caption{\it Feynman diagram for doubly charged scalar production at LHC.}
  \label{fig:lhc}
 \end{figure}
The $W^\pm$ decays to leptons and/or jets giving rise to the following different final states:
\begin{itemize}
 \item $4 \, l + \slashed{{E}_T}$
 \item $3 \, l + 2 \, {\rm jets} + \slashed{{E}_T}$
 \item $2 \, l + 4 \, {\rm jets} + \slashed{{E}_T}$
 \item $1 \, l + 6 \, {\rm jets} + \slashed{{E}_T}$
\end{itemize}
We note that the most promising signal would involve the larger multiplicity of charged leptons in the final state 
which will also be suggestive of the doubly charged scalar as the parent particle. 
The final states with increasing jet 
multiplicities would provide complementary signals, hitherto with reduced sensitivity as the SM background 
would be large compared to an all lepton final state. We therefore restrict ourselves to the first two channels involving $n \geq 3$ charged leptons in the final state for our analysis. We also note that this signal overlaps with the $4W$ final state coming from pair produced doubly charged Higgs when the $H^{\pm\pm} \to W^\pm W^\pm$ is the dominant decay channel, which could give us an idea on the improvement of the signal over that of the 
traditional $4W$ signal.

We  use the publicly available package  {\tt SARAH}~\cite{Staub:2013tta} to write the model files and create the
{\tt Universal Feynman Object~(UFO)}~\cite{Degrande:2011ua} files. The mass spectrum and mixings are 
generated using {\tt SPheno}~\cite{Porod:2003um,Porod:2011nf}.  We have used the 
package {\tt MadGraph5@aMCNLO}~(v2.6.7)~\cite{Alwall:2011uj,Alwall:2014hca} to calculate the scattering process 
and generate parton-level events at LHC with $\sqrt{s}=14$ TeV which were then showered with the help of {\tt Pythia \!8}~\cite{Sjostrand:2014zea}. 
We simulate detector effects using the fast detector simulation in {\tt Delphes-3}~\cite{deFavereau:2013fsa} and have used the default 
ATLAS card. The reconstructed events were finally analyzed using the analysis package {\tt MadAnalysis5}~\cite{Conte:2012fm}.


\subsection{$4l+\slashed{E_T}$ signal}
The $4l+\slashed{{E}_T}$ is one of the cleanest signal because of the low SM 
background. 
In our study this signal will appear when all four $W$ bosons produced in the cascade 
decay of the pair of double charged Higgs, decays leptonically as shown in Fig.~\ref{fig:lhc}. 
The dominant background for the above final state would come 
from the SM subprocesses producing $t\overline{t}Z$, $ZZ$ and $VVV$~\cite{delAguila:2008cj}.
Additional sources of background events could also emerge from $t\overline{t}$ and $WZ$
production, where additional charged leptons can come from misidentification 
of jets. Although such events would be small, the sheer size of the cross section of the 
aforementioned processes could lead to significant events mimicking the signal. 
However, these backgrounds can be eliminated by choosing specific selection cuts. The signal and the background process are 
generated using the same Monte Carlo event generator and then the cross section of the backgrounds are scaled with their 
respective $k$-factors. The $k$-factor for $ZZ$ , $t\bar{t}Z$, $VVV$ and $WZ$ considered here are 
$\simeq$ 1.72, 1.38, 2.27 and 2.01 respectively \cite{Cascioli:2014yka,Kardos:2011na,Wang:2016fvj,Grazzini:2016swo}. Here the $k$-factor 
for $ZZ$ scales it to \textit{next-to-next-to-leading order} (NNLO), while the rest of the 
backgrounds are at \textit{next-to-leading order} (NLO) cross section.

To consider the four charged lepton final state coming from the $4W$ we choose events which have 
exactly $N_l = 4$ isolated charged leptons~($l = e, \mu$) in the final state. As the 
final state will still be littered with jets coming from initial state radiations,
we therefore choose a more inclusive final state where 
all jets are vetoed with a relatively large transverse momenta of $40$ GeV. 
As basic acceptance cuts, we therefore demand that all 
reconstructed objects are isolated~($\Delta R_{ab} > 0.4$). In addition,
\begin{itemize}
\item all charged leptons  must have $p_{T_l} > 10$~GeV and lie within the rapidity gap satisfying $|\eta_l| < 2.5$. 
\item We impose additional conditions to demand a hadronically quite environment by putting veto on events with light jets 
and $b$ jets  with $p_{T_{b/j}} > 40$~GeV and $|\eta_{b/j}| < 2.5$. This helps in suppressing a significant part of the background coming from $t \, \bar{t} (Z)$ production. 
\end{itemize}
 
\begin{table}[h]
\resizebox{\linewidth}{!}{
 \begin{tabular}{|c|c|c|c|c|c|c|}
\hline
 Cuts (GeV)& $\slashed{E_T}<30$ & $82<M_{e^+e^-}<100$  &$ 82<M_{\mu^+\mu^-}<100$ &$ p_T[l_2]<30$ 
 & $M_{l_i^+l_j^+}>110$ & $M_{l_i^-l_j^-}>110$ \\ \hline\hline
BPC1&15.4&13.6&10.4&9.5&9.4&9.3\\ \hline
BPC2&18.2&15.9&11.4&10.2&9.9&9.9\\ \hline
Background&687.7&381.0&53.0&20.0&8.5&1.0\\ 

\hline
\end{tabular}
}
\caption{\it Rejection cut-flow chart of $4l+\slashed{E_T}$ signal analysis for BPC1 \& BPC2 benchmark points at 
3000 fb$^{-1}$ integrated luminosity.}
    \label{tab:tab7}
\end{table}

\begin{figure}[htb!]
 $$
   \includegraphics[scale=0.30]{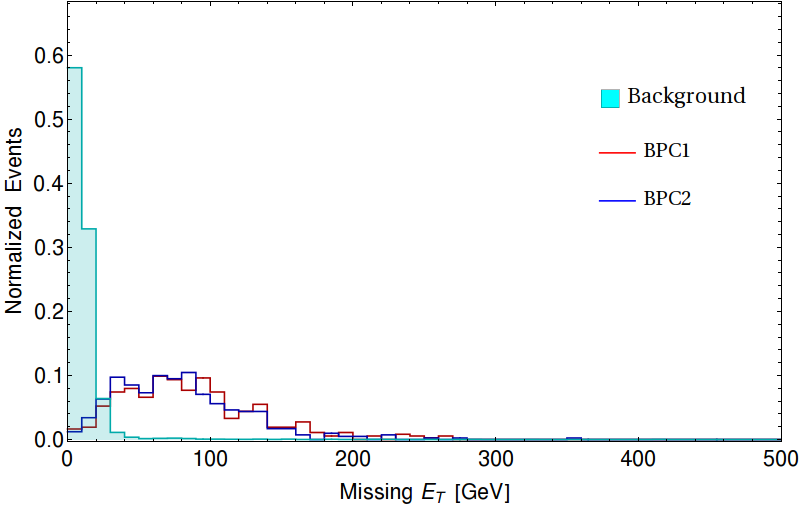}
$$
 
  \caption{\it Normalized distributions of missing transverse energy ($\slashed{E_T}$) for signal (BPC1 and BPC2) and 
  total SM background.}
  \label{fig:4l_os}
 \end{figure}
\begin{itemize}
\item The largest contribution to the SM background comes from $ZZ$. To suppress it and bring it down, 
we choose a missing $\slashed{E_T} > 30$ GeV selection cut. Since the $ZZ$ decaying 
to give four leptons will have very little missing energy in the final state the cut will throw away a 
significant part of the background events. 
This cut does not affects our signal much since it has decay products consisting of dark matter and neutrinos leading to 
a larger $\slashed{E_T}$ in the signal events. Hence this cut becomes very efficient in improving the signal sensitivity.
\end{itemize}

\begin{figure}[htb!]
 $$
   \includegraphics[scale=0.30]{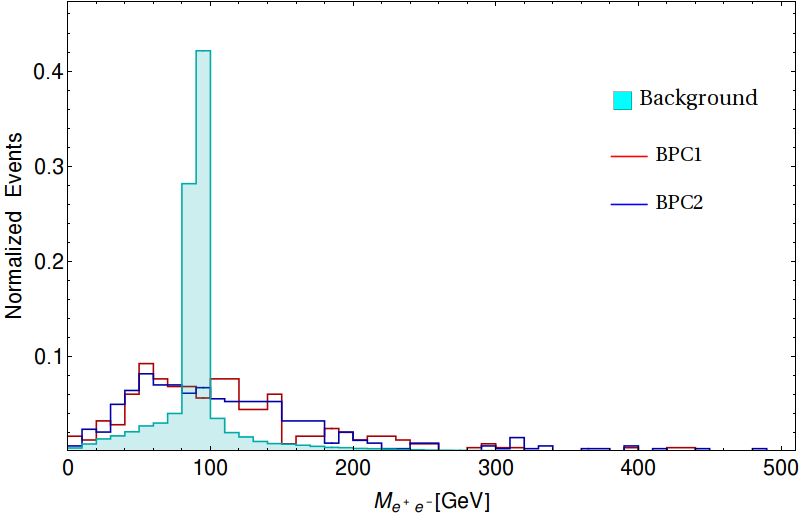}
 \includegraphics[scale=0.30]{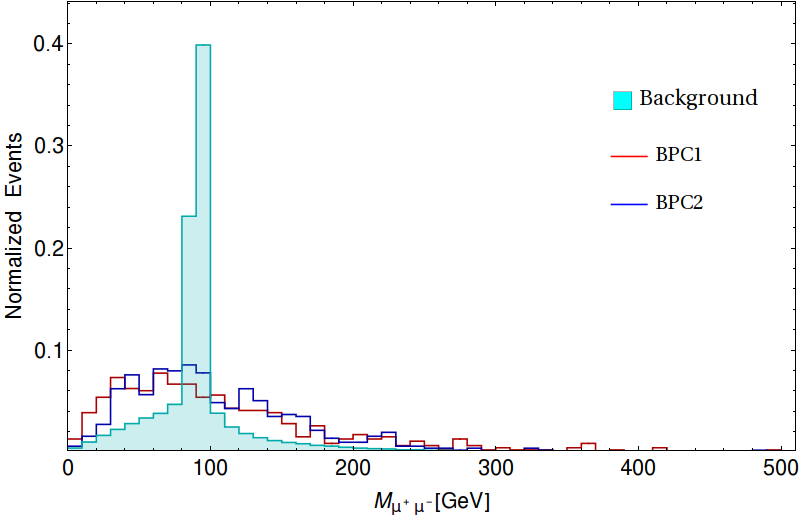}
 $$
  \caption{\it Normalized invariant mass distribution $M_{e^+e^-}$ [left] and $M_{\mu^+\mu^-}$ [right] for 
  signal (BPC1 and BPC2) and total SM background.}
  \label{fig:4l_os}
 \end{figure}

\begin{itemize}
\item To reduce the background further, which may produce final state charged leptons (i.e $WZ$, $VVV$ and $ZZ$)  
from $Z$ decay but have some missing transverse energy which lets them escape the $\slashed{E_T}$ cut, we put a 
cut on the invariant mass of same flavor opposite charge leptons around the $Z$ mass 
pole ($82 {~\rm GeV} < M_{\ell^+\ell^-} < 100~ {\rm GeV} $). In our signal the leptons come from the $W^\pm$ decay 
and no resonant feature can be attributed in the decay as the charged leptons come from 
different parent particles. So the signal is not affected by a cut used to remove the resonant $Z$ peak in the background.
\end{itemize}

\begin{figure}[htb!]
 $$
   \includegraphics[scale=0.30]{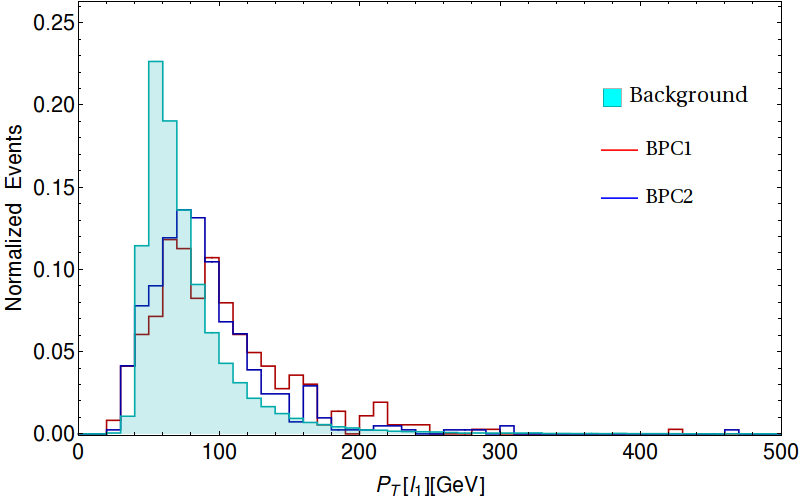}
 \includegraphics[scale=0.30]{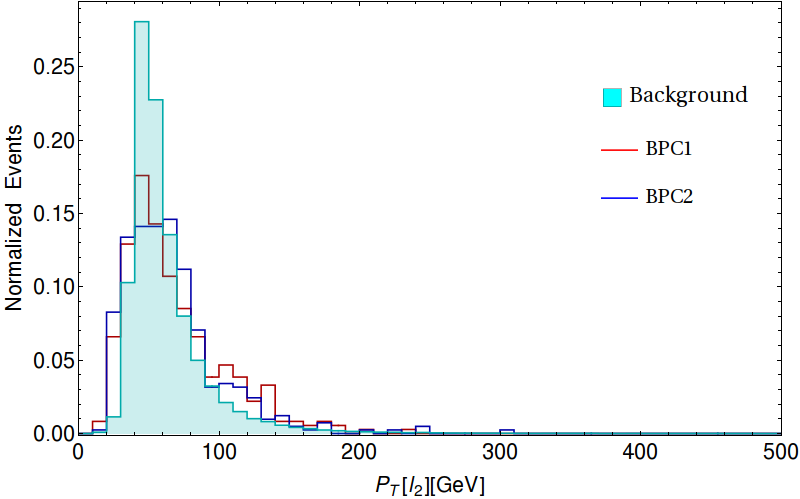}
 $$
  \caption{\it Normalized $p_T$ distribution of leading charged lepton [left] and sub-leading charged lepton [right]
  for signal (BPC1 and BPC2) and  total SM background.}
  \label{fig:4l_pt}
 \end{figure}

\begin{itemize}
\item At this point the background is almost at a comparable level with the signal and most of the remaining SM background 
contribution is from the $WZ$ channel where additional jets/photons can be misidentified as an additional charged lepton. 
But the events from $WZ$ will give softer decay products and we find the use a strong $p_T$ cut helpful in suppressing them 
significantly. In our case the sub-leading lepton with $p_T[l_2]$ separates the signal from background 
when compared to the same observable for other charged leptons. Thus we choose a  $p_T[l_2]>30$ GeV selection 
cut to help reduce the $WZ$ background.  
\end{itemize}

\begin{figure}[htb!]
 $$
 \includegraphics[scale=0.30]{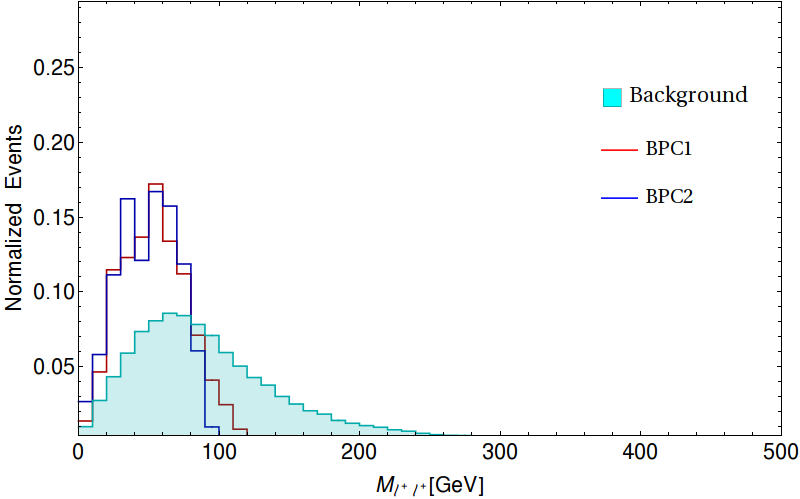}
 \includegraphics[scale=0.30]{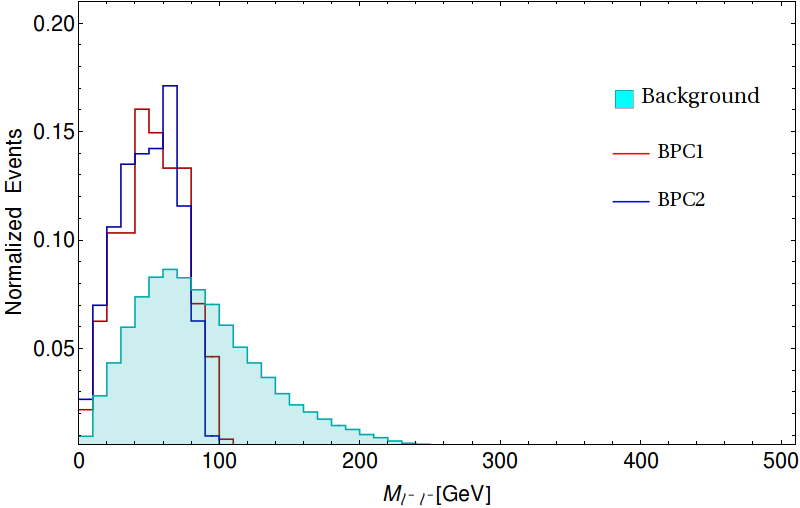}
 $$
  \caption{Normalized invariant mass distribution for same sign leptons ($M_{l_i^\pm l_j^\pm}$) 
  for signal (BPC1 and BPC2) and total SM background.}
  \label{fig:4l_ss}
 \end{figure}

\begin{itemize}
\item There is one more kinematic variable of interest which can be used to distinguish 
the signal from the background. It is the invariant mass of 
same-sign (SS) charged lepton pairs which can be used to reduce 
the background even further. Even though the doubly charged scalar does not decay 
directly to SS leptons, we expect that in the all lepton channel the SS leptons 
would come from the same primary scalar. As the SS leptons in our signal come from 
the decay of same 
parent particle ($H_L^{\pm\pm}$), we expect a maximum invariant mass for such lepton 
pair dictated by the difference in mass of $H_L^{\pm\pm}$ and the DM pair. But in the background there is no 
clear possibility of a kinematic edge and therefore the tail of this 
observable ($M_{l_i^\pm l_j^\pm}$) for the background will be much broader compared to the signal. We can remove 
this tail in the observable for the SM background without rejecting any significant signal events.
\end{itemize}
 All the cuts used above have been shown through a cut-flow chart in 
 Table\ref{tab:tab7}. The final surviving events (after the selection cuts) are shown in Table\ref{tab:tab8} 
 for an integrated luminosity of $3000$ fb$^{-1}$ and the signal significance is estimated using
\begin{equation}
\sigma = [2\{(b+s) \, \log\left( \frac{b+s}{b}\right) -s\}]^{1/2}
\label{eq:sign}
\end{equation}
where $b$ stands for the SM background and $s$ represents the new physics signal 
events respectively. 
\begin{table}[h]
\centering
\resizebox{7 cm}{!}{
 \begin{tabular}{|c|c|c|c|}
\hline
 Benchmark & Signal & Background &Significance  \\ \hline\hline
BPC1&9.4&1.0&5.50\\ \hline

BPC2&9.9&1.0&5.75\\ 

\hline
\end{tabular}
}
\caption{Significance of BPC1 and BPC2 at integrated luminosity of $3000$ fb$^{-1}$}
    \label{tab:tab8}
\end{table}
 \subsection{$3l+2j+\slashed{E_T}$ signal} 
The $3l + 2j + \slashed{E_T}$ is the next cleanest signal for our model after 
$4l + \slashed{E_T}$ but its features can be studied with lower integrated luminosity 
compared to the $4l + \slashed{E_T}$ case. This is because we allow any one of the
$W$ boson to decay hadronically (which has a larger branching ratio over the 
leptonic channel). Hence the effective cross-section of this signal is much larger than 
the $4l + \slashed{E_T}$ case. The prominent SM subprocesses contributing as 
background to our signal are $t\bar{t}$, $WZ$, $VVV$, $t \bar{t}Z$ and $ZZ$. 
For our analysis of the final state we choose only those events as signals which have exactly three charged leptons ($e^\pm$ and $\mu^\pm$) and exactly two jets.  
Similar to the
$4l + \slashed{E_T}$ case we consider the basic acceptance cuts for all isolated 
objects (i.e., $\Delta R_{ab} >0.4$) as given below:
\begin{itemize}
\item all charged leptons  must have $p_{T_\ell} > 5$~GeV and lie within the rapidity gap satisfying $|\eta_\ell| < 2.5$. 
\item We impose additional conditions and demand no $b$-tagged jets by putting veto on events with $b$ jets with 
$p_{T_{b}} > 40$~GeV and $|\eta_{b}| < 2.5$. This helps in suppressing a significant part of the 
background coming from $t \bar{t}$ and $t \, \bar{t} (Z)$ production. 

\end{itemize}

\begin{figure}[htb!]
 $$
 \includegraphics[scale=0.30]{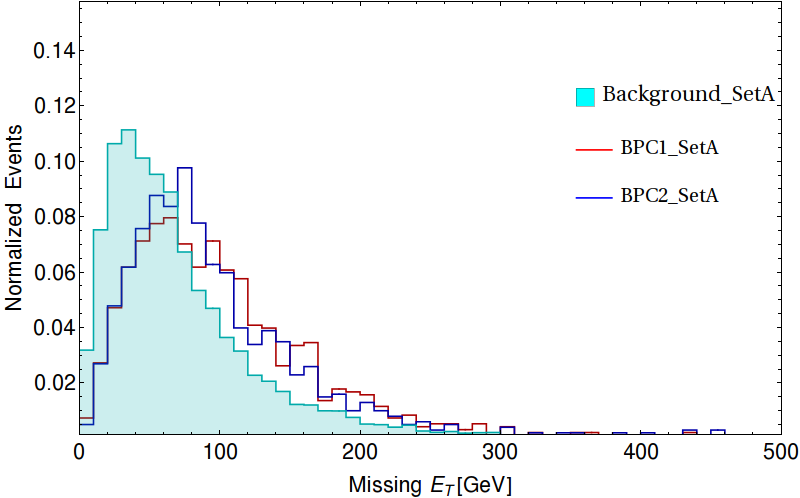}
 \includegraphics[scale=0.30]{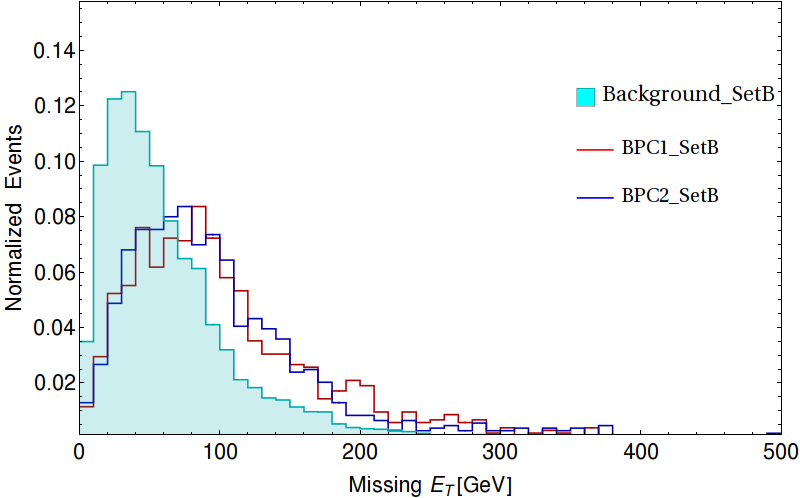}
 $$
  \caption{\it Normalized distributions of missing transverse energy ($\slashed{E_T}$) for both SetA [left] and SetB [right]
  signal events and the total SM background.}
  \label{fig:4l_ss}
 \end{figure}

\begin{figure}[htb!]
 $$
 \includegraphics[scale=0.30]{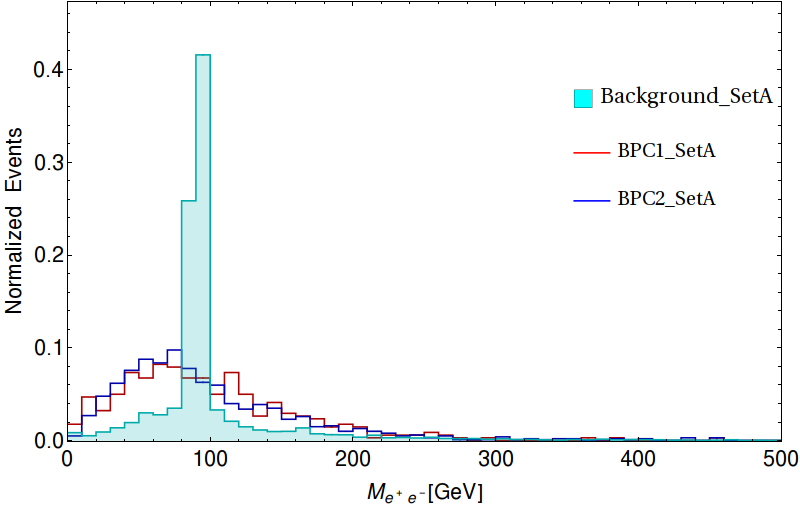}
 \includegraphics[scale=0.30]{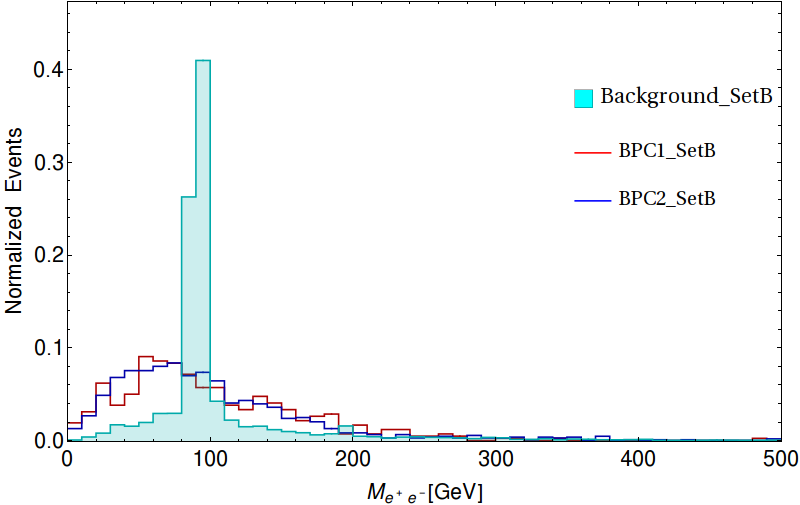}
 $$
  \caption{\it Normalized distributions of invariant mass ($M_{e^+ e^-}$) for both SetA [left] and SetB [right]
  signal events and the total SM background.}
  \label{fig:4l_ss}
 \end{figure}

\begin{figure}[htb!]
 $$
 \includegraphics[scale=0.30]{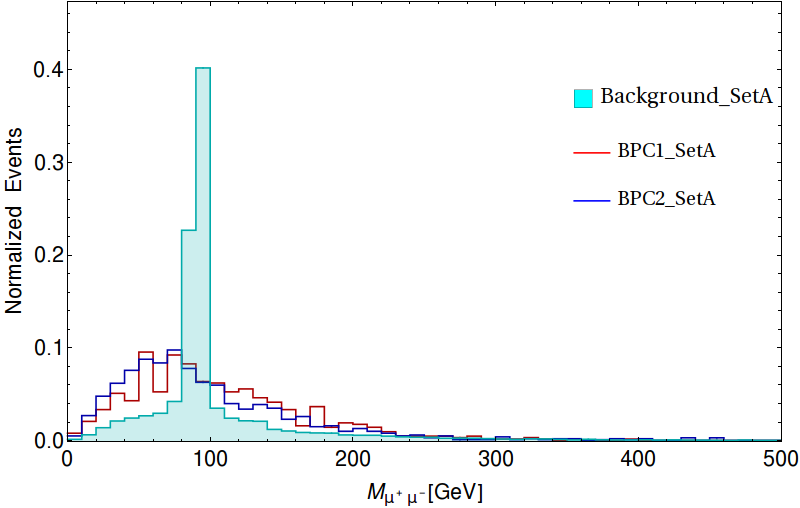}
 \includegraphics[scale=0.30]{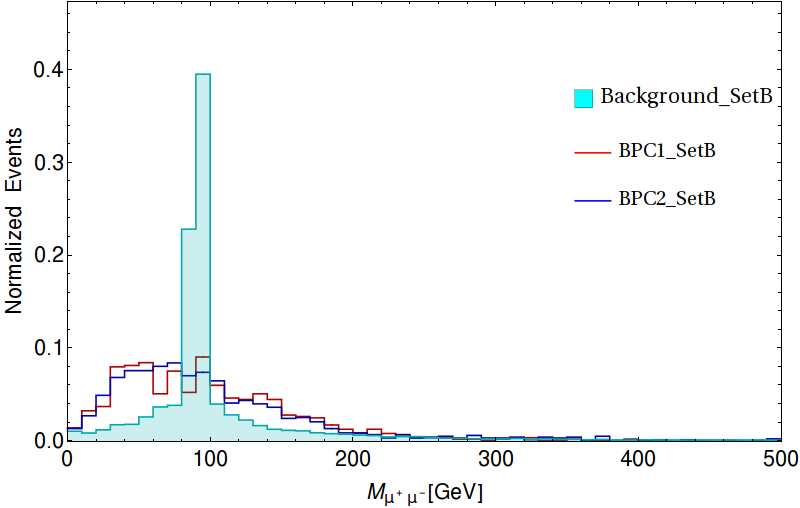}
 $$
  \caption{\it Normalized distributions of invariant mass ($M_{\mu^+ \mu^-}$) for both SetA [left] and SetB [right]
  signal events and the total SM background.}
  \label{fig:4l_ss}
 \end{figure}

\begin{figure}[htb!]
 $$
 \includegraphics[scale=0.30]{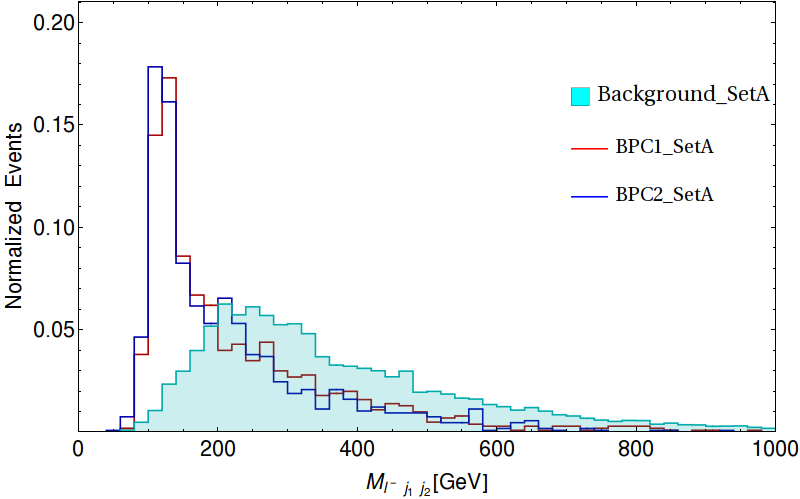}
 \includegraphics[scale=0.30]{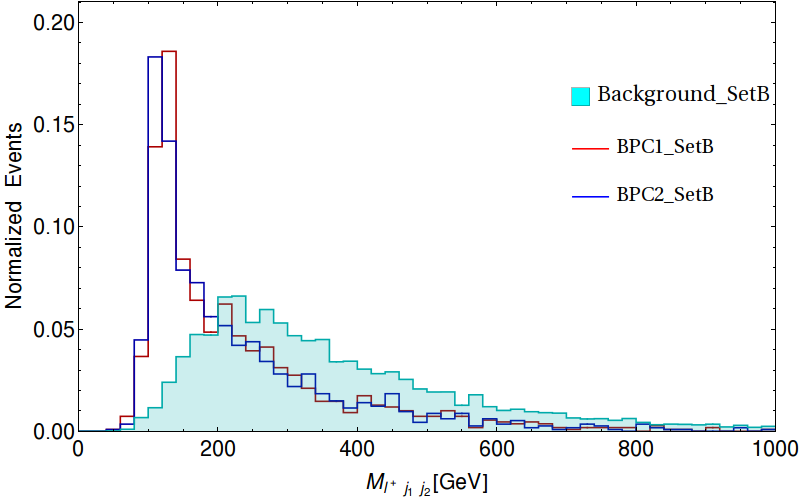}
 $$
  \caption{\it Normalized distributions of invariant mass of lepton-jets configurations ($M_{l^- j_1 j_2}$)
  for both SetA [left] and SetB [right] signal events and the total SM background.}
  \label{fig:4l_ss}
 \end{figure}

\begin{figure}[htb!]
 $$
 \includegraphics[scale=0.30]{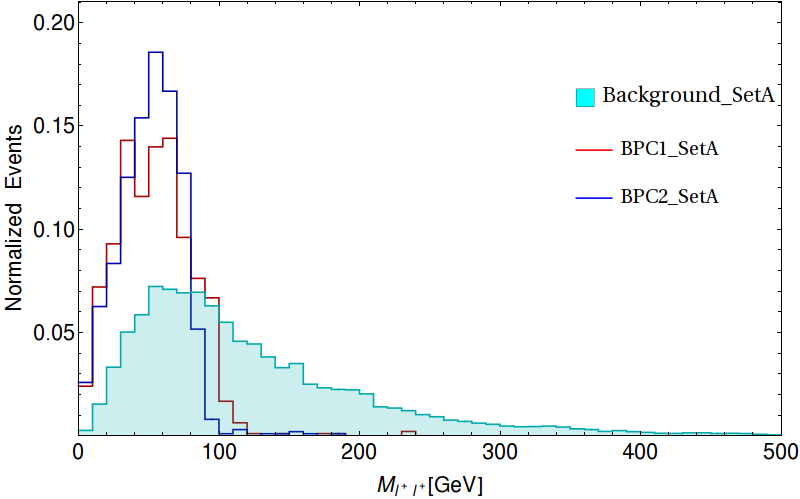}
 \includegraphics[scale=0.30]{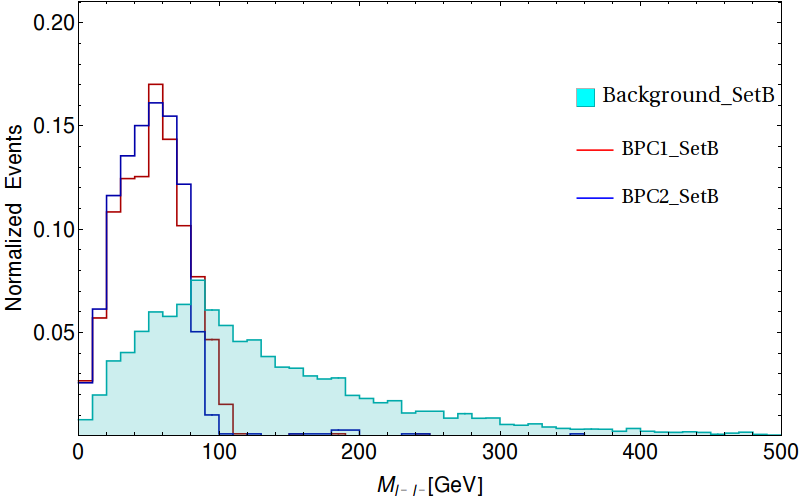}
 $$
  \caption{\it Normalized distributions of invariant mass of same sign leptons ($M_{l_i^+l_j^+}$) 
  for both SetA [left] and SetB [right] signal events and the total SM background.}
  \label{fig:4l_ss}
 \end{figure} 

\begin{figure}[htb!]
 $$
 \includegraphics[scale=0.30]{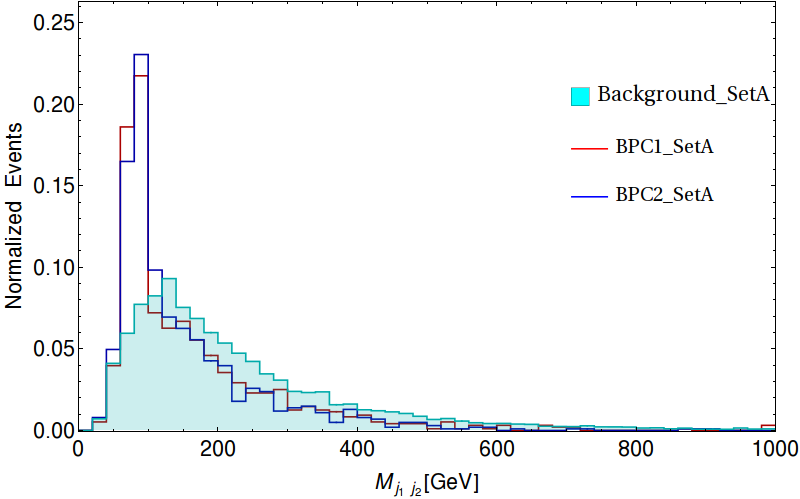}
 \includegraphics[scale=0.30]{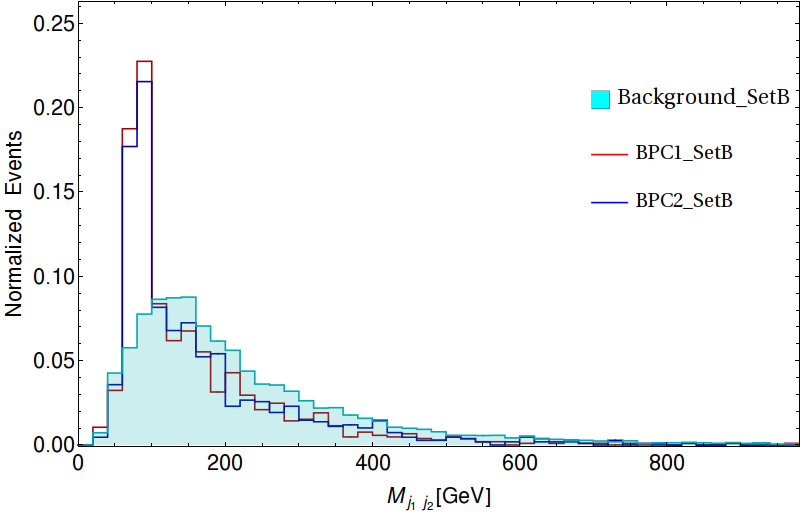}
 $$
  \caption{\it Normalized distributions of invariant mass of jet pairs ($M_{j_1j_2}$) for 
  both SetA [left] and SetB [right] signal events and the total SM background.}
   \end{figure}

\begin{table}[h]
\resizebox{\linewidth}{!}{
 \begin{tabular}{|c|c|c|c|c|c|c|}
\hline
 Cuts (GeV) & $\slashed{E_T}<30$  & $ 75<M_{e^+e^-}<100$ &$ 80<M_{\mu^+\mu^-}<100$ 
 &$M_{l^\pm j_1 j_2}>130$ & SetA: \, $M_{l_i^+ l_j^+}>100$ & SetA: \, $65<M_{j_1 j_2}<90$ \\
& & & & &SetB: \, $M_{l_i^- l_j^-}> \,\, 90$ & SetB: \, $60<M_{j_1 j_2}<90$ \\ \hline\hline
BPC1&28.5&26.3&23.0&7.5&7.0&5.5\\ \hline
BPC2&31.9&29.6&25.7&9.2&9.1&6.5\\ \hline
Background&40395&26293&3974&56.7&24.1&1.7\\ 

\hline
\end{tabular}
}
\caption{\it Rejection (selection for final cut) cut-flow chart of $3l+2j+\slashed{E_T}$ signal analysis for BPC1 and BPC2 
benchmark points at 1000 fb$^{-1}$ luminosity.}
    \label{tab:tab90}
\end{table}
At this stage the background coming from $WZ$ and $t\bar{t}$ is massive compared to other background sources and signal. Hence to suppress these two, we divide our 
event data set for both signal and background in two mutually exclusive sets.\\
\textbf{SetA :} This corresponds to events which have two positively charged leptons 
and one negatively charged lepton. \\
\textbf{SetB :} This set consists of events with one positively charged lepton and two negatively charged leptons.

\begin{itemize}
\item To reduce background sources which do (may) not have any $\slashed{E_T}$, such as $ZZ$ and $VVV$ in contrast 
to our signal which has both neutrinos and dark matter as source for missing transverse energy, we use a selection cut of 
$\slashed{E_T}>30$~GeV. We also use as before, a rejection cut on the invariant mass 
of opposite sign same flavor lepton pairs near the $Z$ mass pole to further suppress the background events.
\item In our signal two same sign leptons come from same parent doubly charged 
Higgs via the $W$ bosons while the other doubly charged Higgs gives the opposite 
sign lepton and two jets in its decay chain. Hence a strong correlation in invariant mass shows up for decay products of these two doubly charged Higgs. So we can put a 
rejection cut on SetA, with $M(l^+ l^+)$ , $M(l^- j_1 j_2)$ $ \gtrsim (m_{H^{\pm\pm}} - 2 m_{DM})$ and similarly cuts with leptons of opposite polarity on SetB. This cut is very effective in reducing the background coming from $t\bar{t}$ and $WZ$.  

\item At this stage we still have significant amount of background events left, mostly from the $WZ$+jets subprocess, 
because of it's huge cross-section compared to the signal and other background 
subprocesses. Since our signal gets two primary jets from the decay of an onshell $W$  while     
the jets coming in $WZ$ are most likely radiative jets as $W$ and $Z$ both decay to leptons, we use a selection cut on the dijet invariant mass around the $W$ mass pole which helps reduce the $WZ$ background.
\end{itemize}
With the above mentioned cuts, shown in Table\ref{tab:tab90}, and using Eqn.\ref{eq:sign} 
for the signal significance we show our results for the two benchmark points in  
Table\ref{tab:tab100}.

\begin{table}[h]
\centering
\resizebox{7 cm}{!}{
 \begin{tabular}{|c|c|c|c|}
\hline
 Benchmark & Signal & Background &Significance  \\ \hline\hline
BPC1&5.5&1.7&3.13\\ \hline

BPC2&6.5&1.7&3.59\\ 

\hline
\end{tabular}
}
\caption{\it Significance of BPC1 and BPC2 for $3l + 2j+\slashed{E_T}$ signal at integrated luminosity of 1000 fb$^{-1}$}
    \label{tab:tab100}
\end{table}

We therefore conclude that with high enough integrated luminosity we can discover a doubly charged Higgs of 
mass around 300 GeV in the multi-lepton final state with at least $N_l=3$ leptons. As the lepton multiplicity decreases 
we find that the large SM background is more difficult to suppress and give enough sensitivity to observe a doubly 
charged Higgs decaying to the dark fermions. 
In addition, a similar analysis of $4l$ and $3l$ final states for the third benchmark point (BPC3) shown in 
Table\ref{tab:tabbp}, which represents a compressed mass spectrum for the lighter states in the dark fermion sector, 
yields a very low ($< 1\sigma$) signal significance for a 300 GeV doubly charged scalar. The 3-body decay of the 
$\psi^\pm$ leads to softer final states which make it more difficult to distinguish from the SM background leading to 
lower signal sensitivity.

\section{Signal comparison at ILC}
\label{sec:coll1}
It is quite clear from the analysis shown in the previous section that signals for a doubly charged Higgs become difficult to 
observe at LHC if they do not decay directly to charged leptons. As is the case for the $H^{\pm\pm}$ of Type-II seesaw model
where the doubly charged scalar has very limited sensitivity if it decays dominantly to a pair of $W$ bosons and where the 
current limits from LHC give a mass bound as low as between 230-350 GeV \cite{ATLAS:2021jol}, the modified decay modes in our model
lead to a much weaker sensitivity at current integrated luminosities. Even with the full high luminosity LHC (HL-LHC), we find that
a discovery of such a doubly charged scalar would still be limited to sub-400 GeV masses. It would therefore seem that 
while the LHC energies would probe a much higher energy scale of models such as LRSM and restrict very heavy $W_R$
and $Z_R$, it would lack in efficiency for these doubly charged exotics. It would be interesting to find out the sensitivity 
for such particles at the proposed ILC which may be restricted by its energy reach but would prove beneficial for such particles
in general which become more elusive at LHC as they develop newer decay channels in their fold. We choose to highlight 
just a simple comparison with one of the signals studied at the LHC here and leave a more dedicated ILC study for later work.
We however present a slight variation in the spectrum to include a compressed scenario, which has very clear challenges in LHC searches.   

\subsection{$3l+2j+\slashed{E}$ signal analysis for BPC3 at ILC}

The benchmark point BPC3 is chosen as it represent a scenario where the mass gap between $\chi^{\pm}_{1}$ and 
dark matter ($\chi_{_1}$) is less than the mass of $W$  such that the decay $\chi^{\pm}_{1} \rightarrow W^\pm \, \chi_{_1}$ is energetically forbidden. Here $\chi^{+}_1$ has a three-body decay to $\l_i^{+} \, \nu_j \, \chi_{_1}$ or $ 2j \, \chi_{_1}$. In 
this case the leptons and jets will be soft  and one can no longer put a strong requirement on the $p_T$ of jets as 
required to avoid large hadronic debris that can affect any analysis at LHC. This makes an analysis for such a 
compressed spectrum leading to soft final states at LHC very challenging. Since ILC has a much cleaner 
environment the jets can be triggered upon with much lower energies and we can put weaker jet tagging conditions 
($P_T(j)>10$ GeV), better suited for benchmarks such as BPC3.
\par For analyzing this signal at ILC we consider the dominant background coming from $WWZ$ which gives us 
$3l+2j+\slashed{E}$ final state where one $W$ decays hadronically while the other $W$ and $Z$ decay leptonically producing 
three leptons and missing energy. As before, our signal comes from $H^{++}_L H^{--}_L$ pair production via photon or 
$Z$ mediator (the heavy $Z_R$ contribution is negligible).  Each $H^{\pm\pm}_L$ then decays to two $\chi^{\pm}_{1}$  
followed by one $\chi^{\pm}_{1}$ decaying to $\chi_{_1}$ and $2~j$ and rest of the dark charged fermion decaying leptonically
via the $W$ boson. For the analysis we choose only those events as signals which have exactly three charged 
leptons (i.e., $e$ and $\mu$) and exactly two jets. 
The basic acceptance cuts for all isolated objects (i.e., $\Delta R_{ab} >0.2$) are chosen similar to that for LHC.

\begin{table}[h]
\resizebox{\linewidth}{!}{
 \begin{tabular}{|c|c|c|c|c|c|}
\hline
 Cuts (GeV) &  $ 85<M_{e^+e^-}<95$ &$ 85<M_{\mu^+\mu^-}<95$ &$M_{l_i^\pm l_j^\pm}>80$
 &$M_{l_3 j_1 j_2}>110$ & significance\\ \hline\hline
BPC3&4.4&4.2&4.2&4.1&4.82\\ \hline

Background&143.7&31.3&3.8&0.1&\\ 

\hline
\end{tabular}
}
\caption{\it Rejection cut-flow chart of $3l+2j+\slashed{E}$ signal analysis for BPC3 benchmark points at 
ILC with $\sqrt{s}=1$ TeV with 1000 $fb^{-1}$ integrated luminosity.}
    \label{tab:tab11}
\end{table}

We list the selection cuts for the signal and background events in Table\ref{tab:tab11} along with the events surviving each cut.
Here the first cut is on invariant mass of opposite sign same flavor lepton at $Z$ mass pole which suppresses the 
background significantly because $WWZ$ has one $Z$ always decaying leptonically. The next cut is on the invariant mass of 
SS leptons which is a prominent characteristic of the signal (as discussed earlier) since these lepton pairs come from 
the decay chain of the same doubly charged Higgs. The final cut is on the invariant mass of visible decay product of one 
doubly charged Higgs where one $\chi^{\pm}_{1}$ decays hadronically while the other decays to give a charged lepton. 

We find that for a mass of $H^{++}$ similar to BPC1 and 
a compressed spectrum in the dark sector which gives a 3-body decay for the charged dark fermion, HL-LHC gives
a signal sensitivity of less than $1\sigma$ for BPC3 while the same final state is able to achieve a much higher sensitivity 
at ILC with $\sqrt{s}=1$ TeV for a much smaller luminosity, and gives a $ \gtrsim 4.5\sigma$ signal with an integrated luminosity of 1000 fb$^{-1}$. In fact, the cleaner environment at ILC would open up the same benchmark for a hadronic rich final state with much 
larger effective cross section. As a multi-particle final state with hadrons improves the size of signal events and one can effectively 
control the background easily since all processes will be through electroweak interactions, ILC will give a much higher 
significance in the final states with smaller lepton multiplicity. The same-sign lepton pairs however provide a symbolism of the
produced doubly charged scalar and therefore to establish its presence, the channel with $2 \ell^{\pm} \, 2j \, \slashed{E}$ 
would be a more appropriate channel for the study at ILC.

\section{Conclusions}
\label{sec:concl}
In this work we have used the well motivated left-right symmetry model and invoke a dark matter candidate 
in the model by including a pair of vector-like lepton doublets which preserve the left-right symmetry but 
are odd under a discrete $\mathcal{Z}_2$ symmetry. This extended model therefore has its own dark sector which
speaks with the SM matter fields via gauge and Yukawa interactions. The dark fermions of the model are neutral 
and some also carry electric charge, where the lightest neutral state which is an admixture of the neutral 
components of the two VL doublets after symmetry breaking, acts as the DM candidate. The model 
gives the DM several characteristics to choose from depending upon its composition and the type 
of gauge interactions it prefers. This opens up an interesting DM phenomenology where the different
composition of the DM will lead to different regions of parameter space that satisfy the 
observed DM experimental data in the form of relic density, direct-detection experiments and 
indirect detection experiments. We explore the available parameter space of the model  in all different 
scenarios where the DM is dominantly $SU(2)_L$ like, $SU(2)_R$ like or an admixture of all 
chiralities of the VL doublet pair in the LRSM model. We show the region of parameter space of the mixed DM scenario
which is consistent with relic density observations and also satisfy direct detection constraints as well as 
indirect detection constraints. 

The dark sector also contains charged states which on one hand can play a major role in DM phenomenology by 
contributing to the number density through co-annihilations with the DM when the mass splitting is 
very low between them, while they could be directly produced at experiments through gauge interactions. Their 
phenomenology would be very similar to a pair produced VLL that decays to a $W$ boson and a DM. We
are however more interested in the signal for the more unique doubly charged Higgs present in the model in the
presence of a dark fermion sector which couples to it directly. As in the case of the lepton doublets having a 
Majorana interaction with the triplet scalars of the model leading to a seesaw mechanism for neutrinos, the 
dark fermion sector would also have a similar seesaw mechanism. Thus in a significant region of parameter 
space consistent with DM observations, the doubly charged scalar can decay to the pair of charged dark fermions.
We consider this interesting possibility in our work and perform a detailed collider analysis of its signal at the LHC 
in multi-lepton final state. We find that the bounds on the doubly charged scalar become weaker compared to the 
more standard leptonic and bosonic decay modes and LHC sensitivity for a doubly charged Higgs in such case 
would be sub-400 GeV even with an integrated luminosity of 3000 fb$^{-1}$. We then show how this reach can 
be improved at the proposed ILC with a center of mass energy of 1 TeV by comparing a similar final state 
which gives less than $1\sigma$ sensitivity at LHC but an improved $\simeq 4.8 \sigma$ sensitivity at ILC.   
\section*{Acknowledgments}

The authors would like to acknowledge the support from DAE, India for the Regional Centre for Accelerator based Particle Physics (RECAPP), Harish Chandra Research Institute.


\bibliographystyle{JHEP}
\bibliography{Ref}

\end{document}